\documentclass[aps,amsmath,amssymb,prb,10pt,showpacs,twocolumn,letterpaper,superscriptaddress]{revtex4-1}
\usepackage{graphicx,color,tcolorbox}
\usepackage{bm}
\usepackage{mathrsfs}
\usepackage{comment}
\usepackage{mathtools}
\usepackage{hyperref}
\usepackage{epstopdf}
\usepackage{letltxmacro}
\usepackage{appendix}
\mathtoolsset{showonlyrefs}
\allowdisplaybreaks

\newcommand{\al}{\alpha}
\newcommand{\be}{\beta}
\newcommand{\g}{\gamma}
\newcommand{\de}{\delta}
\newcommand{\e}{\epsilon}
\newcommand{\z}{\zeta}

\newcommand{\thi}{\theta}
\newcommand{\io}{{\rm i}}
\newcommand{\ka}{\kappa}
\newcommand{\la}{\lambda}
\newcommand{\La}{\Lambda}

\newcommand{\p}{\pi}

\newcommand{\s}{\sigma}
\newcommand{\y}{\upsilon}
\newcommand{\f}{\phi}
\newcommand{\x}{\chi}
\newcommand{\w}{\omega}
\newcommand{\W}{\Omega}
\newcommand{\De}{\Delta}
\newcommand{\G}{\Gamma}
\renewcommand{\S}{\Sigma}

\renewcommand{\y}{\psi}

\newcommand{\pd}{\partial}

\newcommand{\round}[1]{\left( #1 \right)}
\renewcommand{\square}[1]{\left[ #1 \right]}
\newcommand{\curly}[1]{\left\{#1\right\}}
\newcommand{\abs}[1]{\left| #1 \right|}
\newcommand{\cvec}[2]{\round{\begin{array}{c} #1 \\ #2 \end{array}}}

\newcommand{\mat}[4]{\left(\begin{array}{cc}#1&#2\\#3&#4\end{array}\right)}

\newcommand{\rvec}[2]{\round{\begin{array}{cc}#1&#2\end{array}}}

\newcommand{\ang}[1]{\left\langle #1 \right\rangle}
\newcommand{\sang}[1]{\langle #1 \rangle}

\newcommand{\ket}[1]{\left| #1\right\rangle}

\newcommand{\beq}{\begin{equation}}
\newcommand{\eeq}{\end{equation}}
\newcommand{\Beq}{\begin{eqnarray}}
\newcommand{\Eeq}{\end{eqnarray}}
\newcommand{\bml}{\begin{multline}}

\newcommand{\bea}{\begin{align}}
\newcommand{\ena}{\end{align}}
\newcommand{\bsp}{\begin{split}}
\newcommand{\esp}{\end{split}}
\newcommand{\down}{\downarrow}
\newcommand{\up}{\uparrow}

\newcommand{\tW}{\tilde\W}

\newcommand{\tA}{{\tilde A}}

\newcommand{\br}{{\boldsymbol r}}
\newcommand{\brho}{{\boldsymbol\rho}}

\newcommand{\bS}{{\boldsymbol{S}}}

\newcommand{\bp}{{\boldsymbol p}}

\newcommand{\ez}{\hat{\boldsymbol z}}

\newcommand{\bj}{{\boldsymbol j}}
\newcommand{\bJ}{{\boldsymbol J}}
\newcommand{\bv}{{\boldsymbol v}}

\newcommand{\bk}{{\boldsymbol k}}

\newcommand{\bq}{{\boldsymbol q}}

\newcommand{\ba}{{\bf a}}
\renewcommand{\a}{{\rm a}}

\newcommand{\bB}{{\boldsymbol B}}
\newcommand{\bb}{{\boldsymbol b}}

\DeclareMathOperator{\sgn}{sgn}

\newcommand{\sA}{\mathscr{A}}

\newcommand{\hH}{\hat{H}}
\newcommand{\bH}{\boldsymbol{H}}

\newcommand{\tf}{\tilde{\phi}}
\newcommand{\ty}{{\tilde y}}

\newcommand{\tz}{{\tilde z}}

\newcommand{\bx}{\boldsymbol{x}}

\newcommand{\ve}{\varepsilon}
\newcommand{\bs}{{\boldsymbol{s}}}
\newcommand{\bnab}{{\boldsymbol\nabla}}

\newcommand{\bzero}{\boldsymbol{0}}

\newcommand{\tg}{\tilde{\g}}

\begin{document}
\title{Detecting Fractionalization in Critical Spin Liquids using Color Centers}

\date{\today}
\author{So Takei}
\affiliation{Department of Physics, Queens College of the City University of New York, Queens, New York 11367, USA}
\affiliation{Physics Doctoral Program, The Graduate Center of the City University of New York, New York, New York 10016, USA}
\author{Yaroslav Tserkovnyak}
\affiliation{Department of Physics and Astronomy and Bhaumik Institute for Theoretical Physics, University of California, Los Angeles, California 90095, USA}

\begin{abstract}
Quantum spin liquids are highly entangled ground states of insulating spin systems, in which magnetic ordering is prevented down to the lowest temperatures due to quantum fluctuations. One of the most extraordinary characteristics of quantum spin liquid phases is their ability to support fractionalized, low-energy quasiparticles known as spinons, which carry spin-1/2 but bear no charge. Relaxometry based on color centers in crystalline materials | of which nitrogen-vacancy (NV) centers in diamond are a well-explored example | provides an exciting new platform to probe the spin spectral functions of magnetic materials with both energy and momentum resolution and to search for signatures of these elusive, fractionalized excitations. In this work, we theoretically investigate the color-center relaxometry of two archetypal quantum spin liquids: the two-dimensional U(1) quantum spin liquid with a spinon Fermi surface and the spin-1/2 antiferromagnetic spin chain. The former is characterized by a metallic, spin-split ground state of mobile, interacting spinons, which closely resembles a spin-polarized Fermi liquid ground state but with neutral quasiparticles. We show that the observation of the Stoner continuum and the collective spin wave mode in the spin spectral function would provide a strong evidence for the existence of spinons and fractionalization. In one dimension, mobile spinons form a Luttinger liquid ground state. We show that the spin spectral function exhibits strong features representing the collective density and spin-wave modes, which are broadened in an algebraic fashion with an exponent characterized by the Luttinger parameter. The possibilities of measuring these collective modes and detecting the power-law decay of the spectral weight using NV relaxometry are discussed. We also examine how the transition rates are modified by marginally irrelevant operators in the Heisenberg limit. 


\end{abstract}
\maketitle

\section{Introduction}
Quantum spin liquid (QSL) phases are ground states of certain Mott insulators in which strong quantum fluctuations prevent magnetic ordering down to zero temperature.\cite{savaryRPR17,zhouRMP17,knolleAR19,broholmSCI20} These phases are distinguished among themselves by different quantum orders characterizing their many-body quantum entanglement as opposed to which space-time or spin-rotational symmetries they break.\cite{wenPRB02,wenBOOK07} One of the most striking implications of this quantum non-locality is fractionalization: it is a QSL's ability to support sharp, low-energy excitations that behave as ``fractions" of an electron, even though the physical electrons that form these phases are robust against such splintering. Fractionalization is one of the defining characteristics of QSL phases,\cite{balentsNAT10} and identifying experimental signatures of the phenomenon is indispensable for the discovery of these elusive phases. 

The most prominent examples of these fractionalized excitations are spinons. In certain half-filled Mott insulators, these spinons emerge as gapless, mobile fermionic quasiparticles | akin to conventional Landau quasiparticles | that can transport heat but carry no charge. They also carry spin-$1/2$, in spite of the fact that each site of the lattice contains a single localized electron, in which case any local spin excitation corresponds to a full spin-$1$ insertion. In the one-dimensional (1D) spin-1/2 antiferromagnetic spin chain, such spinons arise as domain walls separating two different configurations of antiferromagnetic order.\cite{affleckBOOK88,mikeskaBOOK04,giamarchiBOOK04,gogolinBOOK04} 
In two dimensions, these spinons may emerge in Mott insulators in the vicinity of the metal-insulator transition.\cite{yamashitaNATP09,yamashitaSCI10}  A canonical model for these so-called ``weak" Mott insulators involves a half-filled, single-band Hubbard model on the triangular lattice with a relatively small Mott gap.  Enhanced charge fluctuations stemming from the small charge gap have been shown to stabilize a QSL state, the low-energy model of which is predicted to involve a Fermi surface of spinons coupled to a fluctuating U(1) gauge field.\cite{leePRL05,motrunichPRB05} Measurements supporting this prediction have been reported on organic salt compounds,\cite{shimizuPRL03,itouPRB08} as well as the triangular antiferromagnet YbMgGaO$_4$.\cite{shenNAT16} 

A promising spin-sensitive probe that may shed fresh light on these exotic quasiparticles is relaxometry based on color center defects in crystalline materials. Of all the color centers that are suggested to exist in various wide band-gap materials,\cite{aharonovichNATP16} one prominent example is nitrogen-vacancy (NV) defects in diamond. With a broad operational regime (temperatures from a few Kelvin to above room temperature and magnetic fields from zero to a few Tesla), a dynamic frequency range up to a few hundred GHz, and nanoscale spatial resolution, NV relaxometry offers unrivaled versatility in the field of magnetic sensing.\cite{rondinRPP14,degenRMP17,casolaNRM18} The technique involves placing an NV defect near the surface of a magnetic material and measuring the relaxation rate of the $S=1$ spin localized on the defect. The rate is sensitive to the magnetic-field power spectral density at the defect site, which in turn depends on the imaginary part of the dynamic spin response function | the spin spectral function | of the magnetic material. Measuring the relaxation rates thus allows one to extract the material's spectral properties.

One notable feature of NV relaxometry is that it probes the response function at a specific probe (or ``ESR") frequency, which is determined by an external bias magnetic field. The technique thus functions as a field-tuned spectrometer of spin fluctuations and is excellently well-suited for evincing the spectral properties of QSL materials under the influence of a magnetic field. 

Another attractive feature is that the relaxation rates are typically most sensitive to spin fluctuations at wavevectors $q\sim d^{-1}$, where $d$ is the defect-sample distance. Therefore, by tuning $d$ and the external field, NV relaxometry can be used to quantify the spin spectral function with both energy {\em and} momentum resolution. 

Recent relaxometry measurements on a ferrimagnetic insulator Y$_3$Fe$_5$O$_{12}$ have achieved energy resolution up to $\sim10^{-6}$~K and wavevector resolution up to $\sim10^3$~m$^{-1}$ for magnons with wavevectors $q\sim10^7$~m$^{-1}$.\cite{leewongNL20} These measurements have also demonstrated the possibility of detecting magnon excitations with wavevectors up to $5\times10^7$~m$^{-1}$; this maximum wavevector lies outside of the accessible wavevector range of current FMR spectroscopy and spin-pumping techniques\cite{sandwegPRL11} and approaches the measurement limit of Brillouin light scattering.\cite{anPRL16,holandaNATP18} The possibility of NV spin sensors to scan the spin spectral landscape over a wide range of frequencies and wavevectors with energy and momentum resolution makes it a highly desirable probe of fractionalization in QSL phases. 

\begin{figure*}[t]
\includegraphics[width=0.95\linewidth]{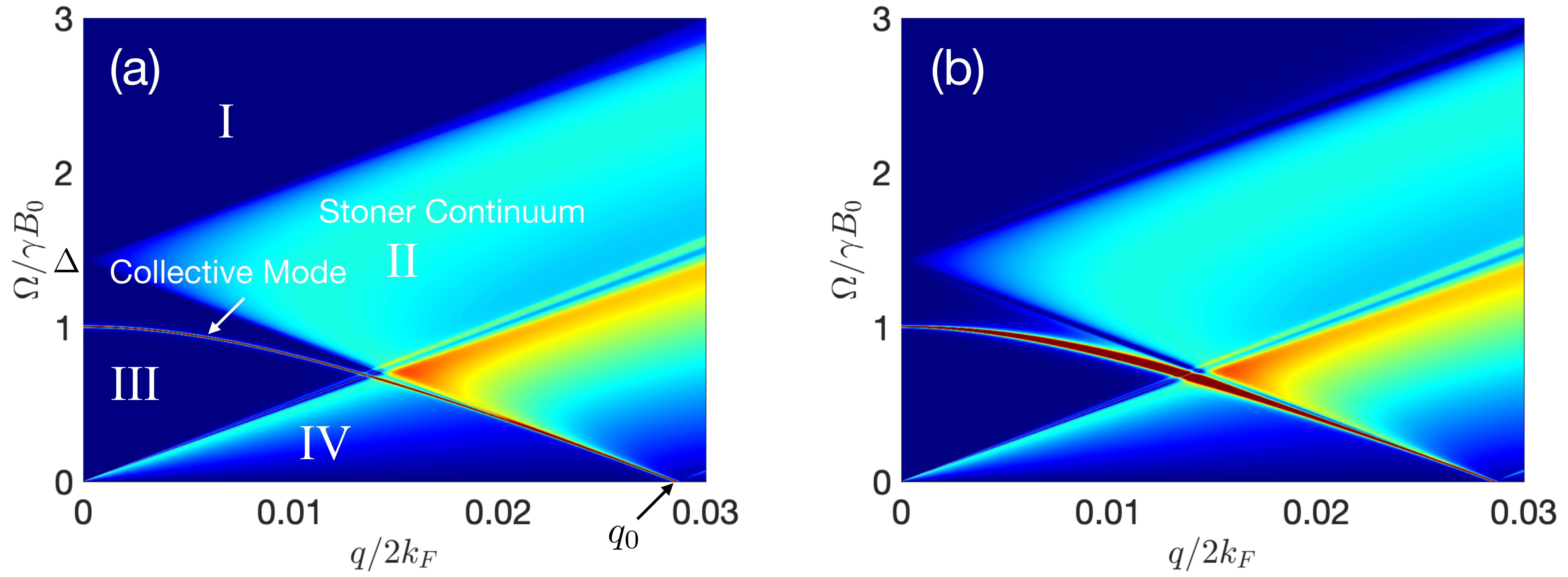}
\caption{Plot of $-{\rm Im}\{\chi^{R}_\perp(q,\W)+\chi^{R}_\parallel(q,\W)\}$ ($q\equiv|\bq|$) without (a) and with (b) corrections due to the fluctuationg U(1) gauge field. The low-frequency spectral weight, i.e., Region IV, comes from the longitudinal component $\chi^{R}_\parallel(q,\W)$, while the Stoner continuum (Region II) and the collective spin wave mode arise from the transverse component $\chi^{R}_\perp(q,\W)$. Correlation strength $\xi\equiv ug_0=0.3$ is used, and $q_0$ is the wavevector at which the lower boundary of the Stoner continuum reaches zero frequency. For the definitions of $\chi^{R}_\perp(q,\W)$ and $\chi^{R}_\parallel(q,\W)$, see Eq.~\eqref{tvlongchir}, and for the definitions of the variables, see Sec.~\ref{rates2d}.}
\label{fig1}
\end{figure*}
Motivated by these latest developments, we theoretically examine the NV relaxometry of two representative QSL phases | the spin-1/2 antiferromagnetic quantum spin chain\cite{mikeskaBOOK04,giamarchiBOOK04} and the 2D QSL with a spinon Fermi surface coupled to a U(1) gauge field.\cite{leePRL05,motrunichPRB05,savaryRPR17,zhouRMP17} With the rising possibility of utilizing NV spin sensors for energy- and momentum-resolved  spectroscopy, our focus will be on the spin spectral functions of these QSL phases and to examine how fractionalization manifests itself in these functions and ultimately in the NV relaxation rates.\cite{dohertyPR13,grinoldsNATP13,rondinRPP14} The ESR frequency of the NV spin is determined by an externally applied bias magnetic field. We therefore evaluate the spectral functions as a function of this bias field by taking account of the effects of this field on the QSLs. We also incorporate the fact that 1D and 2D QSLs are typically embedded in 3D materials. For the 2D QSL, we consider stacking identical 2D QSL layers in the third direction and compute the relaxation rates due to this layered structure. For the 1D case, we compute the rates for a 3D stack of quantum spin chains in which non-local spin correlations extend only along one spatial direction.

\subsection*{Summary: Signatures of Fractionalization in NV Relaxometry}
Before delving into the technical details, we begin by describing the salient signatures of fractionalization in the spin spectral function. This brief summary focuses  on the 2D QSL with a spinon Fermi surface, as the qualitative features in the 2D case carry over to the 1D case. 

In this work, we make some simplifying assumptions that should not impact our main findings qualitatively. First, we do not explicitly take account of the potential due to the background lattice and use effective long-wavelength theories to calculate spin spectral functions. Second, we assume that the quantization axis of the NV spin and the external bias field are collinear to each other and that they are both normal to the magnetic film. If we further assume that the quantum magnet possesses uniaxial spin-rotational symmetry about this axis, NV spin relaxometry effectively probes the imaginary part of the sum of the transverse and longitudinal components of the dynamic spin response function, i.e., ${\rm Im}\{\chi^{R}_\perp(\bq,\W)+\chi^{R}_\parallel(\bq,\W)\}$, where
\beq
\chi^{R}_\perp\equiv\tfrac14(\chi^{R}_{+-}+\chi^{R}_{-+})\,,\ \ \ \chi^{R}_\parallel\equiv\chi^{R}_{zz}\,,
\label{tvlongchir}
\eeq
and the response function reads
\begin{align}
\label{chirs}
\chi^R_{\al\be}(\bq,\W)&=\int dt\int d^2\br\,\chi^R_{\al\be}(\br,t)e^{-i\bq\cdot\br+i\W t}\,,\\
\chi^{R}_{\al\be}(\br,t)&=-i\Theta(t)\sang{[s_{\al}(\br,t),s_{\be}(\bzero,0)]}\,.
\end{align}
Here, $\al,\be=x,y,z,+,-$ label the spin components, $s_\al(\br,t)$ denotes the local spin density in the QSL, and $\br=(x,y)$ and $\bq=(q_x,q_y)$, respectively, denote the position and wavevector within the magnetic plane. 

The low-energy model of the 2D QSL is a ``Fermi liquid" of spinons | spin-1/2, charge-neutral fermionic quasiparticles | coupled to a fluctuating U(1) gauge field, the mathematical structure of which arises in several different contexts of condensed matter physics including the half-filled Landau level\cite{halperinPRB93} and non-Fermi liquid metals\cite{leePRL89,galitskiPRL05,kaulNATP08}. Let us first consider the spin response in the absence of the gauge field. The fractionalization of spin-1 (magnon) excitations into two spinons converts Eq.~\eqref{chirs} into a {\em two-particle} retarded correlation function of spinons. In the presence of the bias magnetic field, ${\rm Im}\{\chi^R_{\perp}(\bq,\W)\}$ then exhibits a two-particle continuum spectral weight, known as the Stoner continuum, representing the kinematically allowed spin-1 particle-hole excitations. Figure~\ref{fig1}(a) is a plot of $-{\rm Im}\{\chi^{R}_\perp(\bq,\W)+\chi^{R}_\parallel(\bq,\W)\}$ computed in the absence of the gauge fluctuations using the time-dependent Hartree-Fock approximation (details are provided in Sec.~\ref{rates2d}); the Stoner continuum is given by the ``fan" region labeled ``II." 

The pole of the transverse component $\chi^{R}_\perp(\bq,\W)$ defines a collective spin-1 mode, i.e., the spin wave mode, of the QSL that appears as a sharp spectral weight below the Stoner continuum and disperses down [see Fig.~\ref{fig1}(a)]. This spin wave mode emerges exactly at the Zeeman energy $\hbar\g B_0$ at $q=0$ (with $\g$ being the gyromagnetic ratio of the QSL and $B_0$ being the bias magnetic field) and remains sharp as long as the damping mechanisms, such as U(1) gauge field fluctuations, are absent. The spin wave mode at $q=0$ describes a uniform spin precession at the Larmor frequency $\g B_0$, a result that is in accordance with the Larmor theorem,\cite{oshikawaPRB02} which states that the only response of any spin system with SU(2) symmetry at zero field is at the Larmor frequency. As we later show, the inclusion of (repulsive) spinon interactions via the Hartree-Fock approximation leads to the detuning of the vertex of the Stoner continuum away from the Larmor frequency, as seen by the detuning of $\De$ away from $\g B_0$ in Fig.~\ref{fig1}(a).

It turns out that the dynamic U(1) gauge field in the 2D QSL has a nearly flat band; therefore, this field acts as an effective momentum sink for the spinons. The addition of the gauge fiuctuations thus generally leads to the smoothening out of the sharp features in Fig.~\ref{fig1}(a) [see Fig.~\ref{fig1}(b)]. The most notable effect is the broadening of the spin wave mode due to the gauge field, i.e., the gauge fluctuations introduce a lifetime to the spin wave mode. This broadening, however, vanishes as $q\rightarrow0$ so that the mode becomes sharp in the uniform limit, in accordance with the Larmor theorem.

The spin wave mode discussed here was studied in the context of a conventional Fermi liquid many decades ago.\cite{silinJETP58,platzmanPRL67} Unlike zero sound, which is undamped at low energies, the collective spin wave mode is typically overdamped by the particle-hole continuum at zero field. However, a finite magnetic field shifts both the continuum and the mode up to Zeeman energy, and the inclusion of repulsive quasiparticle interactions shifts the continuum further up in energy [e.g., up to $\De$ in Fig.~\ref{fig1}(a)]. An important difference between conventional Fermi liquids and the 2D QSL is that in the former case, the Zeeman energy $\hbar\g B_0$ relative to the Fermi energy is essentially zero, while in the latter case the Fermi energy is determined by the exchange constant $J$, so $\hbar\g B_0$ could be a significant fraction of $\ve_F$. As the ratio $\hbar\g B_0/\ve_F$ grows, the ``triangular" region, i.e., Region III in Fig.~\ref{fig1}(a), enlarges, and the interesting collective spin wave dynamics arises over a substantial region in $q$-$\W$ space.\cite{balentsPRB20} 

It is remarkable that a 2D QSL, a half-filled Mott insulator, exhibits a paramagnetic response that resembles that of a weakly correlated metal. This is a striking consequence of strong interactions: strongly correlated electron systems, such as QSLs, often host novel excitations at low energies that bear little resemblance to the constituent electrons. The observation of the Fermi liquid-like signatures, i.e., the Stoner continuum and the collective spin wave mode, in these QSL phases would provide a definitive signature of fractionalization.  

This paper is organized as follows. In Sec.~\ref{nvrates}, derivations of the NV relaxation rates for generic, quasi-1D and quasi-2D quantum magnets are presented. Section~\ref{nvrates} also elucidates how the relaxation rates depend on the spin spectral functions of these quantum magnets. The spin spectral functions for the two representative QSLs are then evaluated in the following two sections. The spectral function for the 2D QSL is derived first in the absence of the gauge field in Secs.~\ref{hshift} and \ref{rpa} using the static and time-dependent Hartree-Fock approximations. It is then re-evaluated by including the gauge fluctuations to lowest order in Sec.~\ref{gaugefield}. The results are plotted, compared, and analyzed in Sec.~\ref{2dresults}. The spectral function for the quantum spin chains is derived in Sec.~\ref{xxzqsc}, where the results are discussed. Conclusions are drawn in Sec.~\ref{conc}.

\section{Relaxation rates}
\label{nvrates}
We begin by evaluating the NV relaxation rates for layered quantum magnets and 3D stacks of quantum spin chains. NV relaxometry involves placing an NV defect near the surface of a magnetic material and measuring the relaxation rate of the $S=1$ spin localized on the defect; see Appendix~\ref{app1} for a brief introduction to the technique.\cite{rondinRPP14,degenRMP17,casolaNRM18} As we show below, the rate is sensitive to the magnetic-field power spectral density at the defect site, which in turn depends on the spin spectral function of the proximate magnetic material. 

In conventional ferromagnetic materials, NV relaxometry has enabled the imaging of single spins\cite{grinoldsNATP13} and magnetic domain walls\cite{tetienneNATC14}, the detection of spin waves\cite{wolfePRB14,vandersarNATC15,wolfNATC16,pageAPL19}, and the extraction of key spin transport quantities like the magnon chemical potential\cite{duSCI17}. Theoretical proposals have also revealed the possibility of detecting magnon condensation in ferromagnets,\cite{flebusPRL18} probing the hydrodynamic modes of a magnon fluid,\cite{rodrigueznievaPRB22} and measuring charge and spin correlations in 1D systems\cite{rodrigueznievaPRB18} through this technique. Recent theoretical works have also shown how the technique can be used to reveal Fermi and non-Fermi liquid behavior of spinons in QSLs.\cite{chatterjeePRB15,khooNJP21} 

\subsection{3D stack of 2D quantum magnets}
\begin{figure}[t]
\includegraphics[width=\linewidth]{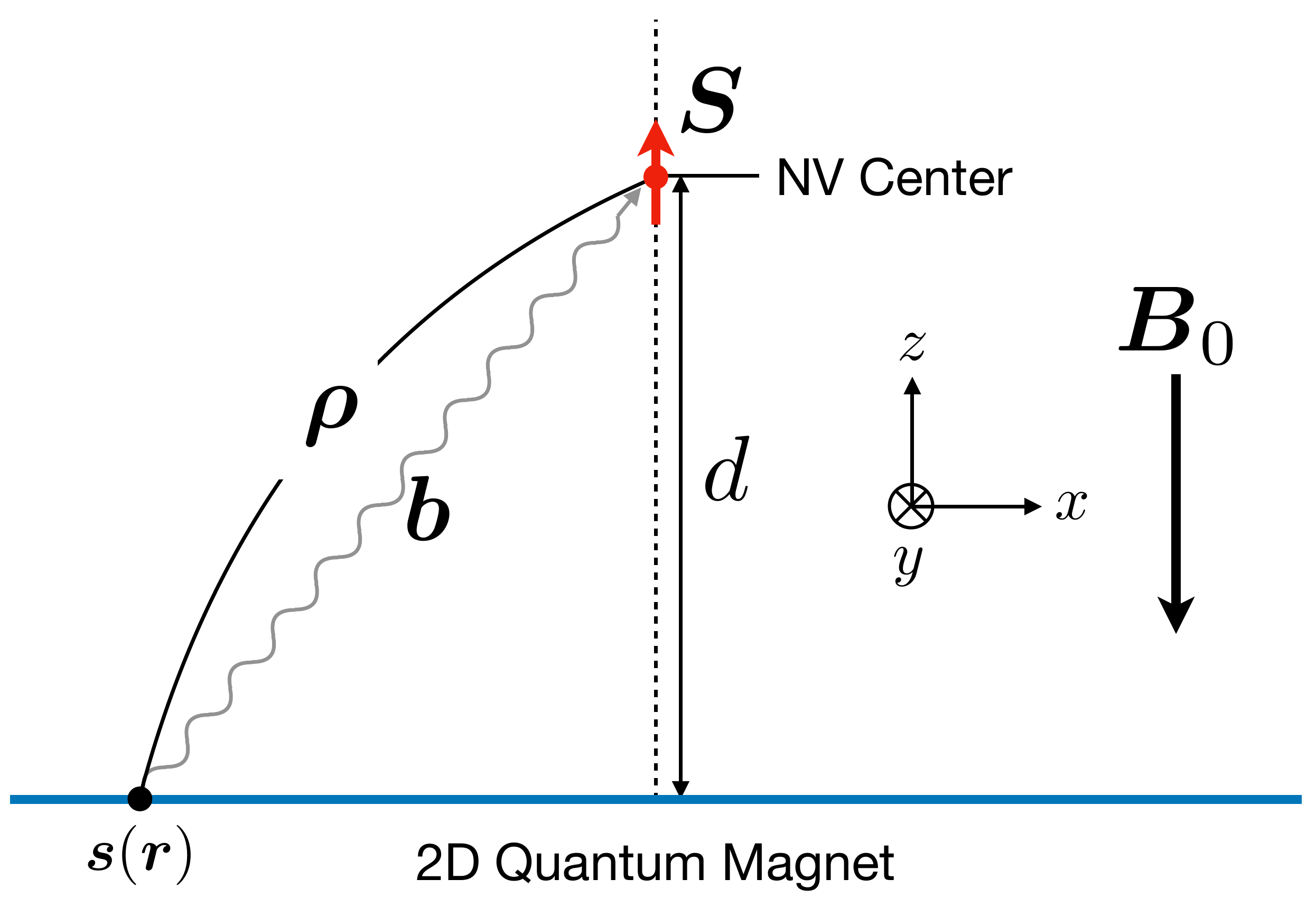}
\caption{NV center placed at distance $d$ above a 2D quantum magnet. Static field $\bB_0$ is applied anti-parallel to the NV spin axis, which is oriented along the $z$ axis.}
\label{fig2}
\end{figure}

Stable NV defect centers can exist within a few nanometers from the surface of diamond, allowing experimentalists to place the sensor close to the sample of interest. Let us consider an NV center located at a distance $d$ above a generic 2D quantum magnet at $\bx=(0,0,d)$; the quantum magnet is placed in the $xy$ plane, and we assume that the NV spin axis lies parallel to the $z$ axis (see Fig.~\ref{fig2}). 

In the presence of a bias magnetic field pointed along the $-z$ direction, i.e., $\bB_0=-B_0\ez$, the NV Hamiltonian can be written as
\beq
\label{h0}
H_0=-\hbar\tg B_0S_z+\De_gS_z^2\,,
\eeq
where $\tg$ is the negative of the gyromagnetic ratio of the impurity spin. The ESR frequencies are then given by $\hbar\W_\pm=\De_g\mp\hbar\tg B_0$.

Spin fluctuations inside the quantum magnet generate a fluctuating magnetic field at the impurity spin. For a single 2D quantum magnet, this field is given through the dipolar formula,\footnote{We use Gaussian units throughout this work.}
\beq
\label{dB}
\bb=-\hbar\g\int d^2\br\square{\frac{3(\bs(\br)\cdot\brho)\brho}{\rho^{5}}-\frac{\bs(\br)}{\rho^{3}}}\,,
\eeq
where $\bs(\br)$ denotes the spin density inside the quantum magnet, $\brho=(-\br,d)$ is the vector joining the source of spin fluctuations to the NV center, and $\g$ is the negative of the gyromagnetic ratio of the quantum magnet (see Fig.~\ref{fig2}). The Zeeman coupling between this fluctuating field and the NV spin then leads to a correction to Eq.~\eqref{h0} of the form
\beq
\label{v}
V=\hbar\tg\bb\cdot\bS\,.
\eeq

The fluctuating dipolar field generates transitions between the $\ket{0}$ and $\ket{\pm1}$ states (for details on the internal states of the NV spin, see Appendix~\ref{app1}). Using Eq.~\eqref{v} and the Fermi golden rule, the symmetrized transition rates between the $\ket{0}$ and $\ket{\pm1}$ states are given by
\beq
\label{ratesgen}
\G_\pm=\frac{\tg^2}{4}\square{C_{\mp\pm}(\W_\pm)+C_{\pm\mp}(-\W_\pm)}\,,
\eeq
where $C_{\mp\pm}(t)=\sang{b_\mp(t)b_\pm(0)}$ and $b_\pm=b_x\pm ib_y$. 

The spin response function of the quantum magnet has been introduced in Eq.~\eqref{chirs}. 
For a spatially isotropic quantum magnet, Eq.~\eqref{chirs}
should depend on the wavevector magnitude only, i.e., $\chi^R_{\al\be}(\bq,\W)=\chi^R_{\al\be}(q,\W)$. If we further assume the presence of uniaxial spin-rotational symmetry about the $z$ axis, the transition rates become
\begin{multline}
\label{rates}
\G_\pm=-\frac{\p(\hbar\g\tg)^2}{d^3}\coth\round{\frac{\hbar\W_\pm}{2k_BT}}\int_0^\infty dq\bar f_2(qd)\\
\times{\rm Im}\curly{\chi^R_{\perp}(q,\W_\pm)+\chi^R_{\parallel}(q,\W_\pm)}\,,
\end{multline}
where $\bar f_2(z)=z^3e^{-2z}$ and $T$ is the temperature of the quantum magnet; the transverse and longitudinal response functions have been defined in Eq.~\eqref{tvlongchir}. Similar result has been obtained in Refs.~\onlinecite{flebusPRL18,chatterjeePRB19}. Since the NV spin couples to the magnet material via dipolar coupling, both longitudinal and transverse components of the response function enter the expressions for the rates.

Equation~\eqref{rates} gives the rates due to a single magnetic layer. If identical layers are now stacked below the $z=0$ plane with uniform inter-layer spacing $a_\perp$ and if these layers are uncorrelated, the total rates can be obtained by summing over the layers incoherently, i.e., 
\begin{multline}
\G_\pm=-\p(\hbar\g\tg)^2\coth\round{\frac{\hbar\W_\pm}{2k_BT}}\sum_{n=0}^\infty\int_0^\infty dq\\
\times q^3e^{-2qd_n}{\rm Im}\curly{\chi^R_{\perp}(q,\W_\pm)+\chi^R_{\parallel}(q,\W_\pm)}\,,
\end{multline}
where $d_n=d+na_\perp$. For $qa_\perp\ll1$, we obtain
\begin{multline}
\label{rates2d2}
\G_\pm=-\frac{d}{a_\perp}\frac{\p(\hbar\g\tg)^2}{2d^3}\coth\round{\frac{\hbar\W_\pm}{2k_BT}}\int_0^\infty dq\\
\times f_2(qd){\rm Im}\curly{\chi^R_{\perp}(q,\W_\pm)+\chi^R_{\parallel}(q,\W_\pm)}\,,
\end{multline}
where the filtering function is now given by
\beq
\label{f2mod}
f_2(z)=z^2e^{-2z}\,.
\eeq

\begin{figure}[t]
\includegraphics[width=\linewidth]{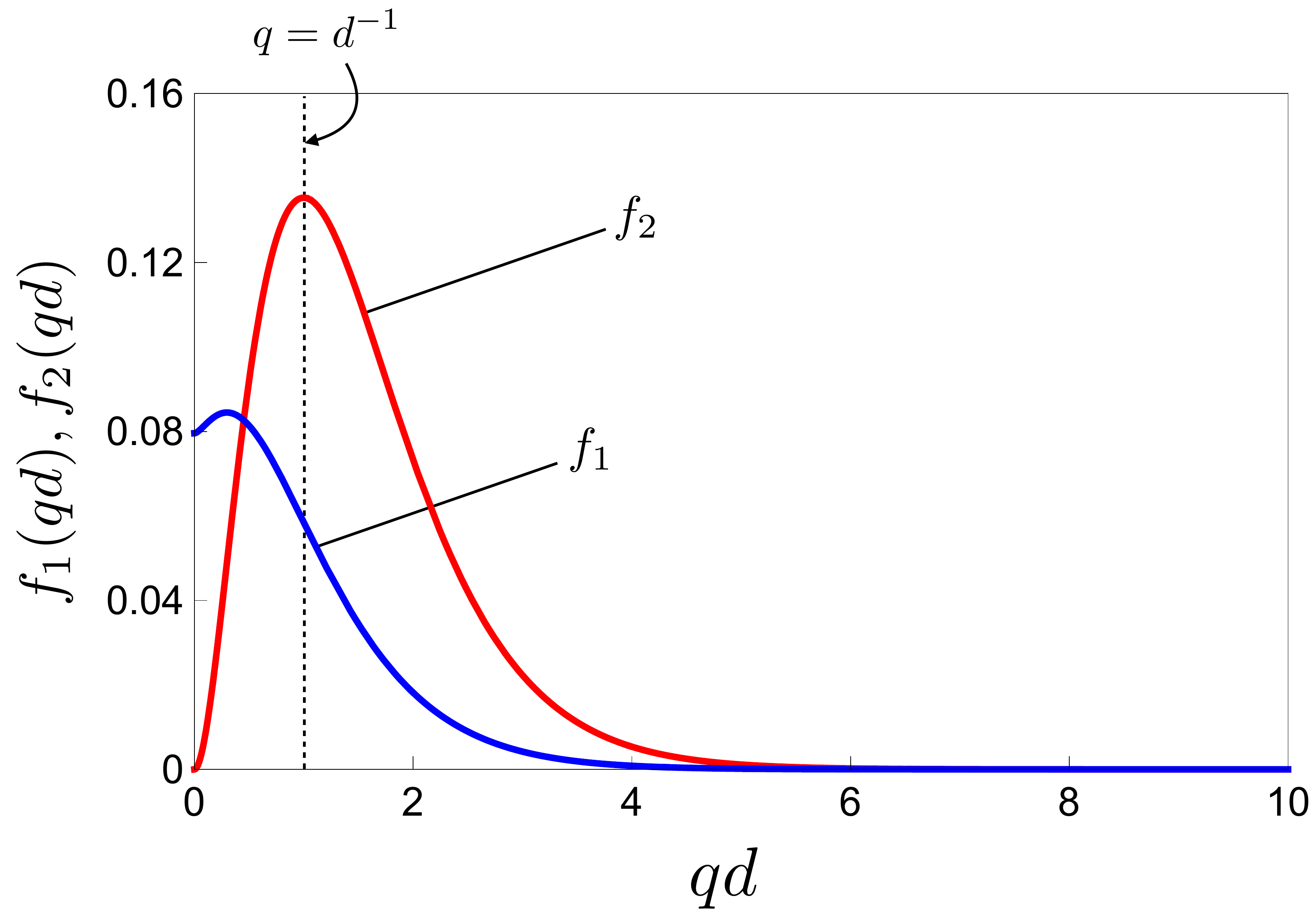}
\caption{The filtering functions in 1D and 2D.}
\label{fig3}
\end{figure}
Comparing Eq.~\eqref{rates2d2} to Eq.~\eqref{rates}, we find two differences. First is the difference between the filtering functions $\bar f_2(z)$ and $f_2(z)$, where we find that summing over the layers in the third direction slightly shifts the weight of the function to smaller $q$. Second is the introduction of an overall enhancement factor $d/a_\perp$ in Eq.~\eqref{rates2d2}, which indicates that layers within a distance $d$ from the top surface contribute most strongly to the rates. The function $f_2(z)$ is plotted in Fig.~\ref{fig3}. A clear peak is obtained at $qd\sim1$, indicating that the relaxation rates are most sensitive to spin fluctuations with wavevectors $q\sim d^{-1}$. 

\subsection{3D stack of quantum spin chains}
\label{3DstackQSC}
\begin{figure}[t]
\includegraphics[width=\linewidth]{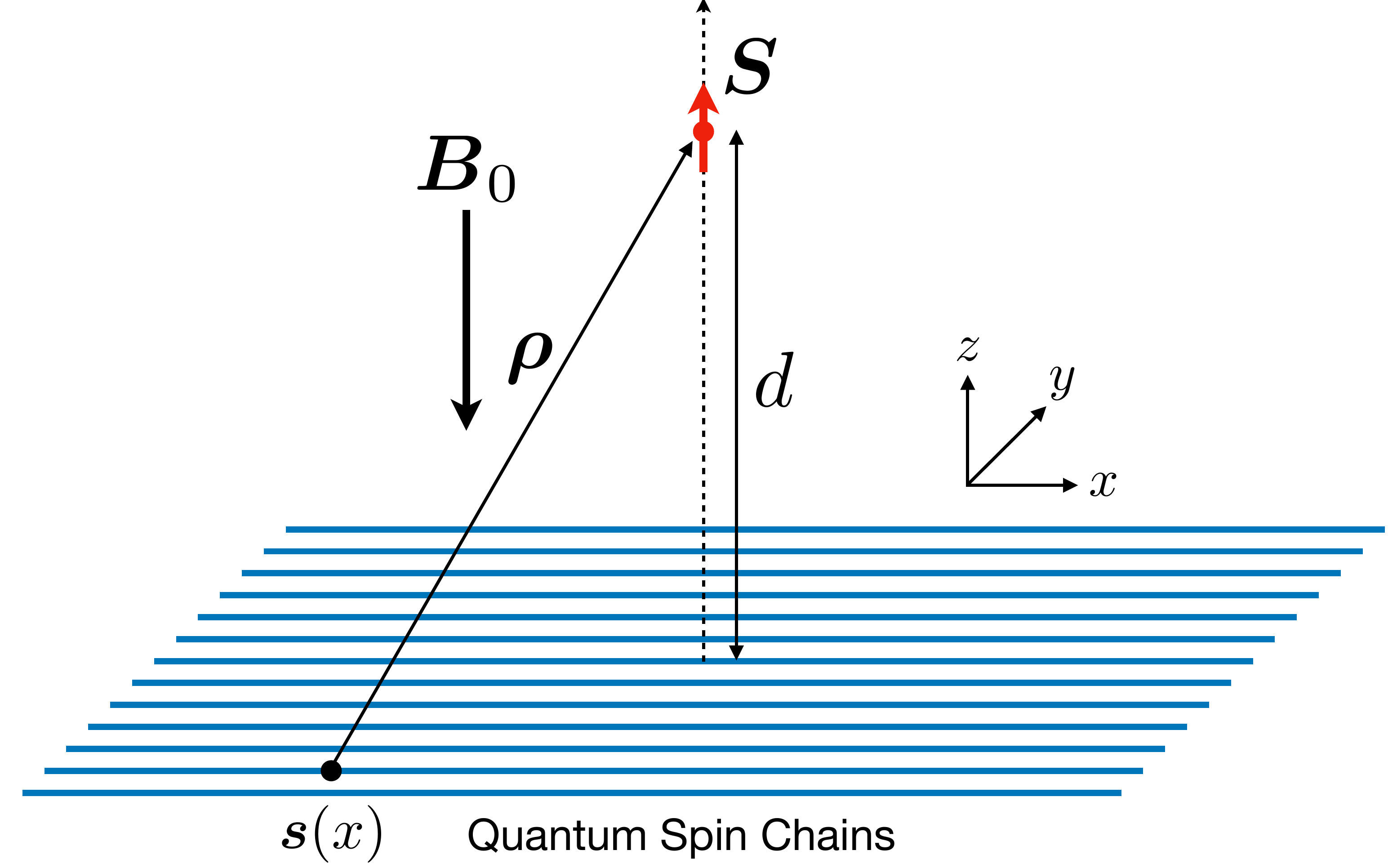}
\caption{NV center placed at a distance $d$ above a sequence of quantum spin chains. The chains extend along the $x$ axis and are stacked in the $y$ direction. As before, a static field $\bB_0$ is applied anti-parallel to the NV spin axis.}
\label{fig4}
\end{figure}

To obtain the relaxation rates for a 3D stack of quantum spin chains, we begin by evaluating the rates due a set of quantum spin chains arranged on a plane, extended along the $x$ axis and stacked in the $y$ direction, as shown in Fig.~\ref{fig4}. The dipolar field generated by the chains is given by
\beq
\label{dpf1d}
\bb=-\hbar\g\sum_{n=-\infty}^\infty\int dx\square{\frac{3(\bs(\br_n)\cdot\brho_n)\brho_n}{\rho_n^{5}}-\frac{\bs(\br_n)}{\rho_n^{3}}}\,,
\eeq
where $\bs(\br_n)$ is now the local spin density (per unit length) on the $n$-th spin chain, $\br_n=(x,na_y)$, $\brho_n=(-\br_n,d)$, $a_y$ is the inter-chain spacing, and the integer $n$ indexes the chains. Here, we ignore inter-chain correlations, i.e., 
\begin{align}
\sang{s_\al(x,na_y,t)s_\be(x',n'a_y,0)}&=\de_{nn'}\sang{s_\al(x,t)s_\be(x',0)}\\
&\equiv\de_{nn'}i\chi^>_{\al\be}(x-x',t)\,.
\end{align}

As in the 2D case, we assume that the spin chains have a uniaxial spin-rotational symmetry about the $z$ axis and that they are inversion symmetric, such that $\chi_{\al\be}(q,t)=\chi_{\al\be}(-q,t)$, where 
\beq
\chi_{\al\be}(q,t)=\int d(x-x') e^{-iq(x-x')}\chi_{\al\be}(x-x',t)\,.
\eeq
If we now consider stacking these layers in the vertical direction with a uniform inter-layer distance $a_\perp$ and summing over the layers, the rates become (see Appendix~\ref{rates1ddetails} for details)
\begin{multline}
\label{rates1d}
\G_\pm=-\frac{d^2}{a_ya_\perp}\frac{\p(\hbar\g\tg)^2}{d^4}\coth\round{\frac{\hbar\W_\pm}{2k_BT}}\int_0^\infty dq\\
\times f_1(qd){\rm Im}\curly{\chi^R_{\perp}(q,\W_\pm)+\chi^R_{\parallel}(q,\W_\pm)}\,,
\end{multline}
where the 1D filtering function $f_1(z)$ is explicitly derived in Appendix~\ref{rates1ddetails} and plotted in Fig.~\ref{fig3}. We note a strong resemblance between Eq.~\eqref{rates2d2} and Eq.~\eqref{rates1d}: we find that the rates for the 1D case can simply be obtained by replacing the spin spectral functions from the 2D result, i.e., Eq.~\eqref{rates2d2}, by those for the 1D spin chains and using a modified filtering function, which accounts for the fact that non-local spin correlations extend only along one spatial direction. As in the quasi-2D case, an enhancement factor of $d/a_\perp$ is obtained here. 

The filtering functions in the 1D and 2D cases both show increased sensitivity at wavevectors $q\approx d^{-1}$. However, they have contrasting behavior as $q\rightarrow0$. In the 2D case, the $q=0$ contribution vanishes because an infinite, uniformly polarized magnetic film does not produce any stray field. In the 1D case, however, the dipolar field does not vanish for each spin chain, and once the contributions from all the spin chains are summed incoherently, a finite $q=0$ component remains.



\section{Spin response at finite magnetic field: Two dimensions}
\label{rates2d}
As shown in Sec.~\ref{nvrates}, NV relaxometry probes the spin response function of a QSL at the probe frequencies $\W_\pm$. The filtering functions Fig.~\ref{fig3} also show that the relaxation rates are most sensitive to spin fluctuations at wavevectors $q\sim d^{-1}$. This opens the possibility of utilizing NV relaxometry to measure the spin spectral function with both energy and wavevector resolution.

The ESR frequencies $\W_\pm$ are determined by the bias magnetic field $B_0$, which should affect the quantum magnet. Therefore, over the next two sections, we analyze the spin response of the two QSLs under the influence of this bias magnetic field. We begin by discussing the 2D QSL in this section and address the 1D QSL scenario in the following section. Many of the features obtained in the 2D case carry over to the 1D case. 

The effective low-energy description of the 2D U(1) QSL consists of weakly interacting, charge-neutral, spin-1/2 fermions (i.e., spinons) coupled to a transverse U(1) gauge field.\cite{leePRL05,motrunichPRB05,zhouPRB13} The effective spinon Hamiltonian can be written as
\begin{multline}
\label{spinonh}
H_s=\frac{1}{2m^*}\sum_\s\int d^2\br\,\y^\dag_\s(\br)(-i\hbar\bnab-\ba(\br))^2\y_\s(\br)\\
+u\int d^2\br\,\y^\dag_\up(\br)\y^\dag_\down(\br)\y_\down(\br)\y_\up(\br)\\
-\sum_\s\int d^2\br\,\y^\dag_\s(\br)\frac{\s b_0}{2}\y_\s(\br)\,,
\end{multline}
where $\y_\s(\br)$ is the field for spin-$\s$ spinons (with effective mass $m^*$), $\ba$ is the transverse gauge field, and $b_0=\hbar\g B_0$ (with $\g$ being the gyromagnetic ratio of the QSL) is the Zeeman energy (see Appendix~\ref{u1qsl} for a phenomenological derivation of this starting theory). The spinons couple to the magnetic field only through Zeeman coupling because they are charge neutral. Notably, the Hamiltonian Eq.~\eqref{spinonh} with $\ba=0$ resembles that of a {\em conventional} Fermi liquid where the Landau quasiparticles interact through contact interaction and are subjected to the Zeeman field. We will therefore start by analyzing the spin response of the U(1) QSL in the limit of $\ba=0$, drawing on the intuition we have for the paramagnetic response of standard metals. We will subsequently incorporate the effects of the transverse gauge field perturbatively.

\subsection{Static Hartree-Fock approximation}
\label{hshift}
Let us set $\ba=0$ and treat the effects of the interaction $u$ first within the static Hartree-Fock (HF) approximation. According to Eq.~\eqref{spinonh}, we obtain the following HF spectrum for the spinons:
\beq
\label{xih}
\xi^{\rm HF}_{\bk\s}=\ve_{\bk}-\s b_0/2-\s um-\ve_F\,,
\eeq
where $\ve_{\bk}=\hbar^2k^2/2m^*$ is the bare spinon dispersion, $\ve_F$ is the spinon Fermi energy, and $m$ is the spin-polarization in the $z$ direction that is to be determined self-consistently. Mean-field decoupling of the quartic Hamiltonian with respect to all particle-hole channels gives (fixing the total spinon density)
\beq
\label{qterm}
u\y^\dag_\up(\br)\y^\dag_\down(\br)\y_\down(\br)\y_\up(\br)\rightarrow-2u\bs(\br)\cdot\sang{\bs(\br)}\,,
\eeq
where $s_\al=\sum_{\s\s'}\y^\dag_\s\s^\al_{\s\s'}\y_{\s'}/2$ is the spinon spin density and we assume that the total spinon density remains fixed. At zero temperature, this leads to the following self-consistent equation for $m$,
\beq
\label{sceq}
m=\frac{1}{2A}\sum_{\bk\s}\,\s\Theta(\ve_F+\s b_0/2+\s um-\ve_{\bk})\,,
\eeq
where $A$ is the area of the QSL. This equation can be solved analytically, giving
\beq
\label{nandm}
m=\frac{1}{2}\frac{g_0b_0}{1-ug_0}\,,
\eeq
where $g_0=m^*/2\p\hbar^2$ is the spinon density of states (per spin projection) at the Fermi level. 

Equation~\eqref{nandm} is an exact result for $m$ for the quadratic dispersion $\ve_\bk=\hbar^2k^2/2m^*$. However, the result also holds true for any dispersion relation $\ve_\bk$ as long as $b_0,um\ll\ve_F$ and the spinon density of states is finite and smooth in the vicinity of the Fermi energy. For $u=0$, Eq.~\eqref{nandm} gives the Pauli susceptibility, akin to the free Fermi gas; the inclusion of spinon interactions gives rise to the Stoner enhancement factor $(1-ug_0)^{-1}$.

\subsection{Time-dependent Hartree-Fock approximation}
\label{rpa}
The dynamic spin response function for the QSL can now be computed using the {\em time-dependent} self-consistent HF approximation, which is equivalent to the RPA. The details of the calculation are presented in Appendix~\ref{tdhfa}, so we only briefly outline the technical procedure here. Our quantity of interest is [see Eq.~\eqref{chirs}]
\begin{multline}
\chi^R_{\al\be}(\bk,\bk';\bq,t)=-\frac{i\Theta(t)}{4A}\sum_{\{\s_i\}}\s^\al_{\s_1\s_2}\s^\be_{\s_3\s_4}\\
\times\sang{[e^{iH_st/\hbar}\y^\dag_{\bk\s_1}\y_{\bk+\bq\s_2}e^{-iH_st/\hbar},\y^\dag_{\bk'\s_3}\y_{\bk'-\bq\s_4}]}\,,
\end{multline}
which obeys the equation of motion,
\beq
\label{chieom}
\begin{multlined}
i\hbar\pd_t\chi^R_{\al\be}(\bk,\bk';\bq,t)=\frac{i}{4A}\sum_{\{\s_i\}}\s^\al_{\s_1\s_2}\s^\be_{\s_3\s_4}\\
\times\Big\{\Theta(t)\sang{[[H_s,\y^\dag_{\bk\s_1}(t)\y_{\bk+\bq\s_2}(t)],\y^\dag_{\bk'\s_3}\y_{\bk'-\bq\s_4}]}\\
-i\hbar\de(t)\sang{[\y^\dag_{\bk\s_1}\y_{\bk+\bq\s_2},\y^\dag_{\bk'\s_3}\y_{\bk'-\bq\s_4}]}\Big\}\,.
\end{multlined}
\eeq
The inner commutator on the second line of Eq.~\eqref{chieom} gives rise to both quadratic and quartic terms, the latter of which are decoupled in the particle-hole channel as was done in the mean-field analysis of Sec.~\ref{hshift}. This leads to a closed equation of motion for the spin response function that can be solved in frequency space. We therefore arrive at the following results (see Appendix~\ref{tdhfa}):
\begin{align}
\label{chirt}
\chi^R_{\perp}(q,\W)&=\frac14\sum_\s\frac{\chi^{R(0)}_{\s\bar\s}}{1+\frac {u}\hbar\chi^{R(0)}_{\s\bar\s}}\,,\\
\label{chirpp}
\chi^{R}_{\parallel}(q,\W)&=\frac14\sum_\s\frac{\chi^{R(0)}_{\s\s}\round{1-\frac u\hbar\chi^{R(0)}_{\bar\s\bar\s}}}{1-\frac{u^2}{\hbar^2}\chi^{R(0)}_{\s\s}\chi^{R(0)}_{\bar\s\bar\s}}\,,
\end{align}
where
\beq
\label{chir0}
\chi^{R(0)}_{\s\s'}(q,\W)=\frac1{A}\sum_{\bk}\frac{n_F(\xi^{\rm HF}_{\bk\s})-n_F(\xi_{\bk+\bq\s'}^{\rm HF})}{\W+\xi^{\rm HF}_{\bk\s}/\hbar-\xi^{\rm HF}_{\bk+\bq\s'}/\hbar+i0}\,,
\eeq
and $n_F$ is the Fermi-Dirac distribution function. 

\subsection{Gauge field corrections}
\label{gaugefield}
We now discuss how the gauge fluctuations affect the dynamic response. In the low-energy, long-wavelength limit, the frequencies $\W$ and momenta $q$ of these gauge fluctuations obey the scaling $\W\sim q^3$.\cite{nagaosaPRL90,leePRB92,halperinPRB93} This relatively flat spectrum makes the gauge field bath an effective momentum sink for the spinons and leads to an overall smoothening of the spin response function. 

\begin{figure}[t]
\includegraphics[width=\linewidth]{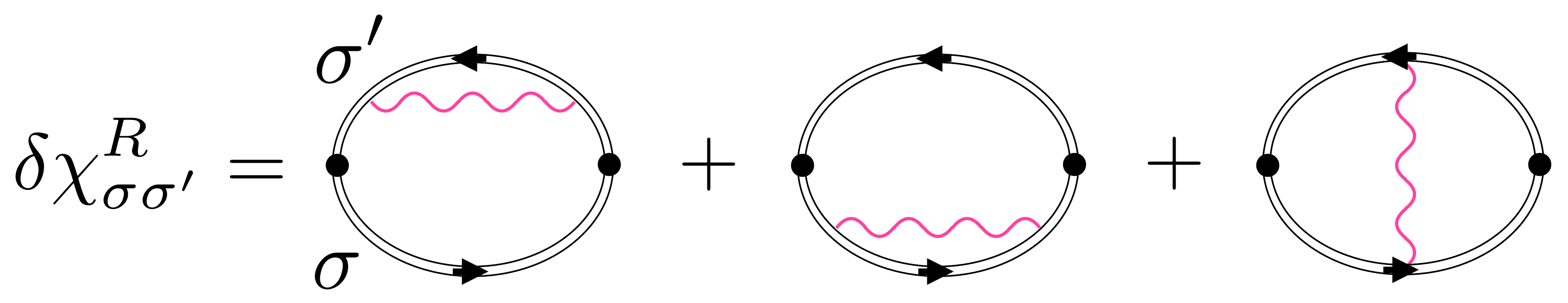}
\caption{Corrections to the bare particle-hole bubble containing one gauge line. The first two are self-energy corrections, and the third diagram is a vertex correction. The red wavy lines represent the gauge propagator, which is given through Eq.~\eqref{seffa}. All three diagrams must be included to preserve gauge invariance.}
\label{fig5}
\end{figure}

We study the extent of this smoothening when a single gauge line is added to the bare bubble, i.e., $\chi^{R(0)}_{\s\s'}$. There are three relevant diagrams for each bubble as shown in Fig.~\ref{fig5}, where the double lines represent spinon propagators with the static HF potential included, and the red wavy lines represent the gauge fluctuations. There are two diagrams corresponding to self-energy corrections and one corresponding to a vertex correction, all of which must be kept to preserve gauge invariance.\cite{kimPRB95,balentsPRB20} Because we are ultimately interested in the imaginary part of the spin response function, we focus solely on the imaginary part of this correction.

The detailed evaluation of the three diagrams are presented in Appendix~\ref{gaugecorr}. If we define dimensionless variables $y=\hbar\W/4\ve_F$, $x=q/2k_F$, and $\de=(b_0+2um)/4\ve_F\equiv\De/4\ve_F$, where $k_F$ is the spinon Fermi wavevector, the corrections to the transverse and longitudinal response functions read 
\begin{align}
\label{gaugecorrection}
\de\x^{R}_{\pm\mp}(q,\W)&=-ic\hbar g_0\frac{x^2y^{7/3}|y\mp\de|}{|(y\mp\de)^2-x^2|^{5/2}}\,,\\
\label{gaugecorrection2}
\de\x^{R}_{\s\s}(q,\W)&=-ic\hbar g_0\frac{x^2y^{10/3}}{|y^2-x^2|^{5/2}}\,,
\end{align}
where $c$ is a real constant of order 1. Accounting for this correction, each of the bare spin response functions is modified as
\beq
\chi^{R(0)}_{\s\s'}(q,\W)\rightarrow\chi^{R(0)}_{\s\s'}(q,\W)+\de\chi^R_{\s\s'}(q,\W)\,,
\eeq
which enter Eq.~\eqref{chirpp}.

\subsection{Results}
\label{2dresults}
A plot of $-{\rm Im}\{\chi^{R}_\perp(q,\W)+\chi^{R}_\parallel(q,\W)\}$ is presented in Fig.~\ref{fig1}, both (a) without and (b) with the gauge field correction: a static field of $b_0/4\ve_F=0.02$ and dimensionless correlation strength $\xi\equiv ug_0=0.3$ are used. We begin by discussing Fig.~\ref{fig1}(a), since most of the qualitative features of the spin response can be understood in the absence of gauge fluctuations. 

The low-frequency continuum, labeled ``IV" in Fig.~\ref{fig1}(a), arises due to the longitudinal response $\chi^R_{\parallel}(q,\W)$. It is bounded from above by
\beq
\W=v_Fq\sqrt{1+2\de}+\frac{\ve_\bq}\hbar\,,
\eeq
where $\de=b_0/4\ve_F(1-\xi)$, as defined earlier, is the Stoner-enhanced magnetic field (see Appendix~\ref{bubbles}) and $\ve_\bq=\hbar^2q^2/2m^*$. Examining the poles of the longitudinal RPA response using Eq.~\eqref{chirpp}, we find that there are no undamped collective modes outside of this continuum. 

The transverse spin response $\chi^R_\perp(q,\W)$ gives rise to the Stoner continuum that emanates from 
\beq
\hbar\W=\De=\frac{b_0}{1-\xi}\,,
\eeq
and expands into Region II. This region is bounded by
\begin{multline}
\label{region2}
\De-\hbar v_Fq\sqrt{1+2\de}+\ve_\bq\le\hbar\W\le \\
\De+\hbar v_Fq\sqrt{1+2\de}+\ve_\bq\,,
\end{multline}
and the lower boundary of this continuum reaches zero frequency at 
\beq
q_0=k_F\round{\sqrt{1+2\de}-\sqrt{1-2\de}}\,.
\eeq

\begin{figure}[t]
\includegraphics[width=0.9\linewidth]{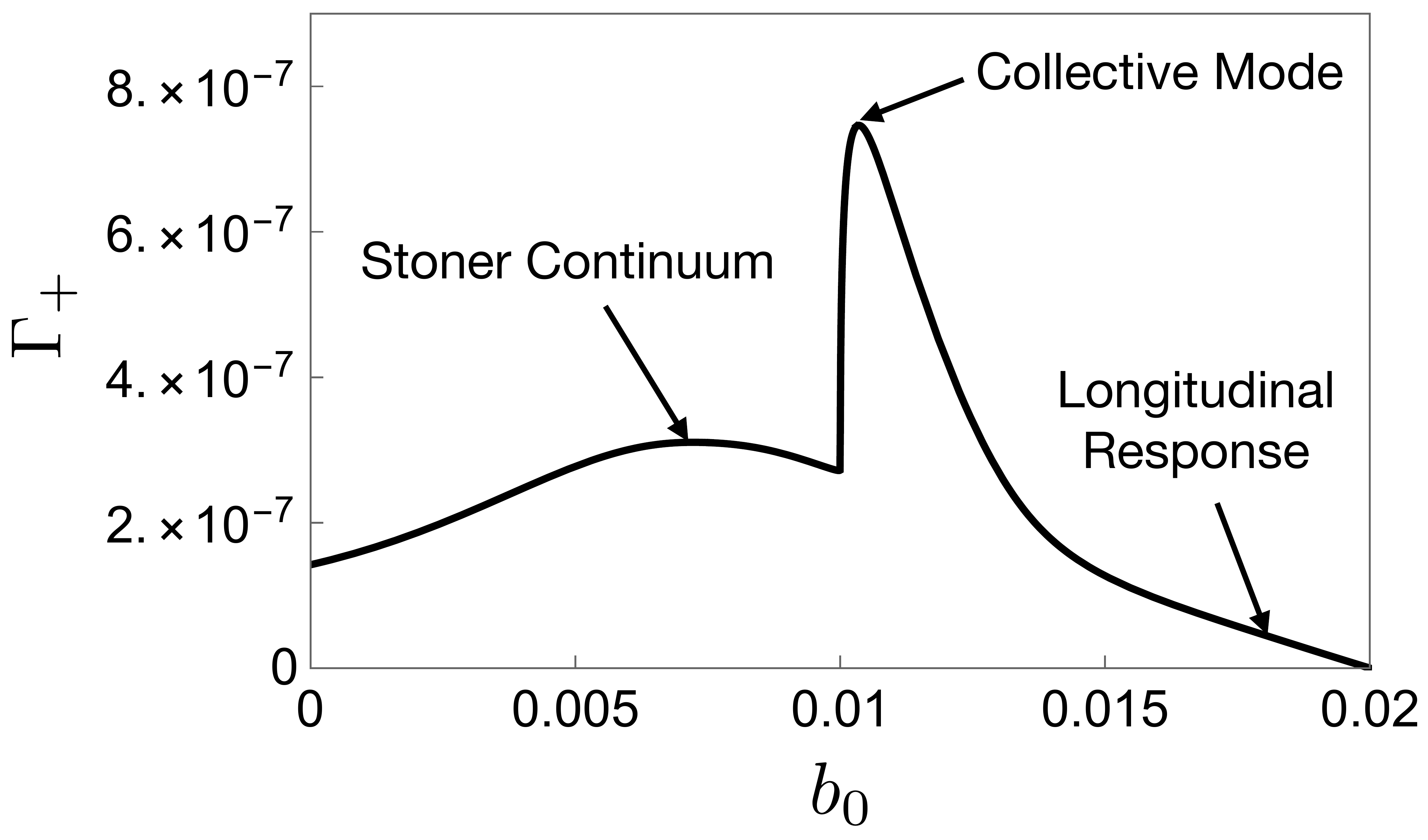}
\caption{Plot of $\G_+$ as a function of the external magnetic field at zero temperature. Here, $\De_g/4\ve_F=0.02$ and $k_Fd=50$ are used, and the rate is plotted in units of $\p\hbar g_0(2k_F)^3(\hbar\g\tg)^2/2a_\perp$.}
\label{fig6}
\end{figure}
The transverse response also contains an undamped collective spin wave mode that remains outside the Stoner continuum for all $q\le q_0$ and sharp in the absence of damping mechanisms, e.g., gauge fluctuations. The dispersion relation $\W_c(q)$ of the spin wave mode is given by the pole of $\chi^{R}_{\perp}(q,\W)$, i.e.,
\begin{multline}
\hbar\W_c(q)=\frac{1-\xi}{1-2\xi}\Bigg\{b_0\\
-\sqrt{\round{\tfrac{\xi}{1-\xi}}^2b_0^2+(\hbar v_Fq)^2(1-2\xi)+\round{\tfrac{1-2\xi}{\xi}}^2\ve_\bq^2}\Bigg\}\,.
\end{multline}
At small wavevectors $q\ll k_F$, 
\beq
\label{omegac}
\hbar\W_c(q)\approx b_0\square{1-\frac{(1-\xi)^2}{2\xi}\round{\frac{v_Fq}{\g B_0}}^2}\,,
\eeq
so the mode emerges from the Zeeman energy and softens quadratically as $q$ increases from zero. The lower boundary of the Stoner continuum [see Eq.~\eqref{region2}] remains larger than $\hbar\W_c(q)$ for $q<q_0$; therefore, this spin wave mode approaches but remains outside of the continuum. In the absence of the gauge field, Regions I and III have zero weight because spinon-hole excitations are kinematically forbidden for those momenta and frequencies. 

The spin wave mode at $q=0$ describes uniform spin precession at the Larmor frequency $\g B_0$, a result that is in accordance with the Larmor theorem,\cite{oshikawaPRB02} which states that the only response of any spin system with SU(2) symmetry at zero field is at the Larmor frequency. The Stoner continuum, whose vertex is off-resonant with $\g B_0$, indeed has vanishing weight as $q\rightarrow0$, so that the theorem remains intact. If we now recall the results from Secs.~\ref{hshift} and \ref{rpa}, we see that the static HF potential (see Sec.~\ref{hshift}) generates an upshift of the Stoner continuum from $b_0$ to $\De$. The subsequent time-dependent HF (RPA) treatment of the spin response (see Sec.~\ref{rpa}) then leads to the spin wave mode that separates off from the continuum and restores the expected Larmor response.\cite{balentsPRB20}

We now discuss how the spin spectral function is modified in the presence of the gauge fluctuations [see Fig.~\ref{fig1}(b)]. These fluctuations open up new scattering processes that are kinematically forbidden in their absence. The addition of Eqs.~\eqref{gaugecorrection} and \eqref{gaugecorrection2} introduces new spectral weight over all four regions of phase space and has a general effect of smoothening out the sharp features in Fig.~\ref{fig1}(a). 

The most visible effect in Fig.~\ref{fig1}(b) is the broadening of the spin wave mode due to the gauge field. This is an intrinsic broadening of the mode due to the internal emergent gauge symmetry of the system.\cite{savaryRPR17} The linewidth of this mode can be understood by studying $\chi^{R(0)}_{+-}(q,\W)$. Since the mode is outside the Stoner continuum, we have ${\rm Im}\{\chi^{R(0)}_{+-}(q,\W)\}=0$; therefore
\begin{multline}
{\rm Im}\{\chi^{R}_{+-}(q,\W)\}\\
=\frac{{\rm Im}\{\de\x^{R}_{+-}(q,\W)\}}{\square{1+\frac{u}{\hbar}\chi^{R(0)}_{+-}(q,\W)}^2+\square{\frac{u}{\hbar}{\rm Im}\{\de\x^{R}_{+-}(q,\W)\}}^2}\ .
\end{multline}
In the limit of $q\ll k_F$, we may write
\beq
1+\frac{u}{\hbar}\chi^{R(0)}_{+-}(q,\W)\approx\frac{\hbar\W-\hbar\W_c(q)}{\hbar\W-\De}\,,
\eeq
where $\W_c(q)$ is given by Eq.~\eqref{omegac}. Using Eq.~\eqref{gaugecorrection}, the linewidth can then be estimated by
\begin{align}
\de\W&\approx\frac{u}{\hbar}{\rm Im}\{\de\x^{R}_{\perp}(q,b_0/\hbar)\}\frac{b_0-\De}{\hbar}\\
\label{hwhm}
&\approx\frac{4c(1-\xi)^3}{\xi^2}\frac{\ve_F}{\hbar}\round{\frac{4\ve_F}{b_0}}^{2/3}\round{\frac{q}{2k_F}}^2\,.
\end{align}
What is notable from Eq.~\eqref{hwhm} is that the broadening vanishes as $q\rightarrow0$ so that the spin wave mode becomes sharp in the uniform limit, again in accordance with the Larmor theorem. This sharpening of the spin wave mode can be observed in Fig.~\ref{fig1}(b).

Comparing Figs.~\ref{fig1}(a) and \ref{fig1}(b), we may say that the gauge fluctuations have a relatively weak effect, as the essential features in Fig.~\ref{fig1}(a) remain intact in Fig.~\ref{fig1}(b). We note, however, that the spinon model Eq.~\eqref{spinonh} is a paradigmatic model of a non-Fermi liquid. The evaluation of the spinon self-energy $\S_\bk(\w)$ due to the transverse gauge propagator is known to give ${\rm Im}\,\S_\bk(\w)\propto\w^{2/3}$ at one-loop level, implying a vanishing quasiparticle weight and the breakdown of the Fermi liquid picture.\cite{nagaosaPRL90,leePRB92,halperinPRB93} An imprint of this non-Fermi liquid scaling is manifested in the anomalous exponent $2/3$ in Eq.~\eqref{hwhm}, which arises due to the presence of the extremely soft gauge fluctuations with a $\W\sim q^3$ spectrum. 


If $\G_\pm$ is computed in the limit of $k_F\gg d^{-1}$, the filtering function $f_2(z)$ [see Eq.~\eqref{f2mod}] places a strong weight at small $q$. Therefore, by sweeping the external magnetic field from zero up to $b_0=\De_g$, $\G_+$ should pick up strong signals when $\W_+$ comes in resonance with the collective spinon mode and the Stoner continuum. The other triplet mode $\W_-$ remains off-resonant, so we focus solely on $\G_+$ here. 

We plot $\G_+$ in units of $\p\hbar g_0(2k_F)^3(\hbar\g\tg)^2/2a_\perp$ in Fig.~\ref{fig6} for the same parameters as above along with zero-field splitting $\De_g/4\ve_F=0.02$ and $k_Fd=50$. For simplicity, we will assume $\g=\tg$ in the following discussion. A peak is obtained at $b_0=\De_g/2$, where the probe frequency $\W_+=(\De_g-b_0)/\hbar$ comes in resonance with the collective spinon mode. A broad peak to the left of the peak is obtained as $\W_+$ sweeps through the Stoner continuum, and a tail is observed to the right due to the longitudinal contribution to the spin response. 

The peak height in Fig.~\ref{fig6} can be estimated by (noting that $\hbar\W_+=\De_g/2$ at the peak)
\begin{align}
\G_+&\approx-\frac{\p(\hbar\g^2)^2}{2a_\perp}\coth\round{\frac{\De_g}{4k_BT}}\\
&\quad\qquad\qquad\times\int_0^\infty dqq^2e^{-2qd}{\rm Im}\curly{\chi^R_{+-}(q,\De_g/2\hbar)}\\
&\approx\frac{\p\hbar g_0(2k_F)^3(\hbar\g\tg)^2}{2a_\perp}\coth\round{\frac{\De_g}{4k_BT}}\\
&\qquad\qquad\qquad\qquad\times\frac{c}{k_Fd}\frac{1}{(1-\xi)^2}\round{\frac{\De_g}{8\ve_F}}^{7/3}\,,
\end{align}
where the peak rate is estimated by dropping the contributions from $\chi^R_{-+}$ and $\chi^R_\parallel$, as they are far off-resonant. The anomalous $7/3$ exponent enters here again due to the coupling to the gauge fluctuations.

\section{Spin response at finite magnetic field: One dimension}
\label{xxzqsc}
The spin-1/2 antiferromagnetic spin chain is a paradigmatic model of a 1D QSL with numerous material realizations.\cite{mikeskaBOOK04,giamarchiBOOK04} Experimental investigations of these materials began with thermal transport measurements and were then followed by the quantification of spin transport via NMR and muon-spin resonance.\cite{bertiniRMP21} More recently, spin Seebeck effect was used to detect spin transport directly in a quasi-1D cuprate material Sr$_2$CuO$_3$.\cite{hirobeNATP16}  

With an external magnetic field applied along the negative $z$ direction, the starting Hamiltonian for the spin chain may be written as
\begin{align}
H&=J\sum_{n}\round{\tfrac{1}{2}S^-_nS^+_{n+1}+h.c.+\z S^z_nS^z_{n+1}}-b_0\sum_nS^z_n\\
\label{hc}
&\equiv H_0+H_Z\,,
\end{align}
where $n$ labels the sites and $J>0$ is the antiferromagnetic exchange constant. We first focus on the gapless critical regime, where the anisotropy parameter $\z$ obeys $|\z|<1$, and comment on the expected NV relaxometry signatures at the isotropic point $\z=1$ later.  

The low-energy properties of the antiferromagnetic spin chain for $|\z|<1$ can be obtained by first mapping $H$ to a one-dimensional system of interacting fermions using the Jordan-Wigner transformation,\cite{jordanZP28}
\beq\begin{aligned}
\label{jwt1}
S^z_n&=\y^\dag_n\y_n-\tfrac{1}{2}\,,\\
S^+_n&=(-1)^n\y^\dag_n\exp\round{i\p\sum_{n'<n}\y^\dag_{n'}\y_{n'}}=\round{S_n^-}^\dag\,,
\end{aligned}\eeq
where the factor $(-1)^n$ is included so that the fermionic spectrum, post-fermionization, has positive concavity at $k=0$. The exponential factor is the Jordan-Wigner string, which ensures commutativity of spins on different sites after fermionization and rotates all the spins located at $n'<n$ by $\p$ about the $z$ axis. If we then start with a state $\ket{\Phi}$, where all the spins are pointed in the $x$ direction, the application of a single $\y^\dag_{\bar n}$ on the state, i.e., $\y^\dag_{\bar n}\ket{\Phi}$, creates a Bloch domain wall, i.e., a quantum kink, at $n=\bar n$ such that $S^x_n=\pm1/2$ for $n\gtrless\bar n$ and $S^z_n=1/2$ at $n=\bar n$. Since a single spin flip is realized by introducing two such domain walls, each Jordan-Wigner fermion should carry spin 1/2. 

With regard to the Jordan-Wigner string, $\p$-rotations about the $z$ axis in the clockwise and anti-clockwise directions are equivalent. This allows for an alternative definition of the transformation, c.f. Eq.~\eqref{jwt1}, where
\beq\begin{aligned}
\label{jwt2}
S^z_n&=\y^\dag_n\y_n-\tfrac{1}{2}\,,\\
S^+_n&=(-1)^n\y^\dag_n\cos\round{\p\sum_{n'<n}\y^\dag_{n'}\y_{n'}}\,.
\end{aligned}\eeq
Physical observables are unaffected by this formal change, and we choose Eq.~\eqref{jwt2} for mathematical convenience. Equation~\eqref{hc} then maps to a theory of interacting fermions on a 1D lattice,
\begin{multline}
\label{hf}
H=-\frac{J}{2}\sum_{n}\round{\y^\dag_n\y_{n+1}+h.c.}+b_0\sum_n\round{\y^\dag_n\y_n-\tfrac{1}{2}}\\
+J\z\sum_{i}\round{\y^\dag_n\y_n-\tfrac{1}{2}}\round{\y^\dag_{n+1}\y_{n+1}-\tfrac{1}{2}}\ .
\end{multline}

In calculating the spin response functions, we first note that the 1D filtering functions, $f^{-+}_1$, $f^{+-}_1$, and $f^{zz}_1$, in Eq.~\eqref{rates1d} restrict most of the integral weight to wavevectors $q\le d^{-1}$, where $d$ is the NV-to-sample distance. For $d\gg a$, where $a$ is the lattice constant of the spin chain,\footnote{A representative quantum spin chain, a copper-oxide material Sr$_2$CuO$_3$, has a lattice constant of $a\approx4\AA$.} and for $\hbar\W_\pm\ll J$, which holds when the relevant frequencies obey $\hbar\W_\pm\lesssim\De_g\ll J$, the spin response functions can be computed within the long-wavelength, low-energy description of $H$. We therefore switch to the Luttinger-liquid representation of Eq.~\eqref{hf} and compute the spin response functions using the method of bosonization. Equation~\eqref{hf} then becomes\cite{giamarchiBOOK04}
\beq
H=\frac{u}{2}\int dx\bigg\{\frac{\p K}{\hbar}\Pi^2(x)+\frac{\hbar }{\p K}\square{\pd_x\f(x)}^2+\frac{b_0}{\p}\pd_x\f\bigg\}\,,
\label{hcb}
\eeq
where $\pd_x\f(x)/\p$ is the continuum variable for the local $z$-polarized spin density, $\Pi(x)$ is the conjugate momentum density obeying 
\beq
[\f(x),\Pi(x')]=i\hbar\de(x-x')\,,
\eeq
and $u$ and $K$ are, respectively, the speed of sound and Luttinger parameter, both of which depend on the anisotropy parameter $\z$.

\begin{figure*}[t]
\includegraphics[width=0.91\linewidth]{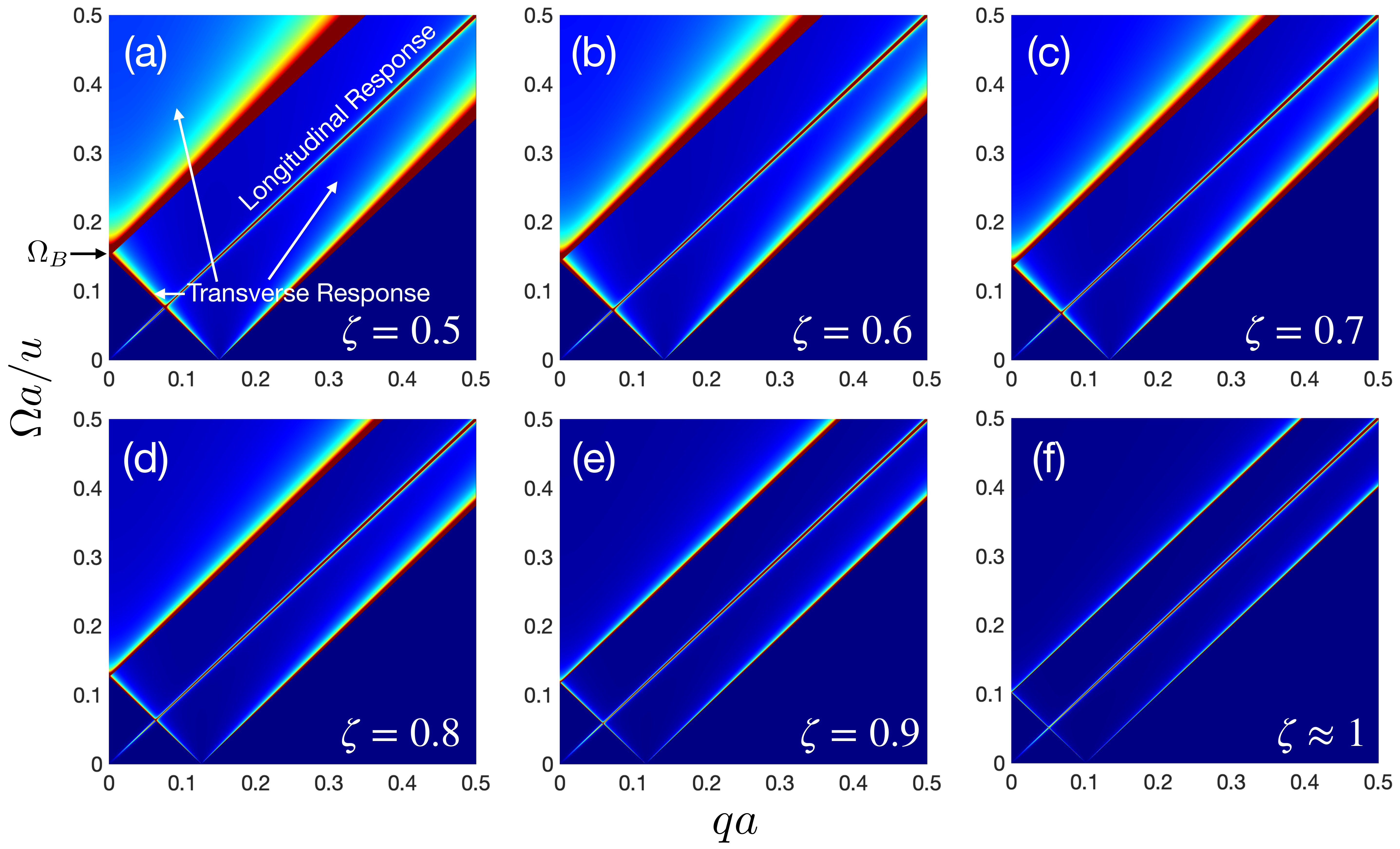}
\caption{Plot of $-{\rm Im}\{\chi^R_{\perp}(q,\W)+\chi^R_{\parallel}(q,\W)\}$ for various values of the anisotropy parameter $\z$ and for Zeeman energy $b_0=0.1$. (a) $\z=0.5$; (b) $\z=0.6$; (c) $\z=0.7$; (d) $\z=0.8$; (e) $\z=0.9$; (f) $\z\approx1$. We have used $C=0.02$. }
\label{fig7}
\end{figure*}
At zero magnetic field, $u$ and $K$ over the entire critical regime are given via the Bethe-ansatz solution,~\cite{johnsonPRA73}
\beq
\label{LLp}
u=\frac{\p Ja\sqrt{1-\z^2}}{2\hbar\cos^{-1}\z}\,,\ \ \ K=\frac{\p}{2(\p-\cos^{-1}\z)}\,.
\eeq
At finite magnetic fields, these parameters are renormalized. However, for NV relaxometry experiments, the external field would reach a maximum magnitude of the order of the zero-field splitting $\De_g\sim0.1$~T. This is a small fraction of the intrinsic antiferromagnetic exchange scale $J$, which in most cases ranges from $1$~T to $1000$~T, so $u$ and $K$ are renormalized negligibly. We will therefore approximate these parameters hereafter with their zero-field values, i.e., Eq.~\eqref{LLp}.

For $\W>0$ and $q>0$, the imaginary parts of the transverse response functions read (see Appendix~\ref{spinchain})
\begin{widetext}
\begin{multline}
-{\rm Im}\{\chi^R_{\pm\mp}(q,\W)\}=\frac{2\p^2C}{u}\frac{(a/u)^{2K+1/2K-2}}{\G(K+1+1/4K)\G(K-1+1/4K)}\\
\times\Big[\Theta\round{uq\pm\be+\W}\Theta\round{\W-uq\mp\be}\round{uq\pm\be+\W}^{K+1/4K}\round{\W-uq\mp\be}^{K+1/4K-2}\\
+\Theta\round{\W-uq\pm\be}\Theta\round{\W+uq\mp\be}\round{\W-uq\pm\be}^{K+1/4K}\round{\W+uq\mp\be}^{K+1/4K-2}\Big]\,,
\end{multline}
\end{widetext}
and the longitudinal response is given by
\beq
-{\rm Im}\{\chi^R_{\parallel}(q,\W)\}=\de\round{\W-uq}\frac{K\W}{2u}\,,
\eeq
where $\W_B=2Kb_0/\hbar$. The constant $C$ is a real dimensionless correlation amplitude, a non-universal constant that cannot be obtained through bosonization. A numerical estimate for $C$ at finite magnetic fields has found $C\sim0.01$ for small fields and $0\le\z<1$.\cite{hikiharaPRB04}

\subsection{Results}
\label{1dresults}
Plots of $-{\rm Im}\{\chi^R_{\perp}(q,\W)+\chi^R_{\parallel}(q,\W)\}$ for various values of $\z$ are presented in Fig.~\ref{fig7}. Following Ref.~\onlinecite{hikiharaPRB04}, a correlation amplitude of $C=0.02$ is used. Comparing these plots with Fig.~\ref{fig1}, we see resemblances with the 2D scenario. Emanating from $q=\W=0$ is the longitudinal response, which in the 1D case corresponds to the collective density waves of the Jordan-Wigner fermions, i.e., zero sound. The transverse response forms a continuum, akin to the Stoner continuum in Fig.~\ref{fig1}, that emanates from $\W_B=2Kb_0/\hbar$ at $q=0$. Away from the Heisenberg point, where $2K=1$, the vertex of the continuum is located away from the Zeeman energy, since the Larmor theorem does not have to be satisfied here. However, as the isotropic point is approached, i.e., $\z\rightarrow1$ (or $K\rightarrow1/2$), the response at $q=0$ sharpens and shifts down toward the Zeeman energy, in agreement with the Larmor theorem.

\begin{figure}[t]
\includegraphics[width=\linewidth]{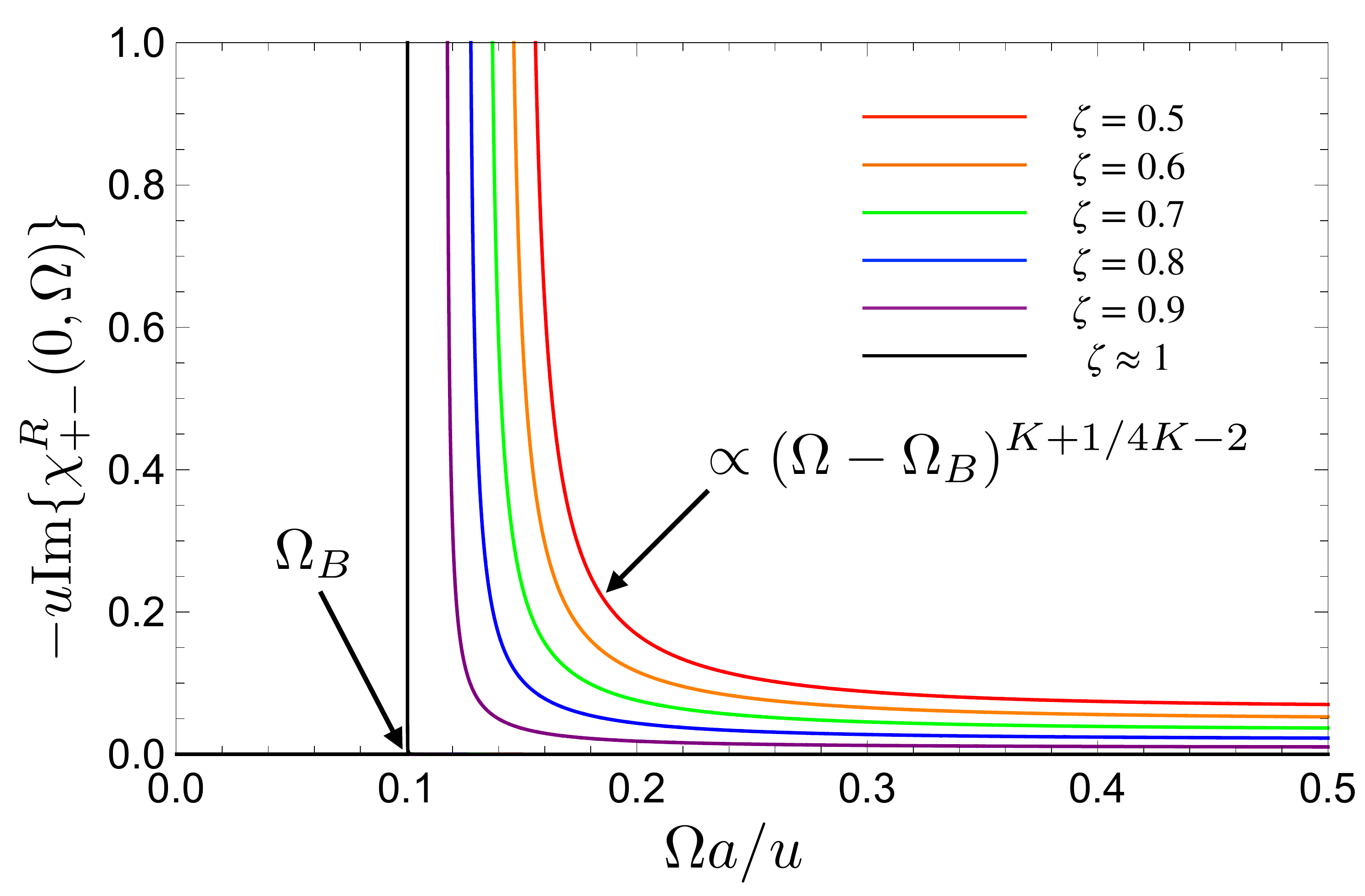}
\caption{Plot of $-{\rm Im}\{\chi^R_{+-}(q=0,\W)\}$ for various values of the anisotropy parameter $\z$ and for Zeeman energy $b_0=0.1$.}
\label{fig8}
\end{figure}
At $q=0$ and for $\W>\W_B$, the spin response comes solely from $\chi^R_{+-}(q,\W)$, and we may therefore ignore the longitudinal component and $\chi^R_{-+}(q,\W)$. A cut of $-{\rm Im}\{\chi^R_{+-}(q,\W)\}$ at $q=0$, given by
\begin{multline}
\label{chirperp1d}
-{\rm Im}\{\chi^R_{+-}(0,\W)\}=\frac{4\p^2C}{u}\\
\times\frac{(a/u)^{2K+1/2K-2}}{\G(K+1+1/4K)\G(K-1+1/4K)}\\
\times\left[\round{\W+\W_B}^{K+1/4K}\round{\W-\W_B}^{K+1/4K-2}\right]\,,
\end{multline}
is plotted as a function of $\W$ in Fig.~\ref{fig8} for the same set of $\z$ values as those used in Fig.~\ref{fig7}. 
As the Heisenberg limit is approached, i.e., $K\rightarrow1/2$, we have $\G(K+1/4K-1)\rightarrow\infty$, and the divergence seen in Fig.~\ref{fig8} comes from Eq.~\eqref{chirperp1d}. Indeed, as $K\rightarrow1/2$, 
\begin{multline}
-{\rm Im}\{\chi^R_{+-}(0,\W)\}\approx\frac{4\p^2C}{u}\\
\times\lim_{K\to1/2}\square{\frac{1}{\G(K-1+1/4K)}\round{\frac{\W-b_0/\hbar}{\W+b_0/\hbar}}^{K+\frac{1}{4K}-2}}\,.
\end{multline}
The limiting expression in the square brackets is a representation of the delta function, so $-{\rm Im}\{\chi^R_{+-}(0,\W)\}\propto\de(\W-b_0/\hbar)$, in agreement with the Larmor theorem. 

Away from the Heisenberg limit, the response peak occurs above the Zeeman energy $b_0$ and there is significant spectral weight at frequencies above the peak. The peak has a power-law divergence, as expected from the Luttinger physics, with an anomalous exponent $K+1/4K-2$. 

This anomalous exponent may be measurable using NV relaxometry. In the limit of $d\gg a$, where $a$ is the spin chain lattice constant, almost all of the weight in the filtering function $f^{+-}_1(qd)$ comes from $q\sim0$, so the transition rates can be estimated as
\begin{align}
\G_\pm&\approx-\frac14{\rm Im}\curly{\chi^R_{+-}(0,(\De_g\mp b_0)/\hbar)}\coth\round{\frac{\De_g\mp b_0}{2k_BT}}\\
&\qquad\qquad\times\frac{d^2}{a_ya_\perp}\frac{\p(\hbar\g\tg)^2}{d^4}\int_0^\infty dqf^{+-}_1(qd)\,.
\end{align}
For $\De_g-b_0\ll k_BT$, we therefore find that $\G_+$ scales according to the anomalous Luttinger scaling with the magnetic field as
\beq
\G_+\propto\round{\frac{2k_BT}{\De_g-b_0}}\square{\De_g-(1+2K)b_0}^{K+1/4K-2}\,,
\eeq
for $b_0<(1+2K)^{-1}\De_g<\De_g$.

\subsection{The isotropic limit: $K=1/2$}
\label{umklapp}
\begin{figure*}[t]
\includegraphics[width=0.9\linewidth]{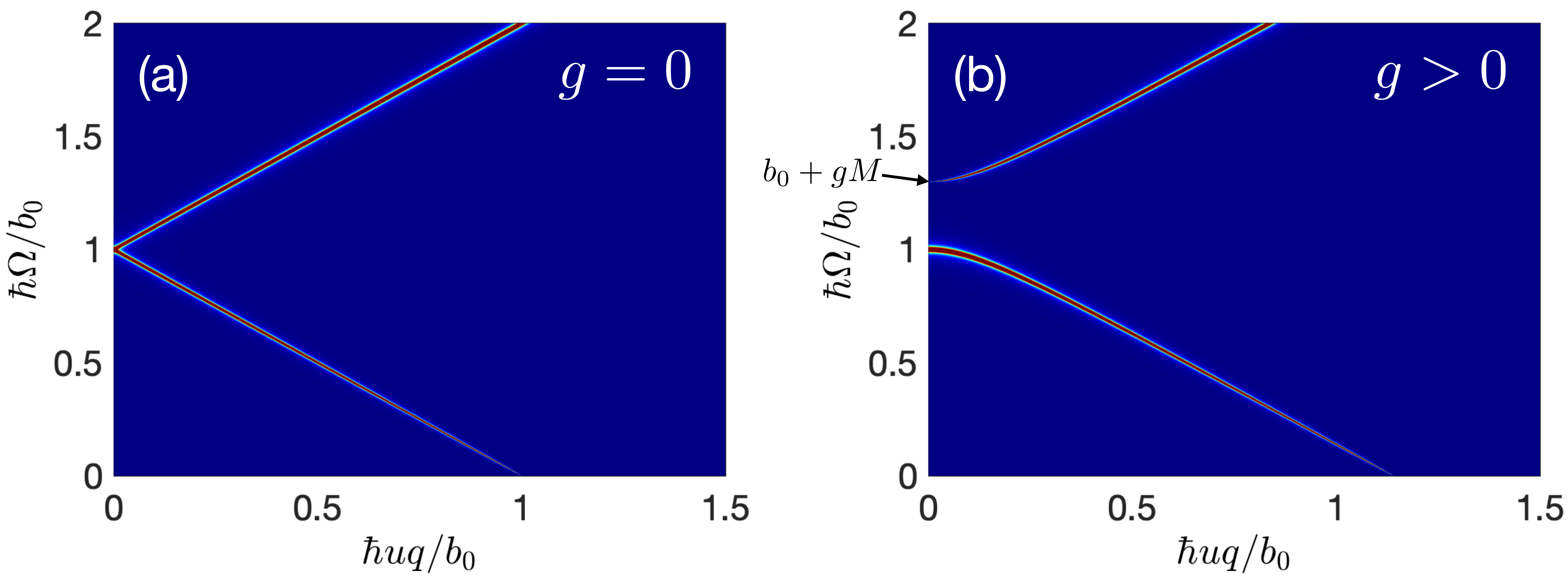}
\caption{Plots of $-{\rm Im}\{\chi^R_{+-}(q,\W)\}$ for the Heisenberg chain (i.e., $K=1/2$) in the long-wavelength regime (taken from Ref.~\onlinecite{keselmanPRL20}). Plot (a) corresponds to the transverse response at $g=0$, which matches the corresponding response shown in Fig.~\ref{fig7}(f). For $g>0$ shown in (b), the upper branch shifts up in energy by an amount proportional to the average magnetization $M$. In contrast, the lower branch remains at $b_0$ and sharp at $q=0$ and the spectral weight of the upper branch vanishes as $q\rightarrow0$. Here, we have used $gM=0.3b_0$.}
\label{fig9}
\end{figure*}

The bosonization of Eq.~\eqref{hf} is known to generate an additional term to Eq.~\eqref{hcb}, the so-called umklapp term,
\beq
H_{\rm umklapp}=\tilde g\int dx\cos\square{4\f(x)}\,,
\eeq
which describes the scattering of two fermions from one Fermi point to the other accompanied by a momentum transfer of $4k_F$. For $|\z|<1$, $\tilde g$ is RG irrelevant and flows to zero at zero magnetic field. At finite fields, the RG flow, in principle, stops at the scale $\ve\sim b_0$, but we expect $\tilde g$ to be algebraically small in $b_0/\La$, where $\La$ is the cutoff scale, such that $H_u$ can be neglected to a good approximation. In the Heisenberg limit, however, $\tilde g$ is only marginally irrelevant and may thus have appreciable magnitude even at the scale $\ve\sim b_0$. This limit must therefore be treated with care. 

A convenient way to examine the effects of marginally irrelevant operators on the spin spectral function [see Fig.~\ref{fig7}(f)] is to utilize a fermionization protocol that is different from the Jordan-Wigner transformation used in Sec.~\ref{xxzqsc}. We note that the fermionized Hamiltonian Eq.~\eqref{hf} is not manifestly SU(2) invariant, although the original Heisenberg model is. Affleck and Haldane developed a fermionic representation of the Heisenberg spin chain that explicitly preserves this SU(2) invariance.\cite{affleckPRB87} The elementary excitations in this representation are charge-neutral, spin-1/2 fermionic quasiparticles, i.e., spinons, which are different from the Jordan-Wigner fermions but more closely resemble the spinons discussed in the context of the 2D QSL. The Affleck-Haldane fermionization procedure thus allows for more unified descriptions of the Heisenberg spin chain and the 2D U(1) QSL. The transverse spin response function of the antiferromagnetic Heisenberg spin chain was studied in Ref.~\onlinecite{keselmanPRL20} using this SU(2) symmetry-preserving fermionization. In this subsection, we briefly discuss how our results from the previous subsection are modified by this fermionization procedure. 

Following Affleck and Haldane, the low-energy effective behavior of the Heisenberg spin chain can be modeled by an interacting gas of spin-1/2 Dirac spinons: $H=H_0+H_Z+H_{\rm back}$,\cite{affleckPRB87,keselmanPRL20,gogolinBOOK04} where
\begin{align}
\label{h0ah}
H_0&=\hbar u\sum_\s\int dx\Big[\y^\dag_{R\s}(x)(-i\pd_x)\y_{R\s}(x)\\
&\qquad\qquad\qquad\qquad+\y^\dag_{L\s}(x)(i\pd_x)\y_{L\s}(x)\Big]\,,\\
H_Z&=-b_0\int dx\square{J^z_R(x)+J^z_L(x)}\,,
\end{align}
describe the right- and left-moving chiral spinons with propagation speed $u$ and their coupling to the external magnetic field, and
\beq
\label{backsc}
H_{\rm back}=-g\int dx\,\bJ_R(x)\cdot\bJ_L(x)\,,
\eeq
models the backscattering interaction; the quantities 
\beq
J^\al_{R,L}(x)=\frac12\sum_{\s\s'}\y^\dag_{R,L\s}(x)\s_{\s\s'}^\al\y_{R,L\s'}(x)
\eeq
repreesent the right- and left-chiral spinon densities. For a qualitative discussion on how $H$ is obtained, see Appendix~\ref{affleckhaldane}. 

In this fermion representation, the marginally irrelevant coupling is the backscattering amplitude $g$, which introduces logarithmic corrections to our previous results for $\z=1$ [see Fig.~\ref{fig7}(f)]. 
Equation~\eqref{backsc} can be decomposed into the transverse and longitudinal parts, 
\begin{align}
H^\perp_{\rm back}&=-\frac g2\int dx\,[j^+_R(x)j^-_L(x)+j^-_R(x)j^+_L(x)]\,,\\
H^\parallel_{\rm back}&=-g\int dx\,j^z_R(x)j^z_L(x)\,,
\end{align}
where $j^\pm_{R,L}(x)=j^x_{R,L}(x)\pm ij^y_{L,R}(x)$. In the presence of the magnetic field in the negative $z$ direction, the spin chain forms a static magnetization $M$. As in the 2D QSL formulation, the effect of this magnetization  can be incorporated by a mean-field decoupling of the longitudinal term, i.e.,
\begin{align}
-gj^z_R(x)j^z_L(x)&\rightarrow-g\sang{j^z_R(x)}j^z_L(x)-g\sang{j^z_L(x)}j^z_R(x)\\
&=-g\frac{M}{2}\round{j^z_R(x)+j^z_L(x)}\,,
\end{align}
where the spinon over-population (magnetization) $M$ is split symmetrically between the chiral channels, $\sang{j^z_R}=\sang{j^z_L}=M/2$, and we see that the backscattering $g$ subjects the spinons to an additional induced field $-gM/2$. This induced field is analogous to the induced field discussed in the 2D QSL scenario, c.f. Sec.~\ref{hshift}. The transverse spin response can then be computed in the presence of this induced field and by treating $H^\perp_{\rm back}$ within the RPA.\cite{keselmanPRL20}

The imaginary part of the transverse spin response function, $\chi^R_{+-}(q,\W)$, is plotted in Fig.~\ref{fig9}; the plots are generated using Eq.~(7) in Ref.~\onlinecite{keselmanPRL20}. We note that $-{\rm Im}\{\chi^R_{+-}(q,\W)\}$ is plotted in Fig.~\ref{fig9} and not $-{\rm Im}\{\chi^R_{\perp}(q,\W)+\chi^R_{\parallel}(q,\W)\}$ as done in Fig.~\ref{fig7}. This difference leads, for example, to the absence of one of the branches in Fig.~\ref{fig9} that emanates from $\hbar uq/b_0=1$ with a positive group velocity. 

Figure~\ref{fig9}(a) shows the result for $g=0$, i.e., result in the absence of the marginally irrelevant backscattering term. There are two branches that converge at the Larmor frequency $b_0$ and exactly match the corresponding branches found in Fig.~\ref{fig7}(f). When one accounts for the backscattering term, i.e., $g>0$, the upper branch moves up in energy by $gM$, much like the up-shift of the Stoner continuum relative to the collective mode observed in the presence of spinon interactions in Fig.~\ref{fig1}. It is important to note that the Larmor theorem is still satisfied in Fig.~\ref{fig9}(b) because the lower branch remains at the Zeeman energy and sharp at $q=0$ and because the spectral weight of the upper branch vanishes as $q\rightarrow0$.\cite{keselmanPRL20} 

The rate $\G_+$ in the limit of $d\gg a$ essentially probes the transverse response at $qa\ll1$. If we now account for the above modification due to the backscattering term, the transition rate $\G_+$ for the Heisenberg chain should have two peaks as the Zeeman field $b_0$ is varied from $0$ to $\De_g$. The first peak should occur at $b_0=\De_g/2$ when $\W_+=\De_g-b_0$ comes in resonance with the lower spectral branch in Fig.~\ref{fig9}(b). The second peak is expected at $b_0\approx(\De_g-gM)/2$ when $\W_+$ comes in resonance with the upper spectral branch located at $b_0+gM$. The height of the latter peak should be smaller than the former because the spectral weight of the upper branch vanishes as $q\rightarrow0$. 

\section{Conclusions}
\label{conc}
Relaxometry based on NV centers in diamond offers exciting new opportunities to non-invasively measure the spin spectral functions of QSL materials with both energy and momentum resolution. This work examines the spin spectral functions of two representative QSLs | the 2D QSL with a spinon Fermi surface coupled to a U(1) gauge field and the spin-1/2 antiferromagnetic quantum spin chain | and elucidates the definitive signatures of fractionalization in these functions that should be directly measurable via NV relaxometry. 

Owing to strong correlations, local spin-1 magnon excitations in these QSLs fractionalize into two neutral, spin-1/2 fermionic quasiparticles called spinons. In 2D, these spinons form a metal-like ground state with a Fermi surface. The emergence of such delocalized, ``electron-like" quasiparticles leads to a paramagnetic response that closely resemble the response of a conventional weakly correlated metal. We have shown, for example, that the spin spectral function should exhibit a spin-1 particle-hole continuum, i.e., the Stoner continuum, along with a collective spin wave mode, both of which are characteristics of conventional Fermi liquids. Such an ``unexpected" response coming from Mott insulators with localized electrons represent one of the most definitive signatures of fractionalization. 

The Luttinger liquid formalism and bosonization are used to compute the dynamic response of the antiferromagnetic quantum spin chain. Resonant spectral weights come from regions in $(q,\W)$-space corresponding to the low-energy collective modes. For an XXZ spin chain, away from the SU(2)-symmetric (Heisenberg) point, there is significant weight away from the mode resonances that decay algebraically with an anomalous critical exponent determined by the Luttinger parameter. NV center relaxometry can be used to measure this exponent. We also examine the effects of marginally irrelevant operators on the spin spectral functions in the Heisenberg limit. 

The 2D QSL model considered in this work may be relevant to certain half-filled Mott insulators on the triangular lattice.\cite{shimizuPRL03,itouPRB08,shenNAT16} However, some of these QSL candidates show evidence for spin-orbit coupling\cite{smithPRB03,winterPRB17} and disorder\cite{maPRB21}. Full characterization of the transition rates in the presence of these departures from the clean, SU(2) limit can be a topic of future work. On a different note, there is now rising interest in utilizing magnetic systems as a resource for quantum entanglement of distant qubits.\cite{yuanCM21} Quantum spin liquids are prototypical spin models with extensive many-body quantum entanglement.\cite{savaryRPR17} It would be interesting to examine how this entanglement can be exploited to couple qubits on the one hand, and how measurements on distant pairs of qubits can be used to quantify the extent and strength of entanglement in QSLs on the other.



%

\acknowledgments
S.~T. acknowledges support by CUNY Research Foundation Project \#90922-07 10 and PSC-CUNY Research Award Program \#63515-00 51. Y.~T. acknowledges support by the US Department of Energy, Office of Basic Energy Sciences under Grant No. DE-SC 0012190.


\onecolumngrid
\appendix
\allowdisplaybreaks
\section{NV relaxometry}
\label{app1}
\begin{figure}[t]
\includegraphics[width=0.62\linewidth]{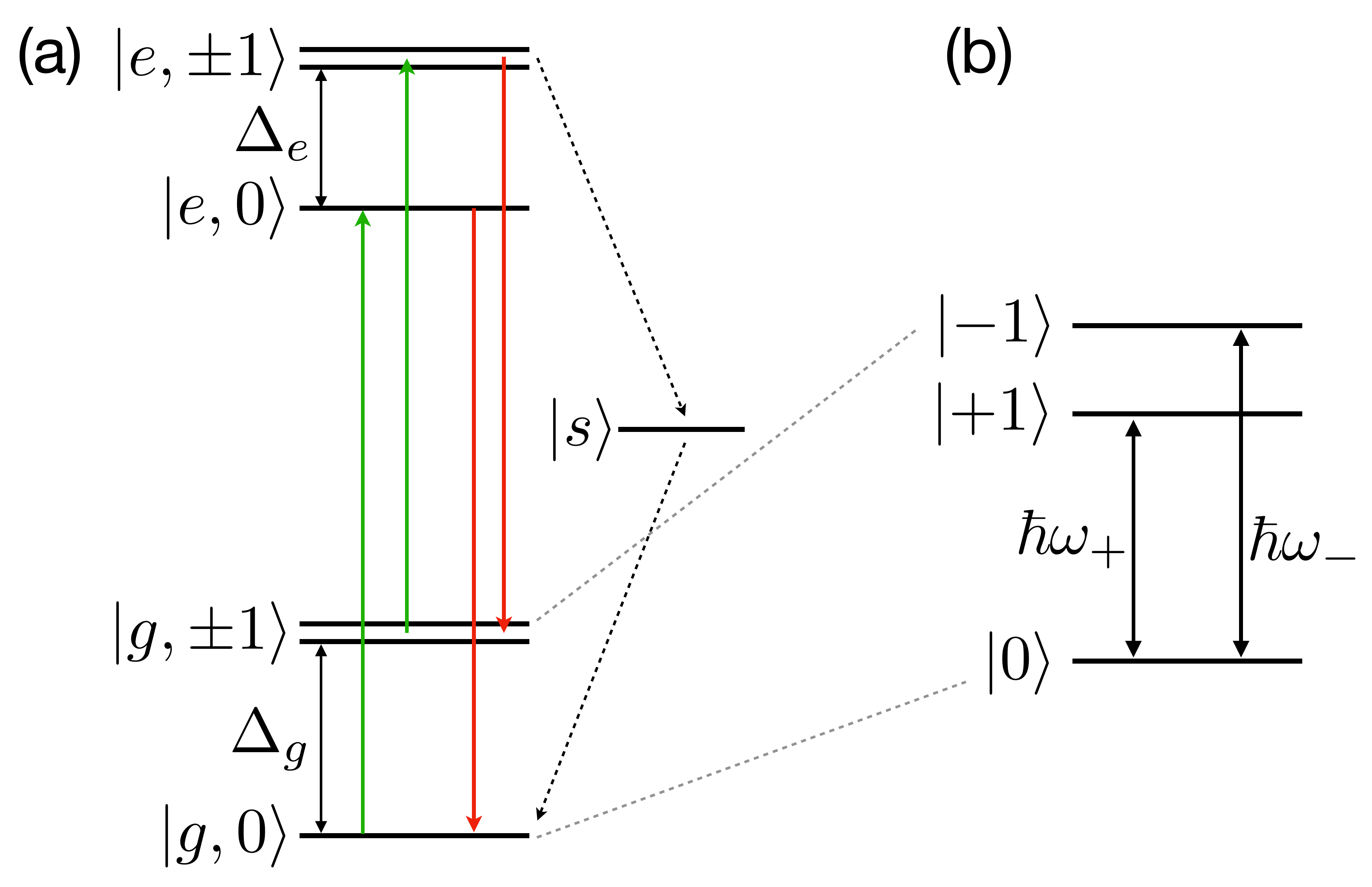}
\caption{(color online) (a) Energy diagram of the NV center consisting of the spin triplet ground state $\ket{g}$, the spin triplet excited state $\ket{e}$, and the intermediate singlet state $\ket{s}$. The zero-field splittings of the ground and excited states are given by $\De_g$ and $\De_e$, respectively. Excitation with a green laser results in two spin-conserving decay paths with red fluorescence and one spin non-conserving decay path via the intermediate state $\ket{s}$ with no fluorescence in the visible range. (b) Eigenstates of the bare NV Hamiltonian $H_0$, i.e., $0,\hbar\W_\pm=\De_g\mp\hbar\tg B_0$, in the presence of a static magnetic field $B_0$ applied along the NV spin axis.}
\label{figs1}
\end{figure}
In this appendix, we give a brief introduction to the physics of NV centers. The NV center is a lattice defect in diamond formed by a substitutional nitrogen atom bound to a lattice vacancy in the diamond lattice.\cite{dohertyPR13,grinoldsNATP13,rondinRPP14} The relevant energy diagram consists of the spin triplet ground state $\ket{g}$, the spin triplet excited state $\ket{e}$, and the intermediate singlet (dark) state $\ket{s}$, all located within the band gap of diamond [see Fig.~\ref{figs1}(a)].\cite{casolaNRM18} A zero-field splitting $\De_g\approx2.87$~GHz separates the $m_s=0$ and $m_s=\pm1$ sub-states of the ground state, while the excited state has a similar triplet level structure with a smaller splitting $\De_e\approx1.42$~GHz. The magnetic resonances of the ground and excited states are both characterized by electron g-factors $g\approx2$.\cite{casolaNRM18}

An off-resonant green laser triggers a spin-conserving excitation of electrons from the ground state up into the excited state. These excited electrons can subsequently decay back to their original spin sub-state by emitting a red photon as shown in Fig.~\ref{figs1}(a). However, the electrons in the $\ket{e,\pm1}$ states have an additional non-radiative decay channel via the intermediate singlet state $\ket{s}$ that competes with the direct optical transition and directs them predominantly to the $\ket{g,0}$ state. This difference in the non-radiative $\ket{e}\rightarrow\ket{g}$ decay pathways between the $m_s=0$ and $m_s=\pm1$ spin projections is at the heart of the NV center's quantum sensing and quantum computing applications. It enables (i) a high-fidelity initialization of the spin state using a green laser and (ii) the identification of the spin state by monitoring the NV fluorescence count-rate, as the competing non-radiative decay path introduces a reduction in the fluorescence count-rate associated with the $\ket{\pm1}$ spin states.

One of the main magnetic-sensing applications of the NV center is relaxometry: the measurement of its longitudinal relaxation rates $\G_\pm$ between the $\ket{0}$ and $\ket{\pm1}$ states, performed by preparing the NV center into a spin eigenstate and tracking the state populations as a function of time.\cite{jakobiNATN17} These rates depend on the magnetic field power spectral density at the quantum impurity site; therefore, if the NV center is placed close to a quantum magnet, spin fluctuations therein generate a fluctuating magnetic field at the NV site and modify these rates.\cite{flebusPRL18,chatterjeePRB19} A direct relationship between the longitudinal relaxation rates and the dynamic spin response function of proximate quantum spin liquids (QSLs) is presented in Sec.~\ref{nvrates} of the main text. 


\section{Transition rates for a 3D stack of quantum spin chains}
\label{rates1ddetails}
In this Appendix, we provide the technical details leading up to Eq.~\eqref{rates1d} in the main text. Using Eq.~\eqref{dpf1d}, we directly obtain
\beq
\langle b_-(t)b_+(0)\rangle=i(\hbar\g)^2\int\frac{dq}{2\p}\sum_{n=-\infty}^\infty\Big\{I^2_{+-}(na_y,d,q)\chi^>_{+-}(q,t)+I^2_{-+}(na_y,d,q)\chi^>_{-+}(q,t)+I^2_{zz}(na_y,d,q)\chi^>_{zz}(q,t)\Big\}\,,
\eeq
where $n$ indexes the spin chains, which are separated by $a_y$,
\beq
\chi^>_{\al\be}(q,t)=-i\int d(x-x')\sang{s_\al(x,t)s_\be(x',0)}e^{-iq(x-x')}\,,
\eeq
and
\begin{align}
I_{+-}(y_n,d,q)&=\int dx\frac32\frac{(x-iy_n)^2}{(d^2+x^2+y_n^2)^{5/2}}e^{iqx}\\
I_{-+}(y_n,d,q)&=\int dx\square{\frac32\frac{(x^2+y_n^2)^2}{(d^2+x^2+y_n^2)^{5/2}}-\frac{1}{(d^2+x^2+y_n^2)^{3/2}}}e^{iqx}\\
I_{zz}(y_n,d,q)&=\int dx\frac{3d(y_n+ix)}{(d^2+x^2+y_n^2)^{5/2}}e^{iqx}\,.
\end{align}
where $y_n=na_y$. Performing these integrals,
\begin{align}
I_{+-}(y_n,d,q)&=\frac{3|q|K_1(|q|\sqrt{d^2+y_n^2})}{\sqrt{d^2+y_n^2}}+\frac{2y_nq|q|K_1(|q|\sqrt{d^2+y_n^2})}{\sqrt{d^2+y_n^2}}-\frac{(d^2+2y_n^2)q^2K_2(|q|\sqrt{d^2+y_n^2})}{d^2+y_n^2}\\
I_{-+}(y_n,d,q)&=\frac{|q|K_1(|q|\sqrt{d^2+y_n^2})}{\sqrt{d^2+y_n^2}}-\frac{d^2q^2K_2(|q|\sqrt{d^2+y_n^2})}{d^2+y_n^2}\\
I_{zz}(y_n,d,q)&=\frac{2y_ndq^2K_2(|q|\sqrt{d^2+y_n^2})}{d^2+y_n^2}-\frac{2dq|q|K_1(|q|\sqrt{d^2+y_n^2})}{\sqrt{d^2+y_n^2}}\,.
\end{align}

In order to consider stacking layers of quantum spin chains along the vertical $z$ axis, we replace $d$ by $d+ma_\perp$ and sum over of $m$ from $0$ to $\infty$. If we approximate both the $m$ and $n$ sums by integrals, we may write
\beq
\langle b_-(t)b_+(0)\rangle=i\frac{(\hbar\g)^2}{a_ya_\perp}\int\frac{dq}{2\p}\int_d^\infty d\la\int_{-\infty}^\infty dy\Big\{I^2_{+-}(y,\la,q)\chi^>_{+-}(q,t)+I^2_{-+}(y,\la,q)\chi^>_{-+}(q,t)+I^2_{zz}(y,\la,q)\chi^>_{zz}(q,t)\Big\}\,,
\eeq
Similar evaluation can be performed for the second term, i.e., $C_{+-}(-\W_{\pm})$ in Eq.~\eqref{ratesgen}. If we finally rescale $y\rightarrow\la y$ and $\la\rightarrow d\nu$, the rates become
\begin{multline}
\G_\pm=-\frac{d^2}{a_ya_\perp}\frac{\p(\hbar\g\tg)^2}{d^4}\coth\round{\frac{\hbar\W_\pm}{2k_BT}}\int_0^\infty dq\\
\times {\rm Im}\curly{\frac14f^{+-}_1(qd)\chi^R_{+-}(q,\W_\pm)+\frac14f^{-+}_1(qd)\chi^R_{-+}(q,\W_\pm)+f^{zz}_1(qd)\chi^R_{zz}(q,\W_\pm)}\,,
\end{multline}
where
\begin{multline}
f^{+-}_1(z)=\frac{2z^4}{\p^2}\int_1^\infty d\nu\nu\int_{-\infty}^\infty dy\Bigg[\frac{9K^2_1(z\nu\sqrt{1+y^2})}{(z\nu)^2(1+y^2)}+\frac{4y^2K^2_1(z\nu\sqrt{1+y^2})}{1+y^2}+\frac{(1+2y^2)^2K^2_2(z\nu\sqrt{1+y^2})}{(1+y^2)^2}\\
-\frac{6(1+2y^2)K_1(z\nu\sqrt{1+y^2})K_2(z\nu\sqrt{1+y^2})}{z\nu(1+y^2)^{3/2}}\Bigg]\,,
\end{multline}
\beq
f^{-+}_1(z)=\frac{2z^4}{\p^2}\int_1^\infty d\nu\nu\int_{-\infty}^\infty dy\Bigg[\frac{K^2_1(z\nu\sqrt{1+y^2})}{(z\nu)^2(1+y^2)}+\frac{K^2_2(z\nu\sqrt{1+y^2})}{(1+y^2)^2}-\frac{2K_1(z\nu\sqrt{1+y^2})K_2(z\nu\sqrt{1+y^2})}{z\nu(1+y^2)^{3/2}}\Bigg]\,,
\eeq
and
\beq
f^{zz}_1(z)=\frac{2z^4}{\p^2}\int_1^\infty d\nu\nu\int_{-\infty}^\infty dy\square{\frac{y^2K^2_2(z\nu\sqrt{1+y^2})}{(1+y^2)^2}+\frac{K^2_1(z\nu\sqrt{1+y^2})}{1+y^2}}\,.
\eeq
Here, $K_n$ is the modified Bessel function of the second kind. Once the integrals over $\nu$ and $y$ are performed, all of the filtering functions, $f^{\pm\mp}_1(z)$ and $f^{zz}_1(z)$, converge to the same function $f_1(z)$ plotted in Fig.~\ref{fig3}.

\section{U(1) quantum spin liquid with a spinon Fermi surface}
\label{u1qsl}
In this appendix, we provide a brief introduction to the physics of U(1) quantum spin liquids (QSLs) and a phenomenological derivation of the starting Hamiltonian~\eqref{spinonh}. 

The U(1) QSL with a spinon Fermi surface first emerged as a possible ground state of half-filled 2D Mott insulators in proximity to a metal-insulator transition. The most well-known candidate materials are the organic molecular crystals, e.g., $\ka$-(ET)$_2$Cu$_2$(CN)$_3$ ($\ka$-ET) and Pd(dmit)$_2$(EtMe$_3$Sb) (dmit),\cite{shimizuPRL03,itouPRB08} both of which can be modeled as an isotropic spin-1/2 system on the triangular lattice. These systems are believed to be close to the Mott transition because of its tendency toward metallic behavior under modest pressures. 

At ambient pressures, these materials satisfy the standard necessary conditions for a QSL, i.e., they are insulating and display no magnetic ordering down to millikelvin temperatures despite its relatively large antiferromagnetic exchange constant of order $J\sim100$~K. Intriguingly, thermodynamic measurements show a linear temperature dependence of the specific heat and Pauli-like spin susceptibility at low temperatures and a Wilson ratio of order 1,\cite{yamashitaNATP08a,yamashitaNATC11} suggesting that the low-energy excitations are nearly-free spin-1/2 fermions with a Fermi surface. This scenario is further corroborated by low-temperature thermal conductivity data, which show a linear-$T$ contribution in addition to the $T^3$ phonon contribution that can be attributed to the spin subsystem.\cite{yamashitaNATP08b,yamashitaSCI10}

Different theoretical approaches exist to understand this QSL state.\cite{florensPRB04,leePRL05,motrunichPRB05} Lee and Lee investigated the single-orbital Hubbard model on the triangular lattice using the parton mean-field and slave-rotor approaches to show the existence of a stable QSL state near the Mott transition and that this state is a U(1) gauge theory coupled to spinons with a Fermi surface.\cite{leePRL05} A similar conclusion was drawn using the strong coupling expansion, where the inclusion of the terms to fourth order in $t/U$, justified for weak Mott insulators, was found to stabilize the spinon Fermi surface state.\cite{motrunichPRB05} 

In the phenomenological model proposed by Zhou and Ng, the U(1) QSL is viewed as a kind of Landau Fermi liquid subjected to specific constraints.\cite{zhouPRB13} The construction of the QSL state begins with the ordinary (metallic) Fermi liquid as the parent state, which becomes unstable to the spinon Fermi surface state as the interaction strength is increased. The phenomenological model therefore suggests a scenario in which the Fermi surface of the parent metal is not destroyed but the Landau quasiparticles are converted into charge-neutral, spin-1/2 spinons at the Mott transition. 
An important ingredient in this phenomenological approach is to place appropriate constraints on the Landau parameters to ensure that the low-energy excitations in the spin liquid state carry heat but no charge. We now briefly sketch this phenomenological model below. 

\begin{figure}[b]
\includegraphics[width=0.65\linewidth]{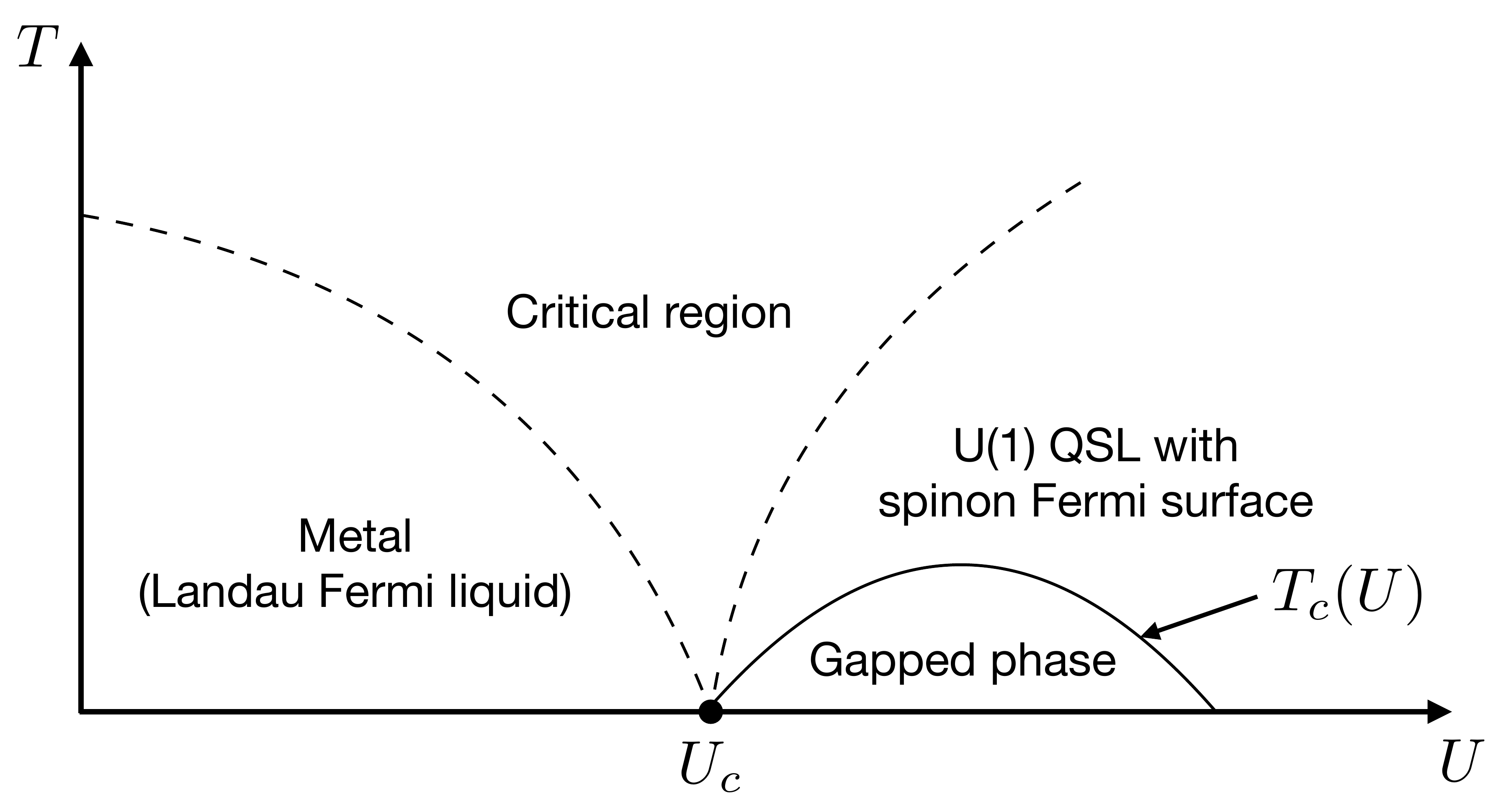}
\caption{A schematic $U$-$T$ phase diagram showing the transition from the metallic Fermi liquid state to the insulating U(1) QSL state as a finite-temperature crossover. The Pomeranchuk instability is denoted by $U_c$, and the possible instability toward a gapped phase at lower temperatures is shown. The phenomenology is not capable of capturing the exact nature of this gapped phase or the finite-temperature ``critical region" around $U_c$. The phase diagram has been adapted directly from Ref.~\onlinecite{zhouPRB13}.}
\label{figs2}
\end{figure}
Central to Landau's Fermi liquid theory is the adiabatic assumption, i.e., the existence of a one-to-one correspondence between the low-lying excited states of the free Fermi gas and those of the interacting system as the interactions are adiabatically switched on. In the phenomenological QSL model, one further assumes that the same labeling scheme holds in the QSL state, so that the low-energy excitations in the spinon Fermi surface state are still described by the same occupation numbers as the free Fermi gas. This one-to-one correspondence between the free Fermi gas and QSL states is a postulate of the model, as it is not guaranteed by adiabaticity. 

Under this assumption, we start with the energy (difference) functional for the ordinary Landau Fermi liquid, 
\beq
\label{edifffunc}
\De E=E-E_0=\sum_\bk\xi_\bk\de n_{\bk\s}+\frac{1}{2}\sum_{\bk\bk'}\sum_{\s\s'}f_{\s\s'}(\bk,\bk')\de n_{\bk\s}\de n_{\bk'\s'}\,,
\eeq
where $\xi_\bk$ is the quasiparticle dispersion, $\de n_{\bk\s}$ is the deviation in the average occupation number from its zero temperature value, i.e., $\de n_{\bk\s}\equiv n_{\bk\s}-\Theta(-\xi_\bk)$, and $f_{\s\s'}(\bk,\bk')$ is the Landau interaction function. For a system with spatial isotropy and spin rotational symmetry, the interaction function can be expanded in terms of the spin-symmetric and spin-asymmetric components, i.e.,
\beq
f_{\s\s'}(\bk,\bk')=f^s(\thi)+\s\s'f^a(\thi)\,,
\eeq
where $\cos\thi=\hat\bk\cdot\hat\bk'$ and we have set $k=k'=k_F$ since we are only interested in states close to the Fermi surface. In 2D, the symmetric and anti-symmetric functions can be expanded in terms of the Chebyshev polynomials $T_\ell(\cos\thi)=\cos(\ell\thi)$, 
\beq
f^{s,a}(\thi)=\sum_{\ell=0}^\infty f^{s,a}_\ell T_\ell(\cos\thi)\,.
\eeq
The dimensionless Landau parameters are then defined by
\beq
F^{s,a}_\ell=g_0Af^{s,a}_\ell\,,
\eeq
where $g_0=m^*/2\p \hbar^2$ is the quasiparticle density of states per spin at the Fermi level ($m^*$ being the quasiparticle effective mass), and $A$ is the system area. 

To distinguish the QSL state from the conventional Fermi liquid, Zhou and Ng note how the quasiparticle interactions renormalize the charge and thermal currents. The Landau theory in two dimensions gives
\beq
\bJ_c=\frac{m}{m^*}\round{1+F^s_1}\bJ_c^{(0)}\,,\ \ \ \ \bJ_Q=\frac{m}{m^*}\bJ_Q^{(0)}\,,
\eeq
where $\bJ_c^{(0)}$ and $\bJ_Q^{(0)}$ denote the charge and thermal currents, respectively, in the corresponding free system.\cite{baymBOOK91} In systems with Galilean invariance, we have the constraint $m^*/m=1+F^s_1$, and the charge current is unrenormalized by interactions. An important point is that in non-Galilean invariant systems, e.g., electrons in crystalline solids, this constraint no longer holds and the charge and thermal currents can be renormalized asymmetrically by interactions. The Mott transition at the critical Hubbard interaction $U_c$ then corresponds to the point at which $1+F^s_1(U_c)=0$, where the original Landau quasiparticles are converted into chargeless, spin-1/2 spinons. As illustrated in Fig.~\ref{figs2}, the quantity $1+F^s_1(U)$ vanishes as $U$ approaches $U_c$, and is assumed to remain zero for $U>U_c$.

As pointed out by Zhou and Ng, the critical point $1+F^s_1(U_c)=0$ coincides with the Pomeranchuk instability point, at which the Fermi surface becomes unstable with respect to deformations.\cite{zhouPRB13,zhouRMP17} Therefore, the U(1) QSL state with a spinon Fermi surface is likely to transition into a more stable QSL phase at lower temperatures that gap out part of or the entire Fermi surface. As shown in Fig.~\ref{figs2}, the system may be driven into a gapped QSL phase for $T<T_c(U)$. The precise nature of the low-temperature QSLs cannot be captured within the phenomenological model because it depends on the microscopic details of the system.

Our basic starting action is therefore given by
\beq
\label{fullS}
S=\int dt\sum_{\bk\s}\bar\y_{\bk\s}(t)\round{i\pd_t-\xi_\bk/\hbar}\y_{\bk\s}(t)-\frac{1}{2\hbar g_0A}\int\frac{d\W}{2\p}\sum_\bq\square{F^s_0(\bq,\W)|\rho(\bq,\W)|^2+\frac{F^s_1(\bq,\W)}{v_F^2}|\bj(\bq,\W)|^2}\,,
\eeq
where $\y_\s(\br,t)$ is the spin-$\s$ quasiparticle field, $v_F=\hbar k_F/m^*$ is the Fermi velocity, and
\beq
\rho(\bq,t)=\sum_{\bk\s}\bar\y_{\bk\s}(t)\y_{\bk+\bq\s}(t)\,,\ \ \ \bj(\bq,t)=\frac{\hbar}{m^*}\sum_{\bk\s}\round{\bk+\frac\bq 2}\bar\y_{\bk\s}(t)\y_{\bk+\bq\s}(t)\,.
\eeq
We now note that if we retain only the $\bq=0$ term in the interaction term and ignore the frequency-dependences of the functions $F^s_0$ and $F^s_1$, Eq.~\eqref{fullS} maps precisely to the standard Landau Fermi-liquid phenomenology with the symmetric, $\ell=0,1$ Landau parameters only. Also under this specific mapping, the QSL state, which we are about to describe below, coincides with the Pomeranchuk instability within the standard Landau phenomenology. 

Here, we assume a renormalized quadratic spectrum, i.e., $\xi_\bk=\hbar^2(k^2-k_F^2)/2m^*$. This choice is not an essential aspect of this model but is made to align with the other microscopic approaches, e.g., Refs.~\onlinecite{leePRL05,motrunichPRB05}. If $F^s_0(\bq,\W)$ is expanded in powers of $\bq$ and $\W$, we may keep the lowest order term, i.e., the constant term, since the constant term is the most RG relevant term. Then the term proportional to $F^s_0$ leads to the quartic term in Eq.~\eqref{spinonh} once we make the identification $u\equiv F^s_0/g_0$. 

The gauge field $\ba$ can be introduced through the Hubbard-Stratonovich decoupling of the current-current interaction term, 
\beq
\frac{1}{2g_0A}\int dt\sum_\bq\frac{F^s_1(\bq)}{v_F^2}|\bj(\bq,t)|^2\rightarrow\int\frac{d\W}{2\p}\frac1A\sum_\bq\square{\bj(\bq,\W)\cdot\ba(-\bq,-\W)-\frac{n}{2m^*}\frac{|\ba(\bq,\W)|^2}{F^s_1(\bq,\W)}}\,,
\label{hst}
\eeq
where $n=k_F^2/2\p$ is the total average quasiparticle density. With the introduction of the gauge field, the gauge-invariant quasiparticle current then gains the ``diamagnetic" term,
\beq
\bj(\bq,t)\rightarrow\frac{\hbar}{m^*}\sum_{\bk\s}\round{\bk+\frac\bq 2}\bar\y_{\bk\s}(t)\y_{\bk+\bq\s}(t)-\frac{n}{m^*}\ba(\bq,t)\,.
\eeq
Combining Eqs.~\eqref{fullS} and \eqref{hst}, the action takes the form of the U(1) gauge theory,
\begin{multline}
\label{fullS2}
S=\frac1\hbar\int dt\int d^2\br\sum_\s\,\square{\bar\y_\s(\br,t)(i\hbar\pd_t+\mu)\y_\s(\br,t)-\frac{1}{2m^*}\bar\y_\s(\br,t)(-i\hbar\bnab-\ba)^2\y_\s(\br,t)}\\
-\frac1\hbar\int\frac{d\W}{2\p}\frac1A\sum_\bq\frac{n}{2m^*}\round{1+\frac{1}{F^s_1(\bq,\W)}}|\ba^2(\bq,\W)|^2-\frac u\hbar\int dt\int d^2\br\,\bar\y_\up(\br,t)\bar\y_\down(\br,t)\y_\down(\br,t)\y_\up(\br,t)\,.
\end{multline}

The Landau parameter $F^s_1(\bq,\W)$ may also be expanded in powers of $\bq$ and $\W$. Then in the spin liquid phase, where $1+F^s_1(0,0)\rightarrow0$, the vector gauge field becomes massless, so higher-order $(\bq,\W)$-dependent terms should be included in the Landau parameter to obtain non-singular results. Therefore, one may write
\beq
1+F^s_1(\bq,\W)\approx\al-\be\W^2+\g_tq_t^2+\g_lq_l^2\,,
\eeq
where $q_t=-i\bnab\times$ and $q_l=-i\bnab$ are the transverse (curl) and longitudinal (gradient) parts, respectively, of the small-$q$ expansion. In a QSL state, $\al=0$. Ref.~\onlinecite{zhouPRB13} also argues that $\g_l$ must be zero to ensure that the system is in an incompressible (insulator) state. The last term in Eq.~\eqref{fullS2} leads to the standard Maxwell Lagrangian involving the vector gauge field,\cite{zhouPRB13} i.e., 
\beq
S_\perp[\ba]=-\frac{n}{2m^*\hbar}\int dt\int d^2\br\square{\be\round{\frac{\pd\ba}{\pd t}}^2-\g_t\round{\bnab\times\ba}^2}\,.
\eeq
We then arrive at the following effective Lagrangian for the spin liquid,
\begin{multline}
\label{fullS3}
S=S_\perp[\ba]+\frac1\hbar\int dt\int d^2\br\sum_\s\,\square{\bar\y_\s(\br,t)(i\hbar\pd_t+\mu)\y_\s(\br,t)-\frac{1}{2m^*}\bar\y_\s(\br,t)(-i\hbar\bnab-\ba)^2\y_\s(\br,t)}\\
-\frac u\hbar\int dt\int d^2\br\,\bar\y_\up(\br,t)\bar\y_\down(\br,t)\y_\down(\br,t)\y_\up(\br,t)\,.
\end{multline}
This is the standard starting action for the U(1) quantum spin liquid that has been derived in various different ways.\footnote{The prefix `U(1)' in U(1) quantum spin liquid does not correspond to any microscopic symmetries of the underlying spin system, which we assume throughout to have full SU(2) symmetry. In the spinon Fermi surface state, the spin-1/2 operator is expressed using the Abrikosov representation, $s_\al=\y^\dag_{\s}\s^\al_{\s\s'}\y_{\s'}/2$, which introduces a symmetry under arbitrary local phase rotations of the fermions, i.e., $\y_\s\rightarrow e^{i\la}\y_\s$. This symmetry is identical to the regular gauge symmetry one encounters in, e.g., quantum electrodynamics. Requiring physical states to be invariant under the associated gauge transformation, Eq.~(\ref{fullS3}) is written in an explicitly gauge-invariant form. Since $\la$ is an arbitrary phase, the state is referred to as a U(1) state.} We emphasize that the field $\y_\s(\br,t)$ in Eq.~\eqref{fullS3} now represents charge-neutral spinons as opposed to charged fermionic quasiparticles as in Eq.~\eqref{fullS2}, since by this point the charge-neutrality condition $1+F^s_1(0,0)\rightarrow0$ has been applied.

We note that the low-energy, long-wavelength behavior of the gauge field is actually not determined by $S_\perp[\ba]$ but by their coupling to the fermion matter field. In the standard approach, the effective theory for the transverse gauge field is obtained by integrating out the fermions and incorporating the effects of the spinons within the random phase approximation (RPA).\cite{leePRB92,polchinskiNPB94} Following the same procedure and fixing ourselves to the Coulomb gauge (i.e., $\bnab\cdot\ba=0$), the effective Euclidean action for the gauge field becomes
\begin{align}
\label{seffa}
S_\perp^{\rm eff}[\ba]=\frac{1}{2}\sum_{\bq n}\round{\chi_dq^2+\frac{|\W_n|}{v_Fq}\frac{2\ve_F}{\p\hbar^3}}|\a(\bq,\W_n)|^2\,,
\end{align}
where $\W_n$ is the Matsubara frequency, $\chi_d=1/12\p\hbar m^*$ is the (2D) spinon diamagnetic susceptibility, and ``$\a$" corresponds to the component of $\ba$ transverse to $\bq$; the term proportional to the frequency represents Landau damping of gauge fluctuations due to the spinon continuum. 

We finally note that the QSL is subjected to the static perpendicular magnetic field $\bB_0=-B_0\ez$, which controls the resonance frequency of the NV center. In the presence of this field, Eq.~\eqref{fullS3} must be amended by the Zeeman term
\beq
\label{zeeman}
S_Z=\frac1\hbar\int dt\int d^2\br\sum_\s\,\bar\y_\s(\br,t)\frac{\s b_0}{2}\y_\s(\br,t)\,,
\eeq
where $b_0=\hbar\g B_0$ is the Zeeman energy. Equations~\eqref{fullS3} and \eqref{zeeman} directly lead to Eq.~\eqref{spinonh} in the main text. The effective action for the vector gauge bosons Eq.~\eqref{seffa} will become important when computing the gauge field corrections to the dynamic spin response functions: see Appendix~\ref{gaugecorr}.

\section{Time-dependent Hartree-Fock approximation}
\label{tdhfa}
In this appendix, we reproduce the standard RPA result by applying the time-dependent Hartree-Fock approximation. The formulation of the approximation presented here is similar to the formulation presented in Ref.~\onlinecite{mukherjeePRB18}. Here, we solely consider the case where spinon-gauge coupling is zero, so the starting spinon Hamiltonian is given by
\beq
H=\sum_{\bk\s}\xi_{\bk\s}f^\dag_{\bk\s}f_{\bk\s}+\frac uA\sum_{\bk_1\bk_2\bp}f^\dag_{\bk_1\up}f^\dag_{\bk_2\down}f_{\bk_2+\bp\down}f_{\bk_1-\bp\up}\equiv H_0+H_{\rm int}\,,
\eeq
where $\xi_{\bk\s}$ is the bare spinon dispersion. 

We are interested in the spin response,
\beq
\chi^R_{\al\be}(\bk,\bk';\bq,t)=-\frac{i}{4A}\Theta(t)\sum_{\s_i}\s^\al_{\s_1\s_2}\s^\be_{\s_3\s_4}\sang{[e^{iHt/\hbar}f^\dag_{\bk\s_1}f_{\bk+\bq\s_2}e^{-iHt/\hbar},f^\dag_{\bk'\s_3}f_{\bk'-\bq\s_4}]}\,.
\eeq
Let us now take the time derivative of the response function to find its equation of motion:
\begin{multline}
\label{chieom2}
i\hbar\pd_t\chi^R_{\al\be}(\bk,\bk';\bq,t)=\frac{i\Theta(t)}{4A}\sum_{\s_i}\s^\al_{\s_1\s_2}\s^\be_{\s_3\s_4}\sang{[[H,f^\dag_{\bk\s_1}(t)f_{\bk+\bq\s_2}(t)],f^\dag_{\bk'\s_3}f_{\bk'-\bq\s_4}]}\\
+\frac{\hbar\de(t)}{4A}\sum_{\s_i}\s^\al_{\s_1\s_2}\s^\be_{\s_3\s_4}\sang{[f^\dag_{\bk\s_1}f_{\bk+\bq\s_2},f^\dag_{\bk'\s_3}f_{\bk'-\bq\s_4}]}\,.
\end{multline}
The last term can be evaluated:
\beq
\frac{\hbar\de(t)}{4A}\sum_{\s_i}\s^\al_{\s_1\s_2}\s^\be_{\s_3\s_4}\sang{[f^\dag_{\bk\s_1}f_{\bk+\bq\s_2},f^\dag_{\bk'\s_3}f_{\bk'-\bq\s_4}]}=\frac{\hbar\de(t)}{4A}\sum_{\s_1\s_2}\s^\al_{\s_1\s_2}\s^\be_{\s_2\s_1}\round{n_{\bk\s_1}-n_{\bk+\bq\s_2}}\de_{\bk',\bk+\bq}\,,
\eeq
where $n_{\bk\s}=\sang{f_{\bk\s}^\dag f_{\bk\s}}$. If we now calculate the internal commutator in the first term of Eq.~\eqref{chieom2}, the first piece is given by
\beq
[H_0,f^\dag_{\bk\s_1}f_{\bk+\bq\s_2}]=\round{\xi_{\bk\s_1}-\xi_{\bk+\bq\s_2}}f^\dag_{\bk\s_1}f_{\bk+\bq\s_2}\,,
\eeq
while the second piece is given by
\begin{align}
[H_{\rm int},f^\dag_{\bk\s_1}f_{\bk+\bq\s_2}]&=\frac uA\sum_{\bk_1\bk_2\bp}\Big[f^\dag_{\bk_1\up}f^\dag_{\bk_2\down}f_{\bk_2+\bp\down}f_{\bk+\bq\s_2}\de_{\bk_1-\bp,\bk}\de_{\up\s_1}+f^\dag_{\bk_1\up}f^\dag_{\bk_2\down}f_{\bk+\bq\s_2}f_{\bk_1-\bp\up}\de_{\bk_2+\bp,\bk}\de_{\down\s_1}\\
&\ \ \ -f^\dag_{\bk_1\up}f^\dag_{\bk\s_1}f_{\bk_2+\bp\down}f_{\bk_1-\bp\up}\de_{\bk_2,\bk+\bq}\de_{\down\s_2}-f^\dag_{\bk\s_1}f^\dag_{\bk_2\down}f_{\bk_2+\bp\down}f_{\bk_1-\bp\up}\de_{\bk_1,\bk+\bq}\de_{\up\s_2}\Big]\\
&=\frac uA\sum_{\bk_1\bp}\Big[f^\dag_{\bk+\bp\up}f^\dag_{\bk_1\down}f_{\bk_1+\bp\down}f_{\bk+\bq\s_2}\de_{\up\s_1}+f^\dag_{\bk_1\up}f^\dag_{\bk-\bp\down}f_{\bk+\bq\s_2}f_{\bk_1-\bp\up}\de_{\down\s_1}\\
&\qquad\qquad\qquad-f^\dag_{\bk_1\up}f^\dag_{\bk\s_1}f_{\bk+\bp+\bq\down}f_{\bk_1-\bp\up}\de_{\down\s_2}-f^\dag_{\bk\s_1}f^\dag_{\bk_1\down}f_{\bk_1+\bp\down}f_{\bk-\bp+\bq\up}\de_{\up\s_2}\Big]\,.
\end{align}
We now apply the Hartree-Fock approximation on these four terms, assuming that the spin order points along the $z$ axis:
\begin{multline}
[H_{\rm int},f^\dag_{\bk\s_1}f_{\bk+\bq\s_2}]\rightarrow\frac uA\sum_{\bk_1}\Big[\sang{f^\dag_{\bk_1\down}f_{\bk_1\down}}f^\dag_{\bk\up}f_{\bk+\bq\s_2}\de_{\up\s_1}+\sang{f^\dag_{\bk+\bq\up}f_{\bk+\bq\up}}f^\dag_{\bk_1\down}f_{\bk_1+\bq\down}\de_{\up\s_1}\de_{\up\s_2}\\
-\sang{f^\dag_{\bk+\bq\down}f_{\bk+\bq\down}}f^\dag_{\bk_1\up}f_{\bk_1+\bq\down}\de_{\up\s_1}\de_{\down\s_2}+\sang{f^\dag_{\bk+\bq\down}f_{\bk+\bq\down}}f^\dag_{\bk_1\up}f_{\bk_1+\bq\up}\de_{\down\s_1}\de_{\down\s_2}+\sang{f^\dag_{\bk_1\up}f_{\bk_1\up}}f^\dag_{\bk\down}f_{\bk+\bq\s_2}\de_{\down\s_1}\\
-\sang{f^\dag_{\bk+\bq\up}f_{\bk+\bq\up}}f^\dag_{\bk_1\down}f_{\bk_1+\bq\up}\de_{\down\s_1}\de_{\up\s_2}-\sang{f^\dag_{\bk\down}f_{\bk\down}}f^\dag_{\bk_1\up}f_{\bk_1+\bq\up}\de_{\down\s_1}\de_{\down\s_2}-\sang{f^\dag_{\bk_1\up}f_{\bk_1\up}}f^\dag_{\bk\s_1}f_{\bk+\bq\down}\de_{\down\s_2}\\
+\sang{f^\dag_{\bk\up}f_{\bk\up}}f^\dag_{\bk_1\up}f_{\bk_1+\bq\down}\de_{\up\s_1}\de_{\down\s_2}-\sang{f^\dag_{\bk_1\down}f_{\bk_1\down}}f^\dag_{\bk\s_1}f_{\bk+\bq\up}\de_{\up\s_2}-\sang{f^\dag_{\bk\up}f_{\bk\up}}f^\dag_{\bk_1\down}f_{\bk_1+\bq\down}\de_{\up\s_1}\de_{\up\s_2}\\
+\sang{f^\dag_{\bk\down}f_{\bk\down}}f^\dag_{\bk_1\down}f_{\bk_1+\bq\up}\de_{\down\s_1}\de_{\up\s_2}\Big]\,.
\end{multline}
Now replacing the expectation values, we obtain
\begin{multline}
[H_{\rm int},f^\dag_{\bk\s_1}f_{\bk+\bq\s_2}]=\frac uA\sum_{\bk_1}\Big[(n_{\bk+\bq\up}-n_{\bk\up})f^\dag_{\bk_1\down}f_{\bk_1+\bq\down}\de_{\up\s_1}\de_{\up\s_2}+(n_{\bk+\bq\down}-n_{\bk\down})f^\dag_{\bk_1\up}f_{\bk_1+\bq\up}\de_{\down\s_1}\de_{\down\s_2}\\
-(n_{\bk+\bq\down}-n_{\bk\up})f^\dag_{\bk_1\up}f_{\bk_1+\bq\down}\de_{\up\s_1}\de_{\down\s_2}-(n_{\bk+\bq\up}-n_{\bk\down})f^\dag_{\bk_1\down}f_{\bk_1+\bq\up}\de_{\down\s_1}\de_{\up\s_2}\Big]\\
+\frac uA\Big[N_{\up}f^\dag_{\bk\down}f_{\bk+\bq\s_2}\de_{\down\s_1}+N_{\down}f^\dag_{\bk\up}f_{\bk+\bq\s_2}\de_{\up\s_1}-N_{\up}f^\dag_{\bk\s_1}f_{\bk+\bq\down}\de_{\down\s_2}-N_{\down}f^\dag_{\bk\s_1}f_{\bk+\bq\up}\de_{\up\s_2}\Big]\,,
\end{multline}
where $N_\s=\sum_\bk n_{\bk\s}$. 

If we now insert this result into Eq.~\eqref{chieom2}, the equation of motion for the $-+$ component becomes
\begin{multline}
i\hbar\pd_t\chi^R_{-+}(\bk,\bk';\bq,t)=\frac{\hbar\de(t)}{A}\round{n_{\bk\down}-n_{\bk+\bq\up}}\de_{\bk',\bk+\bq}-\round{\xi_{\bk\down}-\xi_{\bk+\bq\up}}\chi^R_{-+}(\bk,\bk';\bq,t)\\
+(n_{\bk+\bq\up}-n_{\bk\down})\frac uA\sum_{\bk_1}\chi^R_{-+}(\bk_1,\bk';\bq,t)-(N_{\up}-N_\down)\frac uA\chi^R_{-+}(\bk,\bk';\bq,t)\,.
\end{multline}
Summing over $\bk'$ and Fourier transforming to $\W$,
\begin{multline}
\hbar\W\chi^R_{-+}(\bk;\bq,\W)=\frac{\hbar}{A}\round{n_{\bk\down}-n_{\bk+\bq\up}}-\round{\xi_{\bk\down}-\xi_{\bk+\bq\up}}\chi^R_{-+}(\bk;\bq,\W)\\
+(n_{\bk+\bq\up}-n_{\bk\down})\frac uA\chi^R_{-+}(\bq,\W)-(N_{\up}-N_\down)\frac uA\chi^R_{-+}(\bk;\bq,\W)\,.
\end{multline}
where $\chi^R_{-+}(\bk;\bq,\W)=\sum_{\bk'}\chi^R_{-+}(\bk,\bk';\bq,\W)$. The ``$+-$" component can be computed in a similar way. Solving these equations, we finally obtain
\beq
\chi^R_{-+}(\bq,\W)=\frac{\chi^{R(0)}_{-+}(\bq,\W)}{1+\frac u\hbar\chi^{R(0)}_{-+}(\bq,\W)}\,,\ \ \ \chi^R_{+-}(\bq,\W)=\frac{\chi^{R(0)}_{+-}(\bq,\W)}{1+\frac u\hbar\chi^{R(0)}_{+-}(\bq,\W)}\,,
\eeq
where $\chi^R_{\pm\mp}(\bq,\W)=\sum_{\bk}\chi^R_{\pm\mp}(\bk;\bq,\W)$
\beq
\chi^{R(0)}_{-+}(\bq,\W)=\frac1A\sum_{\bk}\frac{n_F(\xi^{\rm HF}_{\bk\down})-n_F(\xi^{\rm HF}_{\bk+\bq\up})}{\W+\xi^{\rm HF}_{\bk\down}/\hbar-\xi^{\rm HF}_{\bk+\bq\up}/\hbar+i\eta}\,,\ \ \ \chi^{R(0)}_{+-}(\bq,\W)=\frac1A\sum_{\bk}\frac{n_F(\xi^{\rm HF}_{\bk\up})-n_F(\xi^{\rm HF}_{\bk+\bq\down})}{\W+\xi^{\rm HF}_{\bk\up}/\hbar-\xi^{\rm HF}_{\bk+\bq\down}/\hbar+i\eta}\,.
\eeq

To obtain the longitudinal response function, we define four components $\up\up$, $\up\down$, $\down\up$, and $\down\down$ such that 
\beq
\chi^R_{\parallel}(\bq,\W)=\chi^R_{zz}(\bq,\W)=\frac14\square{\chi^R_{\up\up}(\bq,\W)+\chi^R_{\up\down}(\bq,\W)+\chi^R_{\down\up}(\bq,\W)+\chi^R_{\down\down}(\bq,\W)}\,.
\eeq
Using Eq.~\eqref{chieom2}, the equations of motion for these four components become
\begin{align}
(\hbar\W+\xi_{\bk\up}-\xi_{\bk+\bq\up})\chi^R_{\up\up}(\bk;\bq,\W)&=\frac{\hbar}{A}\round{n_{\bk\up}-n_{\bk+\bq\up}}-\frac{u}{\hbar}\frac\hbar{A}(n_{\bk\up}-n_{\bk+\bq\up})\chi^R_{\down\up}(\bq,\W)\,,\\
(\hbar\W+\xi_{\bk\down}-\xi_{\bk+\bq\down})\chi^R_{\down\down}(\bk;\bq,\W)&=\frac{\hbar}{A}\round{n_{\bk\down}-n_{\bk+\bq\down}}-\frac{u}{\hbar}\frac\hbar{A}(n_{\bk\down}-n_{\bk+\bq\down})\chi^R_{\up\down}(\bq,\W)\,,\\
(\hbar\W+\xi_{\bk\up}-\xi_{\bk+\bq\up})\chi^R_{\up\down}(\bk;\bq,\W)&=-\frac{u}{\hbar}\frac\hbar{A}(n_{\bk\up}-n_{\bk+\bq\up})\chi^R_{\down\down}(\bq,\W)\,,\\
(\hbar\W+\xi_{\bk\down}-\xi_{\bk+\bq\down})\chi^R_{\down\up}(\bk;\bq,\W)&=-\frac{u}{\hbar}\frac\hbar{A}(n_{\bk\down}-n_{\bk+\bq\down})\chi^R_{\up\up}(\bq,\W)\,.
\end{align}
From this, we obtain
\beq
\chi^R_{\parallel}(\bq,\W)=\frac14\frac{\chi^{R(0)}_{\up\up}(\bq,\W)\round{1-\frac{u}{\hbar}\chi^{R(0)}_{\down\down}(\bq,\W)}+\chi^{R(0)}_{\down\down}(\bq,\W)\round{1-\frac{u}{\hbar}\chi^{R(0)}_{\up\up}(\bq,\W)}}{1-\round{\frac{u}{\hbar}}^2\chi^{R(0)}_{\up\up}(\bq,\W)\chi^{R(0)}_{\down\down}(\bq,\W)}\,,
\eeq
where
\beq
\chi^{R(0)}_{\s\s}(\bq,\W)=\frac1A\sum_{\bk}\frac{n_F(\xi^{\rm HF}_{\bk\s})-n_F(\xi^{\rm HF}_{\bk+\bq\s})}{\W+\xi^{\rm HF}_{\bk\s}/\hbar-\xi^{\rm HF}_{\bk+\bq\s}/\hbar+i\eta}\,.
\eeq

\section{Bare spin response functions}
\label{bubbles}
In this appendix, we provide detailed evaluations of the bare spin response functions Eq.~\eqref{chir0}. For the $+-$ component, we have
\beq
\chi^{R(0)}_{+-}(\bq,\W)=\frac{1}{A}\sum_\bk\frac{n_F(\xi^{\rm HF}_{\bk\up})-n_F(\xi^{\rm HF}_{\bk+\bq\down})}{\W+(\xi^{\rm HF}_{\bk\up}-\xi^{\rm HF}_{\bk+\bq\down})/\hbar+i\eta}=\frac{1}{A}\sum_\bk\frac{n_F(\xi^{\rm HF}_{\bk\up})-n_F(\xi^{\rm HF}_{\bk+\bq\down})}{\W+(\e_{\bk}-\e_{\bk+\bq}-\De)/\hbar+i\eta}\ ,
\eeq
where $\De=b_0+2um=b_0/(1-ug_0)$ and $\eta$ is an infinitesimal. If we now go to zero temperature, we have
\beq
\chi^{R(0)}_{+-}(\bq,\W)=\hbar\int\frac{d^2\bk}{(2\p)^2}\frac{\Theta(\ve_{F\up}-\e_\bk)}{\hbar\tW+\e_{\bk}-\e_{\bk+\bq}+i\hbar\eta}-\hbar\int\frac{d^2\bk}{(2\p)^2}\frac{\Theta(\ve_{F\down}-\e_\bk)}{\hbar\tW+\e_{\bk+\bq}-\e_{\bk}+i\hbar\eta}\,,
\eeq
where $\ve_{F\s}=\ve_F+\s\De/2$ and $\tW=\W-\De/\hbar$.
If we now scale the energies in the first integral by $\ve_{F\up}$ and in the second integral by $\ve_{F\down}$, we obtain
\beq
\chi^{R(0)}_{+-}(\bq,\W)=\frac{\hbar g_0}{2}\int_0^1kdk\int\frac{d\thi}{2\p}\frac{1}{\ty_\up-kx_\up\cos\thi-x_\up^2+i\eta}-\frac{\hbar g_0}{2}\int_0^1kdk\int\frac{d\thi}{2\p}\frac{1}{\ty_\down+kx_\down\cos\thi+x_\down^2+i\eta}\ ,
\eeq
where $\ty_\s=\hbar\tW/4\ve_{F\s}$ and $x_\s=q/2k_{F\s}$. Performing the integrals, we obtain
\begin{multline}
\chi^{R(0)}_{+-}(q,\W)=\frac{\hbar g_0}{4x_\up^2}\square{\tz_{\up-}-\Theta(\tz_{\up-}^2-x_\up^2)\sgn(\tz_{\up-})\sqrt{\tz_{\up-}^2-x_\up^2}-i\Theta(x_\up^2-\tz_{\up-}^2)\sqrt{x_\up^2-\tz_{\up-}^2}}\\
-\frac{\hbar g_0}{4x_\down^2}\square{\tz_{\down+}-\Theta(\tz_{\down+}^2-x_\down^2)\sgn(\tz_{\down+})\sqrt{\tz_{\down+}^2-x_\down^2}-i\Theta(x_\down^2-\tz_{\down+}^2)\sqrt{x_\down^2-\tz_{\down+}^2}}\ ,
\end{multline}
where $\tz_{\s\pm}=\ty_\s\pm x_\s^2$. If we define $\ty=(\hbar\W-\De)/4\ve_F$, $x=q/2k_{F}$, and $\tz_\pm=\ty\pm x^2$, 
\begin{multline}
\chi^{R(0)}_{+-}(q,\W)=\frac{\hbar g_0}{2x^2}\square{\Theta\round{\tz_{+}^2-x^2\tfrac{\ve_{F\down}}{\ve_F}}\sgn(\tz_{+})\sqrt{\tz_{+}^2-x^2\tfrac{\ve_{F\down}}{\ve_F}}+i\Theta\round{x^2\tfrac{\ve_{F\down}}{\ve_F}-\tz_{+}^2}\sqrt{x^2\tfrac{\ve_{F\down}}{\ve_F}-\tz_{+}^2}}\\
-\frac{\hbar g_0}{2x^2}\square{\Theta\round{\tz_{-}^2-x^2\tfrac{\ve_{F\up}}{\ve_F}}\sgn(\tz_{-})\sqrt{\tz_{-}^2-x^2\tfrac{\ve_{F\up}}{\ve_F}}+i\Theta\round{x^2\tfrac{\ve_{F\up}}{\ve_F}-\tz_{-}^2}\sqrt{x^2\tfrac{\ve_{F\up}}{\ve_F}-\tz_{-}^2}}-\hbar g_0\,.
\end{multline}
Finally, this can be reexpressed as
\begin{multline}
\label{chiroperp}
\chi^{R(0)}_{+-}(\bq,\W)=\frac{\hbar g_0}{2x^2}\bigg[\Theta\round{\tz_{+}^2-x^2(1-2\de)}\sgn(\tz_{+})\sqrt{\tz_{+}^2-x^2(1-2\de)}+i\Theta\round{x^2(1-2\de)-\tz_{+}^2}\sqrt{x^2(1-2\de)-\tz_{+}^2}\bigg]\\
-\frac{\hbar g_0}{2x^2}\bigg[\Theta\round{\tz_{-}^2-x^2(1+2\de)}\sgn(\tz_{-})\sqrt{\tz_{-}^2-x^2(1+2\de)}+i\Theta\round{x^2(1+2\de)-\tz_{-}^2}\sqrt{x^2(1+2\de)-\tz_{-}^2}\bigg]-\hbar g_0\,,
\end{multline}
where $\de\equiv\De/4\ve_F$. The $-+$ component can be obtained by the replacement $\De\rightarrow-\De$.

Let us now compute the equal-spin component:
\beq
\chi^{R(0)}_{\s\s}(\bq,\W)=\frac{1}{A}\sum_{\bk}\frac{n_F(\xi^{\rm HF}_{\bk\s})-n_F(\xi^{\rm HF}_{\bk+\bq\s})}{\W+\e_{\bk}-\e_{\bk+\bq}+i\eta}\ .
\eeq
A calculation similar to the transverse component results in
\begin{multline}
\chi^{R(0)}_{\s\s}(\bq,\W)=\frac{\hbar g_0}{2x^2}\bigg[\Theta\round{z_{+}^2-x^2\round{1+\s2\de}}\sgn(z_{+})\sqrt{z_{+}^2-x^2\round{1+\s2\de}}\\
+i\Theta\round{x^2\round{1+\s2\de}-z_{+}^2}\sqrt{x^2(1+\s2\de)-z_{+}^2}\bigg]\\
-\frac{\hbar g_0}{2x^2}\bigg[\Theta\round{z_{-}^2-x^2(1+\s2\de)}\sgn(z_{-})\sqrt{z_{-}^2-x^2(1+\s2\de)}\\
+i\Theta\round{x^2(1+\s2\de)-z_{-}^2}\sqrt{x^2(1+\s2\de)-z_{-}^2}\bigg]-\hbar g_0\,,
\end{multline}
where $z_\pm=\hbar\W/4\ve_F\pm(q/2k_F)^2$.

\section{Gauge correction to the bare spin response functions}
\label{gaugecorr}
In this appendix, we provide details for the insertion of a single gauge propagator into the bare spin response functions, i.e., the details behind Eqs.~\eqref{gaugecorrection} and \eqref{gaugecorrection2}. The technicalities presented here are similar to those in Ref.~\onlinecite{balentsPRB20}. 

The real-time action for the spinons coupled to gauge fluctuations is given by
\beq
S=\frac{1}{\hbar}\sum_\s\ \curly{\bar\y_\s(\br,t)\round{i \hbar\pd_t+\frac{\hbar^2\bnab^2}{2m^*}+\ve_F}\y_\s(\br,t)-\bj(\br,t)\cdot\ba(\br,t)}\,,
\eeq
where $\y_\s(\br,t)$ is the spin-$\s$ spinon field, and $\bj(\br,t)$ is the spinon current density. Fourier transforming to momentum space, 
\beq
\label{ssrt}
S=\sum_{\bk\s}\bar\y_{\bk\s}(t)(i\pd_t-\xi_\bk/\hbar)\y_{\bk\s}(t)-\frac{1}{2\hbar\sqrt{\sA}}\sum_{\bk\bk'\s}\bar\y_{\bk\s}(t)\bv_{\bk+\bk'}\cdot\ba_{\bk-\bk'}(t)\y_{\bk'\s}(t)\,,
\eeq
where $\xi_\bk=\hbar^2k^2/2m^*-\ve_F$ and $\bv_\bk=\hbar\bk/m^*$. 

In contrast to Ref.~\onlinecite{balentsPRB20}, which uses the imaginary-time formalism, we proceed by placing the real-time action on the Schwinger-Keldysh time-loop contour,\cite{kamenevBOOK11} i.e., working with real times and frequencies. The $+-$ component on the contour is defined as 
\beq
\x^{\ka\ka'}_{+-}(\bq,t)=-\frac{i}{A}\sum_{\bk_1\bk_2}\ang{\bar\y^{\ka}_{\bk_1\up}(t)\y^{\ka}_{\bk_1+\bq\down}(t)\bar\y^{\ka'}_{\bk_2\down}(0)\y^{\ka'}_{\bk_2-\bq\up}(0)}\,,
\eeq
where the superscripts $\ka,\ka'=\pm$ label the forward and backward branches of the time-loop contour on which the time variables lie. In this representation, $\ka=\ka'=+$, $\ka=-\ka'=+$, $\ka=-\ka'=-$, and $\ka=\ka'=-$ correspond to the time-ordered, lesser, greater, and anti-time ordered correlation functions, respectively, and the retarded component can be obtained via
\beq
\label{rpm}
\chi^R=\chi^{++}-\chi^{--}+\chi^{-+}-\chi^{+-}\,.
\eeq

The corrections with a single gauge propagator emerge at second order in $S_{\rm int}$, i.e.,
\beq
\label{dchiperp}
\de\x^{\ka\ka'}_{+-}(\bq,t)=\frac{i}{2A}\sum_{\bk_1\bk_2}\ang{\bar\y^{\ka}_{\bk_1\up}(t)\y^{\ka}_{\bk_1+\bq\down}(t)\bar\y^{\ka'}_{\bk_2\down}(0)\y^{\ka'}_{\bk_2-\bq\up}(0)S^2_{\rm int}}_0\,,
\eeq
where the interaction contribution to the action on the Keldysh contour reads [c.f. Eq.~\eqref{ssrt}]
\beq
\label{sintk}
S_{\rm int}=-\frac{1}{2\hbar\sqrt{\sA}}\int_{-\infty}^\infty dt\sum_{\ka=\pm}\sum_{\bk\bk'\s}\ka \bar\y^\ka_{\bk\s}(t)\bv_{\bk+\bk'}\cdot\ba^{\ka}_{\bk-\bk'}(t)\y^\ka_{\bk'\s}(t)\,,
\eeq
and the average $\sang{\cdots}_0$ is now taken with respect to the non-interacting spinon action. The sum over $\ka$ in Eq.~\eqref{sintk} encodes the sum over the two branches of the Keldysh contour, where the time integral over the backward branch introduces an overall negative sign, hence the pre-factor $\ka$. 

Equation~\eqref{dchiperp} generates three relevant terms: two diagrams (labeled below by 1 and 2) corresponding to self-energy corrections and one vertex correction (labeled below by 3), shown diagrammatically in Fig.~\ref{fig5} of the main text.  All three diagrams must be accounted for to maintain gauge invariance. These three corrections in Keldysh space read
\begin{align}
\de\x^{\ka\ka'}_{+-1}(\bq,\W)&=\frac{1}{(2\hbar)^2\sA^2}\sum_{ij}\sum_{\bk\bp}\sum_{\ka_1\ka_2}\int\frac{d\w}{2\p}\int\frac{d\nu}{2\p}\ka_1\ka_2\ D^{\ka_1\ka_2,ij}_{-\bp}(-
\nu)v^i_{2\bk+2\bq+\bp}v^j_{2\bk+2\bq+\bp}\\
\label{chi1}
&\qquad\qquad\qquad\qquad\times g^{\ka\ka_1}_{\bk+\bq\up}(\w+\nu)g^{\ka_1\ka_2}_{\bk+\bq+\bp\up}(\w+\W+\nu)g^{\ka_2\ka'}_{\bk+\bq\up}(\w+\nu)g^{\ka'\ka}_{\bk\down}(\w)\,,\\
\de\x^{\ka\ka'}_{+-2}(\bq,\W)&=\frac{1}{(2\hbar)^2\sA^2}\sum_{ij}\sum_{\bk\bp}\sum_{\ka_1\ka_2}\int\frac{d\w}{2\p}\int\frac{d\nu}{2\p}\ka_1\ka_2\ D^{\ka_1\ka_2,ij}_{-\bp}(-\nu)v^i_{2\bk+\bp}v^j_{2\bk+\bp}\\
\label{chi2}
&\qquad\qquad\qquad\qquad\times g^{\ka\ka'}_{\bk+\bq\up}(\w+\W)g^{\ka'\ka_1}_{\bk\down}(\w)g^{\ka_1\ka_2}_{\bk+\bp\down}(\w+\nu)g^{\ka_2\ka}_{\bk\down}(\w)\,,\\
\de\x^{\ka\ka'}_{+-3}(\bq,\W)&=\frac{1}{(2\hbar)^2\sA^2}\sum_{ij}\sum_{\bk\bp}\sum_{\ka_1\ka_2}\int\frac{d\w}{2\p}\int\frac{d\nu}{2\p}\ka_1\ka_2\ D^{\ka_1\ka_2,ij}_{-\bp}(-\nu)v^i_{2\bk+2\bq+\bp}v^j_{2\bk+\bp}\\
\label{chi3}
&\qquad\qquad\qquad\qquad\times g^{\ka\ka_1}_{\bk+\bq\up}(\w+\W)g^{\ka_1\ka'}_{\bk+\bq+\bp\up}(\w+\W+\nu)g^{\ka'\ka_2}_{\bk+\bp\down}(\w+\nu)g^{\ka_2\ka}_{\bk\down}(\w)\,.
\end{align}
Here, $g^{\ka\ka'}_{\bk\s}(\w)$ and $D^{\ka\ka',ij}_\bq(\W)$ are the spinon and gauge field Green functions, where $\ka$ and $\ka'$ once again label the contour branches. These Green functions can be ``rotated" from the $+/-$ basis to the $RAK$ (retarded, advanced, and Keldysh) basis via\cite{kamenevBOOK11}
\beq\begin{aligned}
G^R&=\frac{1}{2}\round{G^{++}-G^{--}+G^{-+}-G^{+-}}\\
G^A&=\frac{1}{2}\round{G^{++}-G^{--}-G^{-+}+G^{+-}}\,,\\
G^K&=\frac{1}{2}\round{G^{++}+G^{--}+G^{-+}+G^{+-}}
\label{rakpm}
\end{aligned}\eeq
where this transformation holds for both fermionic and bosonic propagators. The components of the spinon Green function matrix in the $RAK$ basis are given by
\begin{align}
g^R_{\bk\s}(\w)&=\frac{1}{\w-\xi^{\rm HF}_{\bk\s}/\hbar+i\eta}=g^{A*}_{\bk\s}(\w)\,,\\
g^K_{\bk\s}(\w)&=\tanh\round{\frac{\hbar\w}{2k_BT}}\square{g^R_{\bk\s}(\w)-g^A_{\bk\s}(\w)}\,.
\end{align}
If we place Eq.~\eqref{seffa} on the Keldysh contour, the effective action for the gauge fluctuations in terms of momentum and real frequency reads
\beq
S_{\rm eff}=\frac{i }{2}\int\frac{d\W}{2\p}\sum_{\bq}\sum_{i,j=x,y,z}\rvec{a^{i,c}_{-\bq}(-\W)}{a^{i,q}_{-\bq}(-\W)}\mat{D^K_{ij}(\bq,\W)}{D^R_{ij}(\bq,\W)}{D^A_{ij}(\bq,\W)}{0}^{-1}\cvec{a^{j,c}_{\bq}(\W)}{a^{j,q}_{\bq}(\W)}\,,
\eeq
where the components of the (RPA) gauge propagator read
\begin{align}
D^{R}_{ij}(\bq,\W)&=-\round{\de_{ij}-\frac{q_iq_j}{q^2}}\frac{1}{\chi_dq^2-i\frac{\W}{v_Fq}\frac{2\ve_F}{\p\hbar^3}}\equiv-\round{\de_{ij}-\frac{q_iq_j}{q^2}}d^R_\bq(\W)\\
D^K_{ij}(\bq,\W)&=\coth\round{\frac{\hbar\W}{2k_BT}}\square{D^{R}_{ij}(\bq,\W)-D^{A}_{ij}(\bq,\W)}\,.
\end{align}

Using Eqs.~\eqref{rpm}, \eqref{chi1}, and \eqref{rakpm}, we may write the imaginary part of the first correction as
\begin{multline}
{\rm Im}\{\de\x^{R}_{+-1}(\bq,\W)\}=\frac{1}{4\sA}\sum_{\bk}\int\frac{d\w}{2\p}\square{\tanh\round{\frac{\hbar(\w+\W)}{2k_BT}}-\tanh\round{\frac{\hbar\w}{2k_BT}}}\\
\times\square{g^R_{\bk+\bq\up}(\w+\W)\S^R_{\bk+\bq\up}(\w+\W)g^R_{\bk+\bq\up}(\w+\W)-(R\rightarrow A)}\square{g^R_{\bk\down}(\w)-g^A_{\bk\down}(\w)}\,,
\end{multline}
where the retarded and advanced spinon self-energies are defined by
\begin{align}
\label{sser}
\S^R_{\bk\s}(\w)&=\frac{2i}{(4\hbar)^2\sA}\int\frac{d\nu}{2\p}\sum_{\bp,ij}\Big[g^R_{\bk+\bp\s}(\w+\nu)D^K_{ij,-\bp}(-\nu)+g^K_{\bk+\bp\s}(\w+\nu)D^R_{ij,-\bp}(-\nu)\Big]v^i_{2\bk+\bp}v^j_{2\bk+\bp}\,,\\
\S^A_{\bk\s}(\w)&=\frac{2i}{(4\hbar)^2\sA}\int\frac{d\nu}{2\p}\sum_{\bp,ij}\Big[g^A_{\bk+\bp\s}(\w+\nu)D^K_{ij,-\bp}(-\nu)+g^K_{\bk+\bp\s}(\w+\nu)D^A_{ij,-\bp}(-\nu)\Big]v^i_{2\bk+\bp}v^j_{2\bk+\bp}\,.
\end{align}
Performing the sum over $ij$, we then obtain
\begin{align}
\S^R_{\bk\s}(\w)&\approx\frac{i v^2_F}{2\hbar^2\sA}\int\frac{d\nu}{2\p}\sum_{\bp}\Big[g^R_{\bk+\bp\s}(\w+\nu)d^K_{-\bp}(-\nu)+g^K_{\bk+\bp\s}(\w+\nu)d^R_{-\bp}(-\nu)\Big]\,,\\
\S^A_{\bk\s}(\w)&\approx\frac{i v^2_F}{2\hbar^2\sA}\int\frac{d\nu}{2\p}\sum_{\bp}\Big[g^A_{\bk+\bp\s}(\w+\nu)d^K_{-\bp}(-\nu)+g^K_{\bk+\bp\s}(\w+\nu)d^A_{-\bp}(-\nu)\Big]\,.
\end{align}
Similar evaluation of the second term gives
\begin{multline}
{\rm Im}\{\de\x^{R}_{+-2}(\bq,\W)\}=\frac{1}{4\sA}\sum_{\bk}\int\frac{d\w}{2\p}\square{\tanh\round{\frac{\hbar(\w+\W)}{2k_BT}}-\tanh\round{\frac{\hbar\w}{2k_BT}}}\\
\times\square{g^R_{\bk+\bq\up}(\w+\W)-g^A_{\bk+\bq\up}(\w+\W)}\square{g^R_{\bk\down}(\w)\S^R_{\bk\down}(\w)g^R_{\bk\down}(\w)-(R\rightarrow A)}\ .
\end{multline}

The self-energy turns out to be approximately independent of $\xi_k$,\cite{balentsPRB20} so terms that involve only retarded Green functions or only advanced Green functions do not contribute once $\xi_k$ integral is performed.
Using partial fractions, the sum of the first and second terms can then be rewritten as
\begin{multline}
{\rm Im}\{\de\x^R_{+-12}(\bq,\W)\}=\frac{1}{4\sA}\sum_{\bk}\int\frac{d\w}{2\p}\square{\tanh\round{\tfrac{\hbar\w}{2k_BT}}-\tanh\round{\tfrac{\hbar(\w+\W)}{2k_BT}}}\\
\times\curly{\tfrac{g^R_{\bk+\bq\up}(\w+\W)\square{\S^R_{\bk+\bq\up}(\w+\W)-\S^A_{\bk\down}(\w)}g^A_{\bk\down}(\w)}{\W-\frac{\xi_{\bk+\bq\up}}{\hbar}+\frac{\xi_{\bk\down}}{\hbar}+i\eta}+\tfrac{g^A_{\bk+\bq\up}(\w+\W)\square{\S^A_{\bk+\bq\up}(\w+\W)-\S^R_{\bk\down}(\w)}g^R_{\bk\down}(\w)}{\W-\frac{\xi_{\bk+\bq\up}}{\hbar}+\frac{\xi_{\bk\down}}{\hbar}-i\eta}}\ .
\end{multline}
Now performing the $\xi_k$ integral, we obtain
\beq
{\rm Im}\{\de\x^R_{+-12}(\bq,\W)\}=-\frac{m^*}{2\hbar}\int\frac{d\thi}{2\p}\int\frac{d\w}{2\p}\square{\tanh\round{\tfrac{\hbar\w}{2k_BT}}-\tanh\round{\tfrac{\hbar(\w+\W)}{2k_BT}}}{\rm Im}\curly{\tfrac{\S^R_{\bk+\bq\up}(\w+\W)-\S^A_{\bk\down}(\w)}{\round{\W-\frac{\xi_{\bk+\bq\up}}{\hbar}+\frac{\xi_{\bk\down}}{\hbar}+i\eta}^2}}\ .
\eeq
The $\bp$-integral in the self-energy, e.g., Eq.~\eqref{sser}, can be performed by decomposing $\bp$ into the component parallel to $\bk$ and perpendicular to $\bk$; we denote these components, respectively, by $p_\parallel$ and $p_\perp$. Evaluating the self-energy terms and performing the $p_\parallel$ integral only,\cite{balentsPRB20} we obtain
\begin{multline}
{\rm Im}\{\de\x^R_{+-12}(\bq,\W)\}=-\frac{k_F}{(2\hbar)^2}\int\frac{d\thi}{2\p}\int\frac{d\w}{2\p}\int\frac{dp_\perp}{2\p}\int\frac{d\nu}{2\p}\square{\tanh\round{\tfrac{\hbar(\w+\W)}{2k_BT}}-\tanh\round{\tfrac{\hbar\w}{2k_BT}}}\\
\times{\rm Im}\curly{\tfrac{\coth\round{\frac{\hbar\nu}{2k_BT}}\square{d^R_{-p_\perp}(-\nu)-d^A_{-p_\perp}(-\nu)}-\tanh\round{\frac{\hbar(\w+\W+\nu)}{2k_BT}}d^R_{-p_\perp}(-\nu)+\tanh\round{\frac{\hbar(\w+\nu)}{2k_BT}}d^A_{-p_\perp}(-\nu)}{\round{\W-v_Fq\cos\thi-\De/\hbar+i\eta}^2}}\ .
\end{multline}
Combining this with the vertex correction [see Eq.~\eqref{chi3}] and going to the zero temperature limit, we obtain
\begin{multline}
\label{transcomp}
{\rm Im}\{\de\x^R_{+-}(\bq,\W)\}=\frac{k_F}{(2\hbar)^2}\int\frac{d\thi}{2\p}\int\frac{d\w}{2\p}\int\frac{dp_\perp}{2\p}\int\frac{d\nu}{2\p}\square{\sgn(\w+\W)-\sgn(\w)}\\
\times{\rm Im}\bigg\{\square{\sgn(\nu)\round{d^R_{-p_\perp}(-\nu)-d^A_{-p_\perp}(-\nu)}-\sgn(\w+\W+\nu)d^R_{-p_\perp}(-\nu)+\sgn(\w+\nu)d^A_{-p_\perp}(-\nu)}\\
\times\round{\tfrac{1}{\round{\W-v_Fq\cos\thi-\hbar qp_\perp\sin\thi/m^*-\De/\hbar+i\eta}\round{\W-v_Fq\cos\thi-\De/\hbar+i\eta}}-\tfrac{1}{{\round{\W-v_Fq\cos\thi-\De/\hbar+i\eta}^2}}}\bigg\}\ .
\end{multline}
Let us define the following angle integrals,
\beq
\label{ipm}
I_\pm(z)=\int\frac{d\thi}{2\p}\frac{1}{\ty-\cos\thi\pm i\eta}\frac{1}{\ty-\cos\thi-z\sin\thi\pm i\eta}=\frac{|\ty|}{[(\ty\pm i\eta)^2-1]\sqrt{(\ty\pm i\eta)^2-(z^2+1)}}\ ,
\eeq
where $\ty=\W/v_Fq-\De/\hbar v_Fq$ and $z=p_\perp/k_F$. Inserting this result into the above expression, we may write the total correction due to gauge fluctuations as
\begin{multline}
{\rm Im}\{\de\x^R_{+-}(\bq,\W)\}=\frac{k_F}{(2\hbar v_Fq)^2}\int\frac{d\w}{2\p}\int\frac{d\nu}{2\p}\int\frac{dp}{2\p}\square{\sgn(\w+\W)-\sgn(\w)}\\
\times{\rm Im}\bigg\{\square{\sgn(\nu)\round{d^R_{-z}(-\nu)-d^A_{-z}(-\nu)}-\sgn(\w+\W+\nu)d^R_{-z}(-\nu)+\sgn(\w+\nu)d^A_{-z}(-\nu)}[I_+(z)-I_+(0)]\bigg\}\ .
\end{multline}
where we replaced $p_\perp\rightarrow p$, and
\beq
d^{R(A)}_{-z}(-\nu)=\frac{\p\hbar^3}{\ve_F}\frac{z}{z^3/6\pm i\hbar\nu/\ve_F}\ .
\eeq
This can be reexpressed as
\begin{multline}
{\rm Im}\{\de\x^R_{+-}(\bq,\W)\}=\frac{2k_F^2}{(2\hbar v_Fq)^2}\int\frac{d\w}{2\p}\int\frac{d\nu}{2\p}\int_{-\infty}^\infty\frac{dz}{2\p}\square{\sgn(\w+\W)-\sgn(\w)}\\
\times[\sgn(\w+\nu)-\sgn(\nu)]{\rm Im}\curly{d^A_{-z}(-\nu)\round{I_+(z)-I_+(0)}}\ .
\end{multline}
Doing the $\nu$-integral (and assuming $\W>0$),
\beq
{\rm Im}\{\de\x^R_{+-}(\bq,\W)\}=-\frac{k_F^2}{2\p(v_Fq)^2}\int_0^\W\frac{d\w}{2\p}\int_{0}^1 dzz{\rm Re}\curly{\log\round{1-\frac{6i\hbar\w}{z^3\ve_F}}\round{I_+(z)-I_+(0)}}\ .
\eeq
Let us split this integral into two parts:
\begin{align}
{\rm Im}\{\de\x^{R(1)}_{+-}(\bq,\W)\}&=-\frac{k_F^2}{4\p^2(v_Fq)^2}\int_{0}^\W d\w\int_{0}^1 dz\,z\,{\rm Re}\curly{\round{-\tfrac{6\io\hbar\w}{z^3\ve_F}}\round{I_+(z)-I_+(0)}}\ ,\\
{\rm Im}\{\de\x^{R(2)}_{+-}(\bq,\W)\}&=-\frac{k_F^2}{4\p^2(v_Fq)^2}\int_{0}^\W d\w\int_{0}^1 dz\,z\,{\rm Re}\curly{\square{\log\round{1-\tfrac{6\io\hbar\w}{z^3\ve_F}}+\tfrac{6\io\hbar\w}{z^3\ve_F}}\round{I_+(z)-I_+(0)}}\ .
\end{align}
The first integral is finite because at the lower limit the quantity in the parenthesis involving $I_+$ vanishes like $z^2$. At finite magnetic field, we may assume large $\ty$, and we obtain
\beq
{\rm Im}\{\de\x^{R(1)}_{+-}(\bq,\W)\})\approx-\frac{k_F^2}{2\p^2(2v_Fq)^2}\int_{0}^\W d\w{\rm Re}\curly{\round{-\tfrac{6\io\hbar\w}{\ve_F}}\frac{|\ty|}{[(\ty+ i\eta)^2-1]^{5/2}}}\approx0\ .
\eeq
Let us look at the second integral. Here we cannot expand in $\w$ because the leading term, proportional to $\w^2$, is divergent at small $z$. Hence we rescale instead $z\equiv\z(6\hbar\w/\ve_F)^{1/3}$ to obtain 
\begin{multline}
{\rm Im}\{\de\x^{R(2)}_{+-}(\bq,\W)\}=-\frac{k_F^2}{4\p^2(v_Fq)^2}\int_{0}^\W d\w\round{\frac{6\hbar\w}{\ve_F}}^{1/3}\int_{0}^{(\ve_F/6\hbar\w)^{1/3}} d\z\,\z\\
{\rm Re}\curly{\frac{|\ty|}{(\ty+ i\eta)^2-1}\square{\log\round{1-\frac{\io}{\z^3}}+\frac{\io}{\z^3}}\round{\frac{1}{\sqrt{\z_0^2-\z^2}}-\frac{1}{\z_0}}}\ ,
\end{multline}
where
\beq
\z_0\equiv\frac{\sqrt{(\ty+ i\eta)^2-1}}{(6\hbar\w/\ve_F)^{1/3}}\ .
\eeq
We are interested in the limit $\hbar\w/\ve_F\ll1$. Therefore, we can take the upper limit of the integration to infinity, and Taylor expand the term in the round brackets since $\z_0\gg1$ follows. We therefore arrive at
\begin{align}
{\rm Im}\{\de\x^{R(2)}_{+-}(\bq,\W)\}&=-\frac{k_F^2}{2\p^2(2v_Fq)^2}\int_{0}^\W d\w\round{\frac{6\hbar\w}{\ve_F}}^{4/3}{\rm Re}\curly{\frac{|\ty|}{[(\ty+ i\eta)^2-1]^{5/2}}\int_{0}^{\infty} d\z\,\z^3\square{\log\round{1-\frac{\io}{\z^3}}+\frac{\io}{\z^3}}}\\
&=-\frac{k_F^2}{2\p^2(2v_Fq)^2}\int_{0}^\W d\w\round{\frac{6\hbar\w}{\ve_F}}^{4/3}{\rm Re}\curly{\frac{|\ty|}{[(\ty+ i\eta)^2-1]^{5/2}}\frac{\p(\sqrt{3}+3\io)}{12}}\,.
\end{align}
Finally, performing the $\w$-integral, we obtain
\beq
{\rm Im}\{\de\x^{R(2)}_{+-}(\bq,\W)\}\approx-\frac{6^{4/3}\p\sqrt{3}}{14}\frac{\hbar^3k_F^2\ve_F(v_Fq)^2}{4\p^2}\round{\frac{\hbar\W}{\ve_F}}^{7/3}\frac{|\hbar\W-\De|}{|(\hbar\W-\De)^2-(\hbar v_Fq)^2|^{5/2}}\ .
\eeq
This result leads directly to Eq.~\eqref{gaugecorrection}. The $-+$ component can be obtained by the replacement $\De\rightarrow-\De$.

The technicalities presented for the transverse component can be applied to the longitudinal component. The three diagrams in Fig.~\ref{fig5} are given by
\begin{multline}
{\rm Im}\{\de\x^R_{\s\s}(\bq,\W)\}=\frac{k_F}{(4\hbar)^2}\int\frac{d\thi}{2\p}\int\frac{d\w}{2\p}\int\frac{dp_\perp}{2\p}\int\frac{d\nu}{2\p}\square{\sgn(\w+\W)-\sgn(\w)}\\
\times{\rm Im}\bigg\{\square{\sgn(\nu)\round{d^R_{-p_\perp}(-\nu)-d^A_{-p_\perp}(-\nu)}-\sgn(\w+\W+\nu)d^R_{-p_\perp}(-\nu)+\sgn(\w+\nu)d^A_{-p_\perp}(-\nu)}\\
\times\round{\tfrac{1}{\round{\W-v_Fq\cos\thi-\hbar qp_\perp\sin\thi/m^*+i\eta}\round{\W-v_Fq\cos\thi+i\eta}}-\tfrac{1}{{\round{\W-v_Fq\cos\thi+i\eta}^2}}}\bigg\}\ .
\end{multline}
Comparing with Eq.~\eqref{transcomp}, this longitudinal correction is given by the transverse correction with $\De=0$ and therefore leads directly to Eq.~\eqref{gaugecorrection2}.

\section{Spin response functions for the XXZ antiferromagnetic spin chain}
\label{spinchain}
In this appendix, we provide a derivation of the spin response functions for the XXZ spin chain using Luttinger liquid theory and bosonization. We begin with Eq.~\eqref{hf} in the main text and first in the absence of the magnetic field, i.e., $b_0=0$. The spinons are then in a half-filled state with Fermi points at $\pm k_F=\pm \p/2a$, where $a$ is the lattice constant of the spin chain, and the interactions are responsible for creating particle-hole excitations about this half-filled state. For $|\z|\ll1$, only those states close to the Fermi points are important, and we may linearize the dispersion about $k=\pm k_F$ and split the fermion operator into left- and right-moving fields. The continuum fermion fields can then be written as
\beq
\label{psic}
\y(x_j)=\frac{\y_j}{\sqrt{a}}=R(x_j)e^{ik_Fx_j}+L(x_j)e^{-ik_Fx_j}\,,
\eeq
where the two continuum chiral fields obey
\beq
R(x)=\frac{1}{\sqrt{L}}\sum_{k}R_ke^{ikx}\ ,\ \ L(x)=\frac{1}{\sqrt{L}}\sum_{k}L_ke^{ikx}\,,
\eeq
and $\{R_k,R_{k'}\}=\de_{kk'}$ and $\{L_k,L_{k'}\}=\de_{kk'}$. The normal-ordered density operators together with their associated phonon fields are then defined by
\beq
\label{densityops}
\rho_L(x)\equiv\ :L^\dag(x)L(x):\ =\frac{\pd_x\f_L(x)}{\p}\ ,\ \ \rho_R(x)\equiv\ :R^\dag(x)R(x):\ =\frac{\pd_x\f_R(x)}{\p}\,,
\eeq
and if we also define $\f(x)\equiv\f_L(x)+\f_R(x)$, the total fermion density at any point $x$ can be expressed as
\beq
\rho(x)=\rho_L(x)+\rho_R(x)=\frac{\pd_x\f(x)}{\p}\,.
\eeq
The chiral boson fields are then related to the continuum chiral fermion fields via
\beq
\label{chiralf}
R(x)=\frac{\eta_R}{\sqrt{2\p\al}}e^{2i\f_R(x)}\ ,\ \ \ L(x)=\frac{\eta_L}{\sqrt{2\p\al}}e^{-2i\f_L(x)}\ ,
\eeq
where $\eta_{R,L}$ are Majorana fermion variables obeying $\{\eta_\nu,\eta_{\nu'}\}=2\de_{\nu\nu'}$, and $\al$ is a short-length UV cutoff. From Eqs.~\eqref{jwt2}, \eqref{densityops}, and \eqref{chiralf}, the original spin density operators can now be expressed in terms of the boson fields as
\begin{align}
\label{szb}
S^z&=\frac{\pd_x\f(x)}{\p}+\frac{\eta_R\eta_L}{2\p\al}e^{-2ik_Fx}e^{-2i\f(x)}+\frac{\eta_L\eta_R}{2\p\al}e^{2ik_Fx}e^{2i\f(x)}\,,\\
\label{smb}
S^-&=\frac{e^{-i\thi(x)}}{2\sqrt{2\p\al}}\square{\eta_Re^{2i\f(x)}+(\eta_R+\eta_L)e^{2ik_Fx}+\eta_Le^{-2i\f(x)}}\,,\\
\label{spb}
S^+&=\frac{e^{i\thi(x)}}{2\sqrt{2\p\al}}\square{\eta_Le^{2i\f(x)}+(\eta_R+\eta_L)e^{-2ik_Fx}+\eta_Re^{-2i\f(x)}}\,,
\end{align}
where $\thi(x)=\f_L(x)-\f_R(x)$.

The Fourier components of the left and right density operators are given by
\beq
\rho_R(k)=\frac{1}{\sqrt{L}}\sum_{k'}R^\dag_{k'}R_{k'+k}\ ,\ \ \rho_L(k)=\frac{1}{\sqrt{L}}\sum_{k'}L^\dag_{k'}L_{k'+k}\,,
\eeq
and obey the following commutation relation,
\beq
\label{rrcr}
[\rho_{R,L}(-k),\rho_{R,L}(k')]=\mp\frac{k}{2\p}\de_{kk'}\,.
\eeq
We may then represent the boson fields in terms of Fourier components of the left and right density operators,
\beq
\label{chiralbs}
\begin{aligned}
\f_L(x)&=\frac{\p}{\sqrt{L}}\sum_{k>0}\frac{e^{-\al k/2}}{ik}\square{e^{ikx}\rho_L(k)-e^{-ikx}\rho_L(-k)}\,,\\
\f_R(x)&=\frac{\p}{\sqrt{L}}\sum_{k>0}\frac{e^{-\al k/2}}{ik}\square{e^{ikx}\rho_R(k)-e^{-ikx}\rho_R(-k)}\,.
\end{aligned}
\eeq
If we then interpret $\f(x)$ as the canonical position, the conjugate momentum is given by
\beq
\Pi(x)=\hbar\frac{\pd_x\thi(x)}{\p}\,,
\eeq
and one may readily check that $[\f(x),\Pi(x')]=i\hbar\de(x-x')$. 

In terms of these variables, Eq.~\eqref{hf} can be written in terms of the continuum boson fields as
\beq
\label{hcb2}
H=\frac{u}{2}\int dx\curly{\frac{\p K}{\hbar}\Pi^2(x)+\frac{\hbar }{\p K}\square{\pd_x\f(x)}^2}+\frac{2J\z a}{(2\p\al)^2}\int dx\cos\square{4\f(x)}\equiv H_0+H_u\ ,
\eeq
where the speed of sound and the Luttinger parameter are given by
\beq
u=\frac{Ja}{\hbar}\round{1+\frac{4\z}{\p}}^{1/2}\ ,\qquad K=\round{1+\frac{4\z}{\p}}^{-1/2}\ .
\eeq

The second term in Eq.~\eqref{hcb2} is the so-called umklapp term, which describes the scattering of two fermions from one Fermi point (say $-k_F$) to the other (i.e., $+k_F$) accompanied by a momentum transfer of $4k_F$. For zero magnetic field, the umklapp term flows to zero (i.e., it is RG irrelevant) in the critical ``XY" regime $|\z|<1$. The low-energy XXZ spin chain is then described by the Gaussian Hamiltonian $H_0$, and the exact expressions for $u$ and $K$ over the entire $|\z|<1$ region can be extracted from the Bethe-ansatz solution\cite{johnsonPRA73}
\beq
K=\frac{\p}{2(\p-\cos^{-1}\z)}\ ,\ \ \ u=\frac{\p Ja\sqrt{1-\z^2}}{2\hbar\cos^{-1}\z}\ .
\eeq
At a finite magnetic field, the RG flow terminates at $b_0$. However, since the umklapp term is irrelevant for $|\z|<1$, one may still expect the umklapp term to flow to a finite but very small value as long as $b_0$ is much smaller than the cutoff scale. Therefore, we ignore $H_u$ altogether in the rest of the discussion and simply take
\beq
H\approx H_0=\frac{\hbar u}{2\p}\int dx\curly{K\square{\pd_x\thi(x)}^2+\frac{1}{K}\square{\pd_x\f(x)}^2}\ .
\eeq

Let us now include the magnetic field. The effect of the magnetic field is to introduce a chemical potential term to the Hamiltonian, i.e.,
\beq
H_Z=-b_0\int dx\,S^z(x)=-\frac{b_0}{\p}\int dx\,\pd_x\f\,,
\eeq
and therefore to dope the system away from half-filling. 
In the critical regime (where the umklapp term is irrelevant), changing the chemical potential directly changes the magnetization,
\beq
M=\ang{S^z}=\round{\frac{K}{\p\hbar u}}b_0\,,
\eeq
where the coefficient $K/\p\hbar u$ defines the compressibility of the interacting fermion gas. The magnetization $M$ also implies an increase in the densities of the left and the right chiral spinons by $M/2$, which must now be incorporated in the boson fields, c.f. Eq.~\eqref{chiralbs},
\beq\begin{aligned}
\f_L(x)&=\frac{\p}{\sqrt{L}}\sum_{k>0}\frac{e^{-\al k/2}}{ik}\square{e^{ikx}\rho_L(k)-e^{-ikx}\rho_L(-k)}+M\frac{\p x}{2}\,,\\
\f_R(x)&=\frac{\p}{\sqrt{L}}\sum_{k>0}\frac{e^{-\al k/2}}{ik}\square{e^{ikx}\rho_R(k)-e^{-ikx}\rho_R(-k)}+M\frac{\p x}{2}\,,
\end{aligned}\eeq
where now the last terms describe the uniform density increase of the two species. 
The total Hamiltonian of the spin chain and its Zeeman coupling to the magnetic field can be written as
\beq
\label{hphz}
H+H_Z=\frac{\hbar u}{2\p}\int dx\curly{K\square{\pd_x\thi(x)}^2+\frac{1}{K}\square{\pd_x\bar\f(x)}^2}-\frac{L}{2}\frac{K}{\p\hbar u}b_0^2\,,
\eeq
where the boson field, describing the fluctuations relative to the shifted density, is given by $\bar\f=\f-\p Mx$. 

We now rewrite the spin operators Eqs.~\eqref{szb}, \eqref{smb}, and \eqref{spb} in terms of the new boson field $\bar\f$. Since our interest is in the noise, it is only necessary to retain the {\em fluctuating part} of the spin operators. Furthermore, Friedel terms proportional $e^{\pm2i k_Fx}$ only generates a fast-oscillating response which we ignore in order to focus on the long-wavelength response. We then write
\beq
\label{sbosons}
S^z\approx\frac{\pd_x\bar\f(x)}{\p}\,,\ \ \ S^\pm\approx\frac{e^{i\thi(x)}}{2\sqrt{2\p\al}}\square{\eta_Le^{\pm2i\bar\f(x)}e^{\pm2i\p Mx}+\eta_Re^{\mp2i\bar\f(x)}e^{\mp2i\p Mx}}\,.
\eeq

Combining Eq.~\eqref{sbosons} with Eq.~\eqref{hphz}, we may now compute the zero-temperature ``greater" spin correlation functions. The longitudinal component is given by
\beq
\label{chizz}
i\chi^>_{zz}(x,t)=\sang{S^z(x,t)S^z(0,0)}=\frac{K}{4\p^2}\square{\round{\frac{1}{\al-ix+iut}}^2+\round{\frac{1}{\al+ix+iut}}^2}\,,
\eeq
and the transverse components read
\begin{multline}
\label{chixx}
i\chi^>_{\pm\mp}(x,t)=\sang{S^\pm(x,t)S^\mp(0,0)}=\frac{1}{8\p\al}\Bigg[e^{\mp2i\p Mx}\round{\frac{\al}{\al-ix+iut}}^{K+1+1/4K}\round{\frac{\al}{\al+ix+iut}}^{K-1+1/4K}\\
+e^{\pm2i\p Mx}\round{\frac{\al}{\al-ix+iut}}^{K-1+1/4K}\round{\frac{\al}{\al+ix+iut}}^{K+1+1/4K}\Bigg]\,.
\end{multline}

The bosonization procedure gives the correct (anomalous) exponents for the long-distance, long-time decay of the correlation functions. However, Eq.~\eqref{chixx} has a non-universal amplitude, i.e., pre-factor that depends on the short-distance cutoff $\al$, which cannot be determined by bosonization. Therefore, we introduce a phenomenological coefficient $\tilde C$ for now and write
\begin{multline}
\label{chixx2}
i\chi^>_{\pm\mp}(x,t)=2\tilde C\Bigg[e^{\mp2i\p Mx}\round{\frac{1}{\al-ix+iut}}^{K+1+1/4K}\round{\frac{1}{\al+ix+iut}}^{K-1+1/4K}\\
+e^{\pm2i\p Mx}\round{\frac{1}{\al-ix+iut}}^{K-1+1/4K}\round{\frac{1}{\al+ix+iut}}^{K+1+1/4K}\Bigg]\ .
\end{multline}

We must now Fourier transform the above susceptibilities. In doing so, a useful integral to define is
\begin{multline}
\int dt\int dxe^{-iqx}e^{i\W t}e^{\pm i\be x}\round{\frac{1}{\al-ix+iut}}^\mu\round{\frac{1}{\al+ix+iut}}^\nu\\
=\frac{2\p^2}{u}\frac{\Theta(uq\mp u\be+\W)\Theta(\W-uq\pm u\be)}{\G(\mu)\G(\nu)}\round{\frac{uq\mp u\be+\W}{2u}}^{\mu-1}\round{\frac{\W-uq\pm u\be}{2u}}^{\nu-1}\,.
\end{multline}
Then the Fourier transform of the longitudinal spin correlation function becomes
\beq
i\chi^>_{zz}(q,\W)=\Theta(\W)\de(\W-uq)\frac{K\W}{u}+\Theta(\W)\de(\W+uq)\frac{K\W}{u}\,.
\eeq
Similarly, the Fourier transform of the transverse spin correlation function becomes
\begin{multline}
i\chi^>_{\pm\mp}(q,\W)=\frac{4\p^2C}{u}\frac{(a/u)^{2K+1/2K-2}}{\G(K+1+1/4K)\G(K-1+1/4K)}\\
\times\Big[\Theta\round{uq\pm\be+\W}\Theta\round{\W-uq\mp\be}\round{uq\pm\be+\W}^{K+1/4K}\round{\W-uq\mp\be}^{K+1/4K-2}\\
+\Theta\round{\W-uq\pm\be}\Theta\round{\W+uq\mp\be}\round{\W-uq\pm\be}^{K+1/4K}\round{\W+uq\mp\be}^{K+1/4K-2}\Big]\ ,
\end{multline}
where $C=\tilde C/(2a)^{2K+1/2K-2}$ is a dimensionless constant, and $\be=2Kb_0/\hbar$. Noting that
\beq
{\rm Im}\{\chi^R(q,\W)\}=-\frac{i}{2}\chi^>(q,\W)\,,
\eeq
for $\W>0$ and at zero temperature, we obtain the results quoted in the main text. 

\section{Affleck-Haldane fermionization of the Heisenberg spin chain}
\label{affleckhaldane}
In this Appendix, we provide a qualitative justification for the fermion representation of the Heisenberg spin chain introduced in Sec.~\ref{umklapp}. A more detailed version of the derivation can be found in, e.g., Refs.~\onlinecite{affleckPRB87,gogolinBOOK04}.

To obtain the low-energy effective theory of the Heisenberg spin chain, we begin with the half-filled 1D Hubbard model which is more general than the Heisenberg model. With nearest-neighbor hopping amplitude $t$ and onsite Coulomb repulsion $U>0$, the model is given by
\beq
\label{1dhubbard}
H_H=\sum_{\sang{nm}}\sum_\s\round{t\y^\dag_{n\s}\y_{m\s}+h.c.}+U\sum_n\y^\dag_{n\up}\y_{n\up}\y^\dag_{n\down}\y_{n\down}\equiv H_t+H_U\,,
\eeq
where $n,m$ label the sites of the 1D lattice with lattice constant $a$. 

At large $U$, $H_H$ is known to be equivalent to the antiferromagnetic Heisenberg spin chain,
\beq
\label{heisenbergaf}
H_H^{\rm eff}=J\sum_n\bS_n\cdot\bS_{n+1}\,,
\eeq
where $J=4t^2/U>0$. In~Ref.~\onlinecite{affleckPRB87}, Affleck and Haldane postulate that the equivalence between the two systems | at least as far as the relevant low-energy excitations are concerned | holds even when $U$ is decreased down to the weakly interacting regime, $U\ll|t|$. This is plausible because Eq.~\eqref{1dhubbard} is known to possess a Mott-Hubbard charge gap $\De_c$ for {\em any} positive $U$. Therefore, as long as one is interested in energy scales below $\De_c$, only spin excitations remain, and these excitations should describe the universal dynamical properties of the Heisenberg spin chain~\eqref{heisenbergaf} in the continuum limit.

To make the above statement more concrete, let us begin in the weakly interacting limit $U\ll|t|$ and linearize the free-particle spectrum near the two Fermi points, i.e., $\pm k_F=\pm\p/2a$. If we decompose the electron field into right- and left-moving chiral components, 
\beq
\y_{n\s}\rightarrow\sqrt{a}\square{\y_{R\s}e^{in\p/2}+\y_{L\s}e^{-in\p/2}}\,,
\eeq
the hopping Hamiltonian $H_t$ can be reexpressed as Eq.~\eqref{h0ah}, with the group velocity of the fermions given by $u$. The continuum free fermion theory~\eqref{h0ah} has chiral U(1) and SU(2) symmetries: the charge and spin of the right- and left-moving fermions are separately conserved. The conserved currents corresponding to these symmetries are then given by
\begin{align}
J^c_{L,R}&=\sum_\s:\y^\dag_{L,R\s}\y_{L,R\s}:\,,\\
J^{s,\al}_{L,R}&=\tfrac12\sum_{\s\s'}\y^\dag_{L,R\s}\s^\al_{\s\s'}\y_{L,R\s'}\,,
\end{align}
where the double dots denote normal ordering and $\al=x,y,z$ label the spin components: $J^c_{L,R}$ and $J^{s,\al}_{L,R}$, respectively, correspond to the local charge and $\al$-component spin densities of the left- and right-chiral fermions. 

Witten showed that the free fermion theory~\eqref{h0ah} can be written equivalently as $H_0=H_0^{\rm U(1)}+H_0^{\rm SU(2)}$, where
\begin{align}
\label{u1hubbard}
H_0^{\rm U(1)}&=\frac{u_c}{2}\int dx\,\square{\Pi^2_c(x)+\round{\pd_x\f_c(x)}^2}\,,
\end{align}
and $H_0^{\rm SU(2)}$ is the SU(2)-symmetric level $k=1$ critical Wess-Zumino-Witten (WZW) model.\cite{wittenCMP84} The bosonized Hamiltonian for the charge sector~\eqref{u1hubbard} introduces the identification $J^c_R+J^c_L=\pd_x\f_c(x)/\sqrt{\p}$ and $J^c_R-J^c_L=-\Pi_c(x)/\sqrt{\p}$, where $\Pi_c(x)$ is the momentum conjugate to the field $\f_c(x)$, and $u_c$ is the speed of charge density-wave propagation. 

Affleck and Haldane then add continuum interaction operators to $H_0$ that are compatible with the symmetries of the lattice system.\cite{affleckPRB87} Three such terms emerge, and they all preserve the important property of charge-spin separation. One term leads to the renormalization of the charge propagation speed $u_c$, while the second, umklapp term transforms $H^{\rm U(1)}_0$ into a quantum sine-Gordon model. The umklapp term is relevant at half-filling, so it drives the U(1) charge sector to a massive phase, with the single-soliton mass $m_c$ being the Mott-Hubbard commensurability gap. 

The one remaining term adds to the SU(2) WZW model for the spin sector: this term is the backscattering term Eq.~\eqref{backsc} and is marginally irrelevant. Therefore, the low-energy properties of the antiferromagnetic Heisenberg spin chain~\eqref{heisenbergaf} are described essentially by the level $k=1$ WZW model $H^{\rm SU(2)}_0$, and Eq.~\eqref{backsc} enters this theory as the  marginally irrelevant operator. The refermionization of the WZW model back to the spin-1/2 fermion basis then leads directly to Eq.~\eqref{h0ah}, where $\y_{L,R\s}(x)$ now represent {\em charge-neutral} spin-1/2 fermion fields | spinons | since the U(1) charge sector has now been gapped out during the renormalization process. This last statement points to a certain similarity between the Zhou-Ng picture of the 2D QSL and the Affleck-Haldane picture of the Heisenberg spin chain in that the parent models in both pictures are written in terms of the usual electrons with both charge and spin. However, in approaching the QSL state | in 2D, this is achieved by $1+F^s_1/2\rightarrow0$, and in 1D, by descending down from the lattice scale to the long-wavelength limit | these electronic excitations transmute into chargeless, spin-$1/2$ spinons through the process of charge-spin separation.

\twocolumngrid


\begin{thebibliography}{75}%
\makeatletter
\providecommand \@ifxundefined [1]{%
 \@ifx{#1\undefined}
}%
\providecommand \@ifnum [1]{%
 \ifnum #1\expandafter \@firstoftwo
 \else \expandafter \@secondoftwo
 \fi
}%
\providecommand \@ifx [1]{%
 \ifx #1\expandafter \@firstoftwo
 \else \expandafter \@secondoftwo
 \fi
}%
\providecommand \natexlab [1]{#1}%
\providecommand \enquote  [1]{``#1''}%
\providecommand \bibnamefont  [1]{#1}%
\providecommand \bibfnamefont [1]{#1}%
\providecommand \citenamefont [1]{#1}%
\providecommand \href@noop [0]{\@secondoftwo}%
\providecommand \href [0]{\begingroup \@sanitize@url \@href}%
\providecommand \@href[1]{\@@startlink{#1}\@@href}%
\providecommand \@@href[1]{\endgroup#1\@@endlink}%
\providecommand \@sanitize@url [0]{\catcode `\\12\catcode `\$12\catcode
  `\&12\catcode `\#12\catcode `\^12\catcode `\_12\catcode `\%12\relax}%
\providecommand \@@startlink[1]{}%
\providecommand \@@endlink[0]{}%
\providecommand \url  [0]{\begingroup\@sanitize@url \@url }%
\providecommand \@url [1]{\endgroup\@href {#1}{\urlprefix }}%
\providecommand \urlprefix  [0]{URL }%
\providecommand \Eprint [0]{\href }%
\providecommand \doibase [0]{http://dx.doi.org/}%
\providecommand \selectlanguage [0]{\@gobble}%
\providecommand \bibinfo  [0]{\@secondoftwo}%
\providecommand \bibfield  [0]{\@secondoftwo}%
\providecommand \translation [1]{[#1]}%
\providecommand \BibitemOpen [0]{}%
\providecommand \bibitemStop [0]{}%
\providecommand \bibitemNoStop [0]{.\EOS\space}%
\providecommand \EOS [0]{\spacefactor3000\relax}%
\providecommand \BibitemShut  [1]{\csname bibitem#1\endcsname}%
\let\auto@bib@innerbib\@empty
\bibitem [{\citenamefont {Savary}\ and\ \citenamefont
  {Balents}(2016)}]{savaryRPR17}%
  \BibitemOpen
  \bibfield  {author} {\bibinfo {author} {\bibfnamefont {L.}~\bibnamefont
  {Savary}}\ and\ \bibinfo {author} {\bibfnamefont {L.}~\bibnamefont
  {Balents}},\ }\href@noop {} {\bibfield  {journal} {\bibinfo  {journal} {Rep.
  Prog. Phys.}\ }\textbf {\bibinfo {volume} {80}} (\bibinfo {year}
  {2016})}\BibitemShut {NoStop}%
\bibitem [{\citenamefont {Zhou}\ \emph {et~al.}(2017)\citenamefont {Zhou},
  \citenamefont {Kanoda},\ and\ \citenamefont {Ng}}]{zhouRMP17}%
  \BibitemOpen
  \bibfield  {author} {\bibinfo {author} {\bibfnamefont {Y.}~\bibnamefont
  {Zhou}}, \bibinfo {author} {\bibfnamefont {K.}~\bibnamefont {Kanoda}}, \ and\
  \bibinfo {author} {\bibfnamefont {T.-K.}\ \bibnamefont {Ng}},\ }\href
  {\doibase 10.1103/RevModPhys.89.025003} {\bibfield  {journal} {\bibinfo
  {journal} {Rev. Mod. Phys.}\ }\textbf {\bibinfo {volume} {89}},\ \bibinfo
  {pages} {025003} (\bibinfo {year} {2017})}\BibitemShut {NoStop}%
\bibitem [{\citenamefont {Knolle}\ and\ \citenamefont
  {Moessner}(2019)}]{knolleAR19}%
  \BibitemOpen
  \bibfield  {author} {\bibinfo {author} {\bibfnamefont {J.}~\bibnamefont
  {Knolle}}\ and\ \bibinfo {author} {\bibfnamefont {R.}~\bibnamefont
  {Moessner}},\ }\href@noop {} {\bibfield  {journal} {\bibinfo  {journal}
  {Ann. Rev. Cond. Matt. Phys.}\ }\textbf {\bibinfo {volume}
  {10}},\ \bibinfo {pages} {451} (\bibinfo {year} {2019})}\BibitemShut
  {NoStop}%
\bibitem [{\citenamefont {Broholm}\ \emph {et~al.}(2020)\citenamefont
  {Broholm}, \citenamefont {Cava}, \citenamefont {Kivelson}, \citenamefont
  {Nocera}, \citenamefont {Norman},\ and\ \citenamefont
  {Senthil}}]{broholmSCI20}%
  \BibitemOpen
  \bibfield  {author} {\bibinfo {author} {\bibfnamefont {C.}~\bibnamefont
  {Broholm}}, \bibinfo {author} {\bibfnamefont {R.~J.}\ \bibnamefont {Cava}},
  \bibinfo {author} {\bibfnamefont {S.~A.}\ \bibnamefont {Kivelson}}, \bibinfo
  {author} {\bibfnamefont {D.~G.}\ \bibnamefont {Nocera}}, \bibinfo {author}
  {\bibfnamefont {M.~R.}\ \bibnamefont {Norman}}, \ and\ \bibinfo {author}
  {\bibfnamefont {T.}~\bibnamefont {Senthil}},\ }\href
  {https://science.sciencemag.org/content/367/6475/eaay0668} {\bibfield
  {journal} {\bibinfo  {journal} {Science}\ }\textbf {\bibinfo {volume} {367}}
  (\bibinfo {year} {2020})}\BibitemShut {NoStop}%
\bibitem [{\citenamefont {Wen}(2002)}]{wenPRB02}%
  \BibitemOpen
  \bibfield  {author} {\bibinfo {author} {\bibfnamefont {X.-G.}\ \bibnamefont
  {Wen}},\ }\href {\doibase 10.1103/PhysRevB.65.165113} {\bibfield  {journal}
  {\bibinfo  {journal} {Phys. Rev. B}\ }\textbf {\bibinfo {volume} {65}},\
  \bibinfo {pages} {165113} (\bibinfo {year} {2002})}\BibitemShut {NoStop}%
\bibitem [{\citenamefont {Wen}(2007)}]{wenBOOK07}%
  \BibitemOpen
  \bibfield  {author} {\bibinfo {author} {\bibfnamefont {X.-G.}\ \bibnamefont
  {Wen}},\ }\href {\doibase 10.1093/acprof:oso/9780199227259.001.0001} {\emph
  {\bibinfo {title} {{Quantum field theory of many-body systems: from the
  origin of sound to an origin of light and electrons}}}}\ (\bibinfo
  {publisher} {Oxford University Press},\ \bibinfo {address} {Oxford},\
  \bibinfo {year} {2007})\BibitemShut {NoStop}%
\bibitem [{\citenamefont {Balents}(2010)}]{balentsNAT10}%
  \BibitemOpen
  \bibfield  {author} {\bibinfo {author} {\bibfnamefont {L.}~\bibnamefont
  {Balents}},\ }\href {http://dx.doi.org/10.1038/nature08917} {\bibfield
  {journal} {\bibinfo  {journal} {Nature}\ }\textbf {\bibinfo {volume} {464}},\
  \bibinfo {pages} {199} (\bibinfo {year} {2010})}\BibitemShut {NoStop}%
\bibitem [{\citenamefont {Affleck}(1988)}]{affleckBOOK88}%
  \BibitemOpen
  \bibfield  {author} {\bibinfo {author} {\bibfnamefont {I.}~\bibnamefont
  {Affleck}},\ }in\ \href@noop {} {\emph {\bibinfo {booktitle} {{Les Houches
  Summer School in Theoretical Physics: Fields, Strings, Critical
  Phenomena}}}}\ (\bibinfo {year} {1988})\BibitemShut {NoStop}%
\bibitem [{\citenamefont {Mikeska}\ and\ \citenamefont
  {Kolezhuk}(2004)}]{mikeskaBOOK04}%
  \BibitemOpen
  \bibfield  {author} {\bibinfo {author} {\bibfnamefont {H.-J.}\ \bibnamefont
  {Mikeska}}\ and\ \bibinfo {author} {\bibfnamefont {A.~K.}\ \bibnamefont
  {Kolezhuk}},\ }\enquote {\bibinfo {title} {One-dimensional magnetism},}\ in\
  \href {\doibase 10.1007/BFb0119591} {\emph {\bibinfo {booktitle} {Quantum
  Magnetism}}},\ \bibinfo {editor} {edited by\ \bibinfo {editor} {\bibfnamefont
  {U.}~\bibnamefont {Schollw{\"o}ck}}, \bibinfo {editor} {\bibfnamefont
  {J.}~\bibnamefont {Richter}}, \bibinfo {editor} {\bibfnamefont {D.~J.~J.}\
  \bibnamefont {Farnell}}, \ and\ \bibinfo {editor} {\bibfnamefont {R.~F.}\
  \bibnamefont {Bishop}}}\ (\bibinfo  {publisher} {Springer Berlin
  Heidelberg},\ \bibinfo {address} {Berlin, Heidelberg},\ \bibinfo {year}
  {2004})\ pp.\ \bibinfo {pages} {1--83}\BibitemShut {NoStop}%
\bibitem [{\citenamefont {Giamarchi}(2004)}]{giamarchiBOOK04}%
  \BibitemOpen
  \bibfield  {author} {\bibinfo {author} {\bibfnamefont {T.}~\bibnamefont
  {Giamarchi}},\ }\href@noop {} {\emph {\bibinfo {title} {Quantum Physics in
  One Dimension}}}\ (\bibinfo  {publisher} {Oxford University Press},\ \bibinfo
  {address} {Oxford},\ \bibinfo {year} {2004})\BibitemShut {NoStop}%
\bibitem [{\citenamefont {Gogolin}\ \emph {et~al.}(2004)\citenamefont
  {Gogolin}, \citenamefont {Nersesyan},\ and\ \citenamefont
  {Tsvelik}}]{gogolinBOOK04}%
  \BibitemOpen
  \bibfield  {author} {\bibinfo {author} {\bibfnamefont {A.}~\bibnamefont
  {Gogolin}}, \bibinfo {author} {\bibfnamefont {A.}~\bibnamefont {Nersesyan}},
  \ and\ \bibinfo {author} {\bibfnamefont {A.}~\bibnamefont {Tsvelik}},\ }\href
  {https://books.google.com/books?id=BZDfFIpCoaAC} {\emph {\bibinfo {title}
  {Bosonization and Strongly Correlated Systems}}}\ (\bibinfo  {publisher}
  {Cambridge University Press},\ \bibinfo {year} {2004})\BibitemShut {NoStop}%
\bibitem [{\citenamefont {Yamashita}\ \emph {et~al.}(2009)\citenamefont
  {Yamashita}, \citenamefont {Nakata}, \citenamefont {Kasahara}, \citenamefont
  {Sasaki}, \citenamefont {Yoneyama}, \citenamefont {Kobayashi}, \citenamefont
  {Fujimoto}, \citenamefont {Shibauchi},\ and\ \citenamefont
  {Matsuda}}]{yamashitaNATP09}%
  \BibitemOpen
  \bibfield  {author} {\bibinfo {author} {\bibfnamefont {M.}~\bibnamefont
  {Yamashita}}, \bibinfo {author} {\bibfnamefont {N.}~\bibnamefont {Nakata}},
  \bibinfo {author} {\bibfnamefont {Y.}~\bibnamefont {Kasahara}}, \bibinfo
  {author} {\bibfnamefont {T.}~\bibnamefont {Sasaki}}, \bibinfo {author}
  {\bibfnamefont {N.}~\bibnamefont {Yoneyama}}, \bibinfo {author}
  {\bibfnamefont {N.}~\bibnamefont {Kobayashi}}, \bibinfo {author}
  {\bibfnamefont {S.}~\bibnamefont {Fujimoto}}, \bibinfo {author}
  {\bibfnamefont {T.}~\bibnamefont {Shibauchi}}, \ and\ \bibinfo {author}
  {\bibfnamefont {Y.}~\bibnamefont {Matsuda}},\ }\href {\doibase
  10.1038/nphys1134} {\bibfield  {journal} {\bibinfo  {journal} {Nature
  Phys.}\ }\textbf {\bibinfo {volume} {5}},\ \bibinfo {pages} {44} (\bibinfo
  {year} {2009})}\BibitemShut {NoStop}%
\bibitem [{\citenamefont {Yamashita}\ \emph {et~al.}(2010)\citenamefont
  {Yamashita}, \citenamefont {Nakata}, \citenamefont {Senshu}, \citenamefont
  {Nagata}, \citenamefont {Yamamoto}, \citenamefont {Kato}, \citenamefont
  {Shibauchi},\ and\ \citenamefont {Matsuda}}]{yamashitaSCI10}%
  \BibitemOpen
  \bibfield  {author} {\bibinfo {author} {\bibfnamefont {M.}~\bibnamefont
  {Yamashita}}, \bibinfo {author} {\bibfnamefont {N.}~\bibnamefont {Nakata}},
  \bibinfo {author} {\bibfnamefont {Y.}~\bibnamefont {Senshu}}, \bibinfo
  {author} {\bibfnamefont {M.}~\bibnamefont {Nagata}}, \bibinfo {author}
  {\bibfnamefont {H.~M.}\ \bibnamefont {Yamamoto}}, \bibinfo {author}
  {\bibfnamefont {R.}~\bibnamefont {Kato}}, \bibinfo {author} {\bibfnamefont
  {T.}~\bibnamefont {Shibauchi}}, \ and\ \bibinfo {author} {\bibfnamefont
  {Y.}~\bibnamefont {Matsuda}},\ }\href {\doibase 10.1126/science.1188200}
  {\bibfield  {journal} {\bibinfo  {journal} {Science}\ }\textbf {\bibinfo
  {volume} {328}},\ \bibinfo {pages} {1246} (\bibinfo {year}
  {2010})}\BibitemShut {NoStop}%
\bibitem [{\citenamefont {Lee}\ and\ \citenamefont {Lee}(2005)}]{leePRL05}%
  \BibitemOpen
  \bibfield  {author} {\bibinfo {author} {\bibfnamefont {S.-S.}\ \bibnamefont
  {Lee}}\ and\ \bibinfo {author} {\bibfnamefont {P.~A.}\ \bibnamefont {Lee}},\
  }\href {\doibase 10.1103/PhysRevLett.95.036403} {\bibfield  {journal}
  {\bibinfo  {journal} {Phys. Rev. Lett.}\ }\textbf {\bibinfo {volume} {95}},\
  \bibinfo {pages} {036403} (\bibinfo {year} {2005})}\BibitemShut {NoStop}%
\bibitem [{\citenamefont {Motrunich}(2005)}]{motrunichPRB05}%
  \BibitemOpen
  \bibfield  {author} {\bibinfo {author} {\bibfnamefont {O.~I.}\ \bibnamefont
  {Motrunich}},\ }\href {\doibase 10.1103/PhysRevB.72.045105} {\bibfield
  {journal} {\bibinfo  {journal} {Phys. Rev. B}\ }\textbf {\bibinfo {volume}
  {72}},\ \bibinfo {pages} {045105} (\bibinfo {year} {2005})}\BibitemShut
  {NoStop}%
\bibitem [{\citenamefont {Shimizu}\ \emph {et~al.}(2003)\citenamefont
  {Shimizu}, \citenamefont {Miyagawa}, \citenamefont {Kanoda}, \citenamefont
  {Maesato},\ and\ \citenamefont {Saito}}]{shimizuPRL03}%
  \BibitemOpen
  \bibfield  {author} {\bibinfo {author} {\bibfnamefont {Y.}~\bibnamefont
  {Shimizu}}, \bibinfo {author} {\bibfnamefont {K.}~\bibnamefont {Miyagawa}},
  \bibinfo {author} {\bibfnamefont {K.}~\bibnamefont {Kanoda}}, \bibinfo
  {author} {\bibfnamefont {M.}~\bibnamefont {Maesato}}, \ and\ \bibinfo
  {author} {\bibfnamefont {G.}~\bibnamefont {Saito}},\ }\href {\doibase
  10.1103/PhysRevLett.91.107001} {\bibfield  {journal} {\bibinfo  {journal}
  {Phys. Rev. Lett.}\ }\textbf {\bibinfo {volume} {91}},\ \bibinfo {pages}
  {107001} (\bibinfo {year} {2003})}\BibitemShut {NoStop}%
\bibitem [{\citenamefont {Itou}\ \emph {et~al.}(2008)\citenamefont {Itou},
  \citenamefont {Oyamada}, \citenamefont {Maegawa}, \citenamefont {Tamura},\
  and\ \citenamefont {Kato}}]{itouPRB08}%
  \BibitemOpen
  \bibfield  {author} {\bibinfo {author} {\bibfnamefont {T.}~\bibnamefont
  {Itou}}, \bibinfo {author} {\bibfnamefont {A.}~\bibnamefont {Oyamada}},
  \bibinfo {author} {\bibfnamefont {S.}~\bibnamefont {Maegawa}}, \bibinfo
  {author} {\bibfnamefont {M.}~\bibnamefont {Tamura}}, \ and\ \bibinfo {author}
  {\bibfnamefont {R.}~\bibnamefont {Kato}},\ }\href {\doibase
  10.1103/PhysRevB.77.104413} {\bibfield  {journal} {\bibinfo  {journal} {Phys.
  Rev. B}\ }\textbf {\bibinfo {volume} {77}},\ \bibinfo {pages} {104413}
  (\bibinfo {year} {2008})}\BibitemShut {NoStop}%
\bibitem [{\citenamefont {Shen}\ \emph {et~al.}(2016)\citenamefont {Shen},
  \citenamefont {Li}, \citenamefont {Wo}, \citenamefont {Li}, \citenamefont
  {Shen}, \citenamefont {Pan}, \citenamefont {Wang}, \citenamefont {Walker},
  \citenamefont {Steffens}, \citenamefont {Boehm}, \citenamefont {Hao},
  \citenamefont {Quintero-Castro}, \citenamefont {Harriger}, \citenamefont
  {Frontzek}, \citenamefont {Hao}, \citenamefont {Meng}, \citenamefont {Zhang},
  \citenamefont {Chen},\ and\ \citenamefont {Zhao}}]{shenNAT16}%
  \BibitemOpen
  \bibfield  {author} {\bibinfo {author} {\bibfnamefont {Y.}~\bibnamefont
  {Shen}}, \bibinfo {author} {\bibfnamefont {Y.-D.}\ \bibnamefont {Li}},
  \bibinfo {author} {\bibfnamefont {H.}~\bibnamefont {Wo}}, \bibinfo {author}
  {\bibfnamefont {Y.}~\bibnamefont {Li}}, \bibinfo {author} {\bibfnamefont
  {S.}~\bibnamefont {Shen}}, \bibinfo {author} {\bibfnamefont {B.}~\bibnamefont
  {Pan}}, \bibinfo {author} {\bibfnamefont {Q.}~\bibnamefont {Wang}}, \bibinfo
  {author} {\bibfnamefont {H.~C.}\ \bibnamefont {Walker}}, \bibinfo {author}
  {\bibfnamefont {P.}~\bibnamefont {Steffens}}, \bibinfo {author}
  {\bibfnamefont {M.}~\bibnamefont {Boehm}}, \bibinfo {author} {\bibfnamefont
  {Y.}~\bibnamefont {Hao}}, \bibinfo {author} {\bibfnamefont {D.~L.}\
  \bibnamefont {Quintero-Castro}}, \bibinfo {author} {\bibfnamefont {L.~W.}\
  \bibnamefont {Harriger}}, \bibinfo {author} {\bibfnamefont {M.~D.}\
  \bibnamefont {Frontzek}}, \bibinfo {author} {\bibfnamefont {L.}~\bibnamefont
  {Hao}}, \bibinfo {author} {\bibfnamefont {S.}~\bibnamefont {Meng}}, \bibinfo
  {author} {\bibfnamefont {Q.}~\bibnamefont {Zhang}}, \bibinfo {author}
  {\bibfnamefont {G.}~\bibnamefont {Chen}}, \ and\ \bibinfo {author}
  {\bibfnamefont {J.}~\bibnamefont {Zhao}},\ }\href {\doibase
  10.1038/nature20614} {\bibfield  {journal} {\bibinfo  {journal} {Nature}\
  }\textbf {\bibinfo {volume} {540}},\ \bibinfo {pages} {559} (\bibinfo {year}
  {2016})}\BibitemShut {NoStop}%
\bibitem [{\citenamefont {Rondin}\ \emph {et~al.}(2014)\citenamefont {Rondin},
  \citenamefont {Tetienne}, \citenamefont {Hingant}, \citenamefont {Roch},
  \citenamefont {Maletinsky},\ and\ \citenamefont {Jacques}}]{rondinRPP14}%
  \BibitemOpen
  \bibfield  {author} {\bibinfo {author} {\bibfnamefont {L.}~\bibnamefont
  {Rondin}}, \bibinfo {author} {\bibfnamefont {J.-P.}\ \bibnamefont
  {Tetienne}}, \bibinfo {author} {\bibfnamefont {T.}~\bibnamefont {Hingant}},
  \bibinfo {author} {\bibfnamefont {J.-F.}\ \bibnamefont {Roch}}, \bibinfo
  {author} {\bibfnamefont {P.}~\bibnamefont {Maletinsky}}, \ and\ \bibinfo
  {author} {\bibfnamefont {V.}~\bibnamefont {Jacques}},\ }\href {\doibase
  10.1088/0034-4885/77/5/056503} {\bibfield  {journal} {\bibinfo  {journal}
  {Rep. Prog. Phys.}\ }\textbf {\bibinfo {volume} {77}},\
  \bibinfo {pages} {056503} (\bibinfo {year} {2014})}\BibitemShut {NoStop}%
\bibitem [{\citenamefont {Degen}\ \emph {et~al.}(2017)\citenamefont {Degen},
  \citenamefont {Reinhard},\ and\ \citenamefont {Cappellaro}}]{degenRMP17}%
  \BibitemOpen
  \bibfield  {author} {\bibinfo {author} {\bibfnamefont {C.~L.}\ \bibnamefont
  {Degen}}, \bibinfo {author} {\bibfnamefont {F.}~\bibnamefont {Reinhard}}, \
  and\ \bibinfo {author} {\bibfnamefont {P.}~\bibnamefont {Cappellaro}},\
  }\href {\doibase 10.1103/RevModPhys.89.035002} {\bibfield  {journal}
  {\bibinfo  {journal} {Rev. Mod. Phys.}\ }\textbf {\bibinfo {volume} {89}},\
  \bibinfo {pages} {035002} (\bibinfo {year} {2017})}\BibitemShut {NoStop}%
\bibitem [{\citenamefont {Casola}\ \emph {et~al.}(2018)\citenamefont {Casola},
  \citenamefont {van~der Sar},\ and\ \citenamefont {Yacoby}}]{casolaNRM18}%
  \BibitemOpen
  \bibfield  {author} {\bibinfo {author} {\bibfnamefont {F.}~\bibnamefont
  {Casola}}, \bibinfo {author} {\bibfnamefont {T.}~\bibnamefont {van~der Sar}},
  \ and\ \bibinfo {author} {\bibfnamefont {A.}~\bibnamefont {Yacoby}},\ }\href
  {https://doi.org/10.1038/natrevmats.2017.88} {\bibfield  {journal} {\bibinfo
  {journal} {Nature Rev. Mater.}\ }\textbf {\bibinfo {volume} {3}},\
  \bibinfo {pages} {17088 EP } (\bibinfo {year} {2018})}\BibitemShut {NoStop}%
\bibitem [{\citenamefont {Lee-Wong}\ \emph {et~al.}(2020)\citenamefont
  {Lee-Wong}, \citenamefont {Xue}, \citenamefont {Ye}, \citenamefont {Kreisel},
  \citenamefont {van~der Sar}, \citenamefont {Yacoby},\ and\ \citenamefont
  {Du}}]{leewongNL20}%
  \BibitemOpen
  \bibfield  {author} {\bibinfo {author} {\bibfnamefont {E.}~\bibnamefont
  {Lee-Wong}}, \bibinfo {author} {\bibfnamefont {R.}~\bibnamefont {Xue}},
  \bibinfo {author} {\bibfnamefont {F.}~\bibnamefont {Ye}}, \bibinfo {author}
  {\bibfnamefont {A.}~\bibnamefont {Kreisel}}, \bibinfo {author} {\bibfnamefont
  {T.}~\bibnamefont {van~der Sar}}, \bibinfo {author} {\bibfnamefont
  {A.}~\bibnamefont {Yacoby}}, \ and\ \bibinfo {author} {\bibfnamefont {C.~R.}\
  \bibnamefont {Du}},\ }\href {\doibase 10.1021/acs.nanolett.0c00085}
  {\bibfield  {journal} {\bibinfo  {journal} {Nano Lett.}\ }\textbf {\bibinfo
  {volume} {20}},\ \bibinfo {pages} {3284} (\bibinfo {year}
  {2020})}\BibitemShut {NoStop}%
\bibitem [{\citenamefont {Sandweg}\ \emph {et~al.}(2011)\citenamefont
  {Sandweg}, \citenamefont {Kajiwara}, \citenamefont {Chumak}, \citenamefont
  {Serga}, \citenamefont {Vasyuchka}, \citenamefont {Jungfleisch},
  \citenamefont {Saitoh},\ and\ \citenamefont {Hillebrands}}]{sandwegPRL11}%
  \BibitemOpen
  \bibfield  {author} {\bibinfo {author} {\bibfnamefont {C.~W.}\ \bibnamefont
  {Sandweg}}, \bibinfo {author} {\bibfnamefont {Y.}~\bibnamefont {Kajiwara}},
  \bibinfo {author} {\bibfnamefont {A.~V.}\ \bibnamefont {Chumak}}, \bibinfo
  {author} {\bibfnamefont {A.~A.}\ \bibnamefont {Serga}}, \bibinfo {author}
  {\bibfnamefont {V.~I.}\ \bibnamefont {Vasyuchka}}, \bibinfo {author}
  {\bibfnamefont {M.~B.}\ \bibnamefont {Jungfleisch}}, \bibinfo {author}
  {\bibfnamefont {E.}~\bibnamefont {Saitoh}}, \ and\ \bibinfo {author}
  {\bibfnamefont {B.}~\bibnamefont {Hillebrands}},\ }\href {\doibase
  10.1103/PhysRevLett.106.216601} {\bibfield  {journal} {\bibinfo  {journal}
  {Phys. Rev. Lett.}\ }\textbf {\bibinfo {volume} {106}},\ \bibinfo {pages}
  {216601} (\bibinfo {year} {2011})}\BibitemShut {NoStop}%
\bibitem [{\citenamefont {An}\ \emph {et~al.}(2016)\citenamefont {An},
  \citenamefont {Olsson}, \citenamefont {Weathers}, \citenamefont {Sullivan},
  \citenamefont {Chen}, \citenamefont {Li}, \citenamefont {Marshall},
  \citenamefont {Ma}, \citenamefont {Klimovich}, \citenamefont {Zhou},
  \citenamefont {Shi},\ and\ \citenamefont {Li}}]{anPRL16}%
  \BibitemOpen
  \bibfield  {author} {\bibinfo {author} {\bibfnamefont {K.}~\bibnamefont
  {An}}, \bibinfo {author} {\bibfnamefont {K.~S.}\ \bibnamefont {Olsson}},
  \bibinfo {author} {\bibfnamefont {A.}~\bibnamefont {Weathers}}, \bibinfo
  {author} {\bibfnamefont {S.}~\bibnamefont {Sullivan}}, \bibinfo {author}
  {\bibfnamefont {X.}~\bibnamefont {Chen}}, \bibinfo {author} {\bibfnamefont
  {X.}~\bibnamefont {Li}}, \bibinfo {author} {\bibfnamefont {L.~G.}\
  \bibnamefont {Marshall}}, \bibinfo {author} {\bibfnamefont {X.}~\bibnamefont
  {Ma}}, \bibinfo {author} {\bibfnamefont {N.}~\bibnamefont {Klimovich}},
  \bibinfo {author} {\bibfnamefont {J.}~\bibnamefont {Zhou}}, \bibinfo {author}
  {\bibfnamefont {L.}~\bibnamefont {Shi}}, \ and\ \bibinfo {author}
  {\bibfnamefont {X.}~\bibnamefont {Li}},\ }\href {\doibase
  10.1103/PhysRevLett.117.107202} {\bibfield  {journal} {\bibinfo  {journal}
  {Phys. Rev. Lett.}\ }\textbf {\bibinfo {volume} {117}},\ \bibinfo {pages}
  {107202} (\bibinfo {year} {2016})}\BibitemShut {NoStop}%
\bibitem [{\citenamefont {Holanda}\ \emph {et~al.}(2018)\citenamefont
  {Holanda}, \citenamefont {Maior}, \citenamefont {Azevedo},\ and\
  \citenamefont {Rezende}}]{holandaNATP18}%
  \BibitemOpen
  \bibfield  {author} {\bibinfo {author} {\bibfnamefont {J.}~\bibnamefont
  {Holanda}}, \bibinfo {author} {\bibfnamefont {D.~S.}\ \bibnamefont {Maior}},
  \bibinfo {author} {\bibfnamefont {A.}~\bibnamefont {Azevedo}}, \ and\
  \bibinfo {author} {\bibfnamefont {S.~M.}\ \bibnamefont {Rezende}},\ }\href
  {\doibase 10.1038/s41567-018-0079-y} {\bibfield  {journal} {\bibinfo
  {journal} {Nature Phys.}\ }\textbf {\bibinfo {volume} {14}},\ \bibinfo
  {pages} {500} (\bibinfo {year} {2018})}\BibitemShut {NoStop}%
\bibitem [{\citenamefont {Doherty}\ \emph {et~al.}(2013)\citenamefont
  {Doherty}, \citenamefont {Manson}, \citenamefont {Delaney}, \citenamefont
  {Jelezko}, \citenamefont {Wrachtrup},\ and\ \citenamefont
  {Hollenberg}}]{dohertyPR13}%
  \BibitemOpen
  \bibfield  {author} {\bibinfo {author} {\bibfnamefont {M.~W.}\ \bibnamefont
  {Doherty}}, \bibinfo {author} {\bibfnamefont {N.~B.}\ \bibnamefont {Manson}},
  \bibinfo {author} {\bibfnamefont {P.}~\bibnamefont {Delaney}}, \bibinfo
  {author} {\bibfnamefont {F.}~\bibnamefont {Jelezko}}, \bibinfo {author}
  {\bibfnamefont {J.}~\bibnamefont {Wrachtrup}}, \ and\ \bibinfo {author}
  {\bibfnamefont {L.~C.~L.}\ \bibnamefont {Hollenberg}},\ }\href
  {https://www.sciencedirect.com/science/article/pii/S0370157313000562}
  {\bibfield  {journal} {\bibinfo  {journal} {Phys. Rep.}\ }\textbf
  {\bibinfo {volume} {528}},\ \bibinfo {pages} {1} (\bibinfo {year}
  {2013})}\BibitemShut {NoStop}%
\bibitem [{\citenamefont {Grinolds}\ \emph {et~al.}(2013)\citenamefont
  {Grinolds}, \citenamefont {Hong}, \citenamefont {Maletinsky}, \citenamefont
  {Luan}, \citenamefont {Lukin}, \citenamefont {Walsworth},\ and\ \citenamefont
  {Yacoby}}]{grinoldsNATP13}%
  \BibitemOpen
  \bibfield  {author} {\bibinfo {author} {\bibfnamefont {M.~S.}\ \bibnamefont
  {Grinolds}}, \bibinfo {author} {\bibfnamefont {S.}~\bibnamefont {Hong}},
  \bibinfo {author} {\bibfnamefont {P.}~\bibnamefont {Maletinsky}}, \bibinfo
  {author} {\bibfnamefont {L.}~\bibnamefont {Luan}}, \bibinfo {author}
  {\bibfnamefont {M.~D.}\ \bibnamefont {Lukin}}, \bibinfo {author}
  {\bibfnamefont {R.~L.}\ \bibnamefont {Walsworth}}, \ and\ \bibinfo {author}
  {\bibfnamefont {A.}~\bibnamefont {Yacoby}},\ }\href {\doibase
  10.1038/nphys2543} {\bibfield  {journal} {\bibinfo  {journal} {Nature
  Phys.}\ }\textbf {\bibinfo {volume} {9}},\ \bibinfo {pages} {215} (\bibinfo
  {year} {2013})}\BibitemShut {NoStop}%
\bibitem [{\citenamefont {Halperin}\ \emph {et~al.}(1993)\citenamefont
  {Halperin}, \citenamefont {Lee},\ and\ \citenamefont {Read}}]{halperinPRB93}%
  \BibitemOpen
  \bibfield  {author} {\bibinfo {author} {\bibfnamefont {B.~I.}\ \bibnamefont
  {Halperin}}, \bibinfo {author} {\bibfnamefont {P.~A.}\ \bibnamefont {Lee}}, \
  and\ \bibinfo {author} {\bibfnamefont {N.}~\bibnamefont {Read}},\ }\href
  {\doibase 10.1103/PhysRevB.47.7312} {\bibfield  {journal} {\bibinfo
  {journal} {Phys. Rev. B}\ }\textbf {\bibinfo {volume} {47}},\ \bibinfo
  {pages} {7312} (\bibinfo {year} {1993})}\BibitemShut {NoStop}%
\bibitem [{\citenamefont {Lee}(1989)}]{leePRL89}%
  \BibitemOpen
  \bibfield  {author} {\bibinfo {author} {\bibfnamefont {P.~A.}\ \bibnamefont
  {Lee}},\ }\href {\doibase 10.1103/PhysRevLett.63.680} {\bibfield  {journal}
  {\bibinfo  {journal} {Phys. Rev. Lett.}\ }\textbf {\bibinfo {volume} {63}},\
  \bibinfo {pages} {680} (\bibinfo {year} {1989})}\BibitemShut {NoStop}%
\bibitem [{\citenamefont {Galitski}\ \emph {et~al.}(2005)\citenamefont
  {Galitski}, \citenamefont {Refael}, \citenamefont {Fisher},\ and\
  \citenamefont {Senthil}}]{galitskiPRL05}%
  \BibitemOpen
  \bibfield  {author} {\bibinfo {author} {\bibfnamefont {V.~M.}\ \bibnamefont
  {Galitski}}, \bibinfo {author} {\bibfnamefont {G.}~\bibnamefont {Refael}},
  \bibinfo {author} {\bibfnamefont {M.~P.~A.}\ \bibnamefont {Fisher}}, \ and\
  \bibinfo {author} {\bibfnamefont {T.}~\bibnamefont {Senthil}},\ }\href
  {\doibase 10.1103/PhysRevLett.95.077002} {\bibfield  {journal} {\bibinfo
  {journal} {Phys. Rev. Lett.}\ }\textbf {\bibinfo {volume} {95}},\ \bibinfo
  {pages} {077002} (\bibinfo {year} {2005})}\BibitemShut {NoStop}%
\bibitem [{\citenamefont {Kaul}\ \emph {et~al.}(2008)\citenamefont {Kaul},
  \citenamefont {Kim}, \citenamefont {Sachdev},\ and\ \citenamefont
  {Senthil}}]{kaulNATP08}%
  \BibitemOpen
  \bibfield  {author} {\bibinfo {author} {\bibfnamefont {R.~K.}\ \bibnamefont
  {Kaul}}, \bibinfo {author} {\bibfnamefont {Y.~B.}\ \bibnamefont {Kim}},
  \bibinfo {author} {\bibfnamefont {S.}~\bibnamefont {Sachdev}}, \ and\
  \bibinfo {author} {\bibfnamefont {T.}~\bibnamefont {Senthil}},\ }\href
  {\doibase 10.1038/nphys790} {\bibfield  {journal} {\bibinfo  {journal}
  {Nature Phys.}\ }\textbf {\bibinfo {volume} {4}},\ \bibinfo {pages} {28}
  (\bibinfo {year} {2008})}\BibitemShut {NoStop}%
\bibitem [{\citenamefont {Oshikawa}\ and\ \citenamefont
  {Affleck}(2002)}]{oshikawaPRB02}%
  \BibitemOpen
  \bibfield  {author} {\bibinfo {author} {\bibfnamefont {M.}~\bibnamefont
  {Oshikawa}}\ and\ \bibinfo {author} {\bibfnamefont {I.}~\bibnamefont
  {Affleck}},\ }\href {\doibase 10.1103/PhysRevB.65.134410} {\bibfield
  {journal} {\bibinfo  {journal} {Phys. Rev. B}\ }\textbf {\bibinfo {volume}
  {65}},\ \bibinfo {pages} {134410} (\bibinfo {year} {2002})}\BibitemShut
  {NoStop}%
\bibitem [{\citenamefont {Silin}(1958)}]{silinJETP58}%
  \BibitemOpen
  \bibfield  {author} {\bibinfo {author} {\bibfnamefont {V.~P.}\ \bibnamefont
  {Silin}},\ }\href@noop {} {\bibfield  {journal} {\bibinfo  {journal} {Sov.
  Phys. JETP}\ }\textbf {\bibinfo {volume} {6}},\ \bibinfo {pages} {945}
  (\bibinfo {year} {1958})}\BibitemShut {NoStop}%
\bibitem [{\citenamefont {Platzman}\ and\ \citenamefont
  {Wolff}(1967)}]{platzmanPRL67}%
  \BibitemOpen
  \bibfield  {author} {\bibinfo {author} {\bibfnamefont {P.~M.}\ \bibnamefont
  {Platzman}}\ and\ \bibinfo {author} {\bibfnamefont {P.~A.}\ \bibnamefont
  {Wolff}},\ }\href {\doibase 10.1103/PhysRevLett.18.280} {\bibfield  {journal}
  {\bibinfo  {journal} {Phys. Rev. Lett.}\ }\textbf {\bibinfo {volume} {18}},\
  \bibinfo {pages} {280} (\bibinfo {year} {1967})}\BibitemShut {NoStop}%
\bibitem [{\citenamefont {Balents}\ and\ \citenamefont
  {Starykh}(2020)}]{balentsPRB20}%
  \BibitemOpen
  \bibfield  {author} {\bibinfo {author} {\bibfnamefont {L.}~\bibnamefont
  {Balents}}\ and\ \bibinfo {author} {\bibfnamefont {O.~A.}\ \bibnamefont
  {Starykh}},\ }\href {\doibase 10.1103/PhysRevB.101.020401} {\bibfield
  {journal} {\bibinfo  {journal} {Phys. Rev. B}\ }\textbf {\bibinfo {volume}
  {101}},\ \bibinfo {pages} {020401} (\bibinfo {year} {2020})}\BibitemShut
  {NoStop}%
\bibitem [{\citenamefont {Tetienne}\ \emph {et~al.}(2015)\citenamefont
  {Tetienne}, \citenamefont {Hingant}, \citenamefont {Mart{\'\i}nez},
  \citenamefont {Rohart}, \citenamefont {Thiaville}, \citenamefont {Diez},
  \citenamefont {Garcia}, \citenamefont {Adam}, \citenamefont {Kim},
  \citenamefont {Roch}, \citenamefont {Miron}, \citenamefont {Gaudin},
  \citenamefont {Vila}, \citenamefont {Ocker}, \citenamefont {Ravelosona},\
  and\ \citenamefont {Jacques}}]{tetienneNATC14}%
  \BibitemOpen
  \bibfield  {author} {\bibinfo {author} {\bibfnamefont {J.~P.}\ \bibnamefont
  {Tetienne}}, \bibinfo {author} {\bibfnamefont {T.}~\bibnamefont {Hingant}},
  \bibinfo {author} {\bibfnamefont {L.~J.}\ \bibnamefont {Mart{\'\i}nez}},
  \bibinfo {author} {\bibfnamefont {S.}~\bibnamefont {Rohart}}, \bibinfo
  {author} {\bibfnamefont {A.}~\bibnamefont {Thiaville}}, \bibinfo {author}
  {\bibfnamefont {L.~H.}\ \bibnamefont {Diez}}, \bibinfo {author}
  {\bibfnamefont {K.}~\bibnamefont {Garcia}}, \bibinfo {author} {\bibfnamefont
  {J.~P.}\ \bibnamefont {Adam}}, \bibinfo {author} {\bibfnamefont {J.~V.}\
  \bibnamefont {Kim}}, \bibinfo {author} {\bibfnamefont {J.~F.}\ \bibnamefont
  {Roch}}, \bibinfo {author} {\bibfnamefont {I.~M.}\ \bibnamefont {Miron}},
  \bibinfo {author} {\bibfnamefont {G.}~\bibnamefont {Gaudin}}, \bibinfo
  {author} {\bibfnamefont {L.}~\bibnamefont {Vila}}, \bibinfo {author}
  {\bibfnamefont {B.}~\bibnamefont {Ocker}}, \bibinfo {author} {\bibfnamefont
  {D.}~\bibnamefont {Ravelosona}}, \ and\ \bibinfo {author} {\bibfnamefont
  {V.}~\bibnamefont {Jacques}},\ }\href {\doibase 10.1038/ncomms7733}
  {\bibfield  {journal} {\bibinfo  {journal} {Nature Commun.}\ }\textbf
  {\bibinfo {volume} {6}},\ \bibinfo {pages} {6733} (\bibinfo {year}
  {2015})}\BibitemShut {NoStop}%
\bibitem [{\citenamefont {Wolfe}\ \emph {et~al.}(2014)\citenamefont {Wolfe},
  \citenamefont {Bhallamudi}, \citenamefont {Wang}, \citenamefont {Du},
  \citenamefont {Manuilov}, \citenamefont {Teeling-Smith}, \citenamefont
  {Berger}, \citenamefont {Adur}, \citenamefont {Yang},\ and\ \citenamefont
  {Hammel}}]{wolfePRB14}%
  \BibitemOpen
  \bibfield  {author} {\bibinfo {author} {\bibfnamefont {C.~S.}\ \bibnamefont
  {Wolfe}}, \bibinfo {author} {\bibfnamefont {V.~P.}\ \bibnamefont
  {Bhallamudi}}, \bibinfo {author} {\bibfnamefont {H.~L.}\ \bibnamefont
  {Wang}}, \bibinfo {author} {\bibfnamefont {C.~H.}\ \bibnamefont {Du}},
  \bibinfo {author} {\bibfnamefont {S.}~\bibnamefont {Manuilov}}, \bibinfo
  {author} {\bibfnamefont {R.~M.}\ \bibnamefont {Teeling-Smith}}, \bibinfo
  {author} {\bibfnamefont {A.~J.}\ \bibnamefont {Berger}}, \bibinfo {author}
  {\bibfnamefont {R.}~\bibnamefont {Adur}}, \bibinfo {author} {\bibfnamefont
  {F.~Y.}\ \bibnamefont {Yang}}, \ and\ \bibinfo {author} {\bibfnamefont
  {P.~C.}\ \bibnamefont {Hammel}},\ }\href {\doibase
  10.1103/PhysRevB.89.180406} {\bibfield  {journal} {\bibinfo  {journal} {Phys.
  Rev. B}\ }\textbf {\bibinfo {volume} {89}},\ \bibinfo {pages} {180406}
  (\bibinfo {year} {2014})}\BibitemShut {NoStop}%
\bibitem [{\citenamefont {van~der Sar}\ \emph {et~al.}(2015)\citenamefont
  {van~der Sar}, \citenamefont {Casola}, \citenamefont {Walsworth},\ and\
  \citenamefont {Yacoby}}]{vandersarNATC15}%
  \BibitemOpen
  \bibfield  {author} {\bibinfo {author} {\bibfnamefont {T.}~\bibnamefont
  {van~der Sar}}, \bibinfo {author} {\bibfnamefont {F.}~\bibnamefont {Casola}},
  \bibinfo {author} {\bibfnamefont {R.}~\bibnamefont {Walsworth}}, \ and\
  \bibinfo {author} {\bibfnamefont {A.}~\bibnamefont {Yacoby}},\ }\href
  {https://doi.org/10.1038/ncomms8886} {\bibfield  {journal} {\bibinfo
  {journal} {Nature Commun.}\ }\textbf {\bibinfo {volume} {6}},\
  \bibinfo {pages} {7886 EP } (\bibinfo {year} {2015})}\BibitemShut {NoStop}%
\bibitem [{\citenamefont {Wolf}\ \emph {et~al.}(2016)\citenamefont {Wolf},
  \citenamefont {Badea},\ and\ \citenamefont {Berezovsky}}]{wolfNATC16}%
  \BibitemOpen
  \bibfield  {author} {\bibinfo {author} {\bibfnamefont {M.~S.}\ \bibnamefont
  {Wolf}}, \bibinfo {author} {\bibfnamefont {R.}~\bibnamefont {Badea}}, \ and\
  \bibinfo {author} {\bibfnamefont {J.}~\bibnamefont {Berezovsky}},\ }\href
  {\doibase 10.1038/ncomms11584} {\bibfield  {journal} {\bibinfo  {journal}
  {Nature Commun.}\ }\textbf {\bibinfo {volume} {7}},\ \bibinfo {pages}
  {11584} (\bibinfo {year} {2016})}\BibitemShut {NoStop}%
\bibitem [{\citenamefont {Page}\ \emph {et~al.}(2019)\citenamefont {Page},
  \citenamefont {McCullian}, \citenamefont {Purser}, \citenamefont {Schulze},
  \citenamefont {Nakatani}, \citenamefont {Wolfe}, \citenamefont {Childress},
  \citenamefont {McConney}, \citenamefont {Howe}, \citenamefont {Hammel},\ and\
  \citenamefont {Bhallamudi}}]{pageAPL19}%
  \BibitemOpen
  \bibfield  {author} {\bibinfo {author} {\bibfnamefont {M.~R.}\ \bibnamefont
  {Page}}, \bibinfo {author} {\bibfnamefont {B.~A.}\ \bibnamefont {McCullian}},
  \bibinfo {author} {\bibfnamefont {C.~M.}\ \bibnamefont {Purser}}, \bibinfo
  {author} {\bibfnamefont {J.~G.}\ \bibnamefont {Schulze}}, \bibinfo {author}
  {\bibfnamefont {T.~M.}\ \bibnamefont {Nakatani}}, \bibinfo {author}
  {\bibfnamefont {C.~S.}\ \bibnamefont {Wolfe}}, \bibinfo {author}
  {\bibfnamefont {J.~R.}\ \bibnamefont {Childress}}, \bibinfo {author}
  {\bibfnamefont {M.~E.}\ \bibnamefont {McConney}}, \bibinfo {author}
  {\bibfnamefont {B.~M.}\ \bibnamefont {Howe}}, \bibinfo {author}
  {\bibfnamefont {P.~C.}\ \bibnamefont {Hammel}}, \ and\ \bibinfo {author}
  {\bibfnamefont {V.~P.}\ \bibnamefont {Bhallamudi}},\ }\href {\doibase
  10.1063/1.5083991} {\bibfield  {journal} {\bibinfo  {journal} {J.
  Appl. Phys.}\ }\textbf {\bibinfo {volume} {126}},\ \bibinfo {pages}
  {124902} (\bibinfo {year} {2019})} \BibitemShut {NoStop}%
\bibitem [{\citenamefont {Du}\ \emph {et~al.}(2017)\citenamefont {Du},
  \citenamefont {van~der Sar}, \citenamefont {Zhou}, \citenamefont {Upadhyaya},
  \citenamefont {Casola}, \citenamefont {Zhang}, \citenamefont {Onbasli},
  \citenamefont {Ross}, \citenamefont {Walsworth}, \citenamefont
  {Tserkovnyak},\ and\ \citenamefont {Yacoby}}]{duSCI17}%
  \BibitemOpen
  \bibfield  {author} {\bibinfo {author} {\bibfnamefont {C.}~\bibnamefont
  {Du}}, \bibinfo {author} {\bibfnamefont {T.}~\bibnamefont {van~der Sar}},
  \bibinfo {author} {\bibfnamefont {T.~X.}\ \bibnamefont {Zhou}}, \bibinfo
  {author} {\bibfnamefont {P.}~\bibnamefont {Upadhyaya}}, \bibinfo {author}
  {\bibfnamefont {F.}~\bibnamefont {Casola}}, \bibinfo {author} {\bibfnamefont
  {H.}~\bibnamefont {Zhang}}, \bibinfo {author} {\bibfnamefont {M.~C.}\
  \bibnamefont {Onbasli}}, \bibinfo {author} {\bibfnamefont {C.~A.}\
  \bibnamefont {Ross}}, \bibinfo {author} {\bibfnamefont {R.~L.}\ \bibnamefont
  {Walsworth}}, \bibinfo {author} {\bibfnamefont {Y.}~\bibnamefont
  {Tserkovnyak}}, \ and\ \bibinfo {author} {\bibfnamefont {A.}~\bibnamefont
  {Yacoby}},\ }\href {http://science.sciencemag.org/content/357/6347/195}
  {\bibfield  {journal} {\bibinfo  {journal} {Science}\ }\textbf {\bibinfo
  {volume} {357}},\ \bibinfo {pages} {195} (\bibinfo {year}
  {2017})}\BibitemShut {NoStop}%
\bibitem [{\citenamefont {Flebus}\ and\ \citenamefont
  {Tserkovnyak}(2018)}]{flebusPRL18}%
  \BibitemOpen
  \bibfield  {author} {\bibinfo {author} {\bibfnamefont {B.}~\bibnamefont
  {Flebus}}\ and\ \bibinfo {author} {\bibfnamefont {Y.}~\bibnamefont
  {Tserkovnyak}},\ }\href {\doibase 10.1103/PhysRevLett.121.187204} {\bibfield
  {journal} {\bibinfo  {journal} {Phys. Rev. Lett.}\ }\textbf {\bibinfo
  {volume} {121}},\ \bibinfo {pages} {187204} (\bibinfo {year}
  {2018})}\BibitemShut {NoStop}%
\bibitem [{\citenamefont {Rodriguez-Nieva}\ \emph {et~al.}(2022)\citenamefont
  {Rodriguez-Nieva}, \citenamefont {Podolsky},\ and\ \citenamefont
  {Demler}}]{rodrigueznievaPRB22}%
  \BibitemOpen
  \bibfield  {author} {\bibinfo {author} {\bibfnamefont {J.~F.}\ \bibnamefont
  {Rodriguez-Nieva}}, \bibinfo {author} {\bibfnamefont {D.}~\bibnamefont
  {Podolsky}}, \ and\ \bibinfo {author} {\bibfnamefont {E.}~\bibnamefont
  {Demler}},\ }\href {\doibase 10.1103/PhysRevB.105.174412} {\bibfield
  {journal} {\bibinfo  {journal} {Phys. Rev. B}\ }\textbf {\bibinfo {volume}
  {105}},\ \bibinfo {pages} {174412} (\bibinfo {year} {2022})}\BibitemShut
  {NoStop}%
\bibitem [{\citenamefont {Rodriguez-Nieva}\ \emph {et~al.}(2018)\citenamefont
  {Rodriguez-Nieva}, \citenamefont {Agarwal}, \citenamefont {Giamarchi},
  \citenamefont {Halperin}, \citenamefont {Lukin},\ and\ \citenamefont
  {Demler}}]{rodrigueznievaPRB18}%
  \BibitemOpen
  \bibfield  {author} {\bibinfo {author} {\bibfnamefont {J.~F.}\ \bibnamefont
  {Rodriguez-Nieva}}, \bibinfo {author} {\bibfnamefont {K.}~\bibnamefont
  {Agarwal}}, \bibinfo {author} {\bibfnamefont {T.}~\bibnamefont {Giamarchi}},
  \bibinfo {author} {\bibfnamefont {B.~I.}\ \bibnamefont {Halperin}}, \bibinfo
  {author} {\bibfnamefont {M.~D.}\ \bibnamefont {Lukin}}, \ and\ \bibinfo
  {author} {\bibfnamefont {E.}~\bibnamefont {Demler}},\ }\href {\doibase
  10.1103/PhysRevB.98.195433} {\bibfield  {journal} {\bibinfo  {journal} {Phys.
  Rev. B}\ }\textbf {\bibinfo {volume} {98}},\ \bibinfo {pages} {195433}
  (\bibinfo {year} {2018})}\BibitemShut {NoStop}%
\bibitem [{\citenamefont {Chatterjee}\ and\ \citenamefont
  {Sachdev}(2015)}]{chatterjeePRB15}%
  \BibitemOpen
  \bibfield  {author} {\bibinfo {author} {\bibfnamefont {S.}~\bibnamefont
  {Chatterjee}}\ and\ \bibinfo {author} {\bibfnamefont {S.}~\bibnamefont
  {Sachdev}},\ }\href {\doibase 10.1103/PhysRevB.92.165113} {\bibfield
  {journal} {\bibinfo  {journal} {Phys. Rev. B}\ }\textbf {\bibinfo {volume}
  {92}},\ \bibinfo {pages} {165113} (\bibinfo {year} {2015})}\BibitemShut
  {NoStop}%
\bibitem [{\citenamefont {Khoo}\ \emph {et~al.}(2021)\citenamefont {Khoo},
  \citenamefont {Pientka},\ and\ \citenamefont {Sodemann}}]{khooNJP21}%
  \BibitemOpen
  \bibfield  {author} {\bibinfo {author} {\bibfnamefont {J.~Y.}\ \bibnamefont
  {Khoo}}, \bibinfo {author} {\bibfnamefont {F.}~\bibnamefont {Pientka}}, \
  and\ \bibinfo {author} {\bibfnamefont {I.}~\bibnamefont {Sodemann}},\ }\href
  {\doibase 10.1088/1367-2630/ac2dab} {\bibfield  {journal} {\bibinfo
  {journal} {New J. Phys.}\ }\textbf {\bibinfo {volume} {23}},\
  \bibinfo {pages} {113009} (\bibinfo {year} {2021})}\BibitemShut {NoStop}%
\bibitem [{Note1()}]{Note1}%
  \BibitemOpen
  \bibinfo {note} {We use Gaussian units throughout this work.}\BibitemShut
  {Stop}%
\bibitem [{\citenamefont {Chatterjee}\ \emph {et~al.}(2019)\citenamefont
  {Chatterjee}, \citenamefont {Rodriguez-Nieva},\ and\ \citenamefont
  {Demler}}]{chatterjeePRB19}%
  \BibitemOpen
  \bibfield  {author} {\bibinfo {author} {\bibfnamefont {S.}~\bibnamefont
  {Chatterjee}}, \bibinfo {author} {\bibfnamefont {J.~F.}\ \bibnamefont
  {Rodriguez-Nieva}}, \ and\ \bibinfo {author} {\bibfnamefont {E.}~\bibnamefont
  {Demler}},\ }\href {\doibase 10.1103/PhysRevB.99.104425} {\bibfield
  {journal} {\bibinfo  {journal} {Phys. Rev. B}\ }\textbf {\bibinfo {volume}
  {99}},\ \bibinfo {pages} {104425} (\bibinfo {year} {2019})}\BibitemShut
  {NoStop}%
\bibitem [{\citenamefont {Zhou}\ and\ \citenamefont {Ng}(2013)}]{zhouPRB13}%
  \BibitemOpen
  \bibfield  {author} {\bibinfo {author} {\bibfnamefont {Y.}~\bibnamefont
  {Zhou}}\ and\ \bibinfo {author} {\bibfnamefont {T.-K.}\ \bibnamefont {Ng}},\
  }\href@noop {} {\bibfield  {journal} {\bibinfo  {journal} {Phys. Rev. B}\
  }\textbf {\bibinfo {volume} {88}},\ \bibinfo {pages} {165130} (\bibinfo
  {year} {2013})}\BibitemShut {NoStop}%
\bibitem [{\citenamefont {Nagaosa}\ and\ \citenamefont
  {Lee}(1990)}]{nagaosaPRL90}%
  \BibitemOpen
  \bibfield  {author} {\bibinfo {author} {\bibfnamefont {N.}~\bibnamefont
  {Nagaosa}}\ and\ \bibinfo {author} {\bibfnamefont {P.~A.}\ \bibnamefont
  {Lee}},\ }\href {\doibase 10.1103/PhysRevLett.64.2450} {\bibfield  {journal}
  {\bibinfo  {journal} {Phys. Rev. Lett.}\ }\textbf {\bibinfo {volume} {64}},\
  \bibinfo {pages} {2450} (\bibinfo {year} {1990})}\BibitemShut {NoStop}%
\bibitem [{\citenamefont {Lee}\ and\ \citenamefont {Nagaosa}(1992)}]{leePRB92}%
  \BibitemOpen
  \bibfield  {author} {\bibinfo {author} {\bibfnamefont {P.~A.}\ \bibnamefont
  {Lee}}\ and\ \bibinfo {author} {\bibfnamefont {N.}~\bibnamefont {Nagaosa}},\
  }\href {\doibase 10.1103/PhysRevB.46.5621} {\bibfield  {journal} {\bibinfo
  {journal} {Phys. Rev. B}\ }\textbf {\bibinfo {volume} {46}},\ \bibinfo
  {pages} {5621} (\bibinfo {year} {1992})}\BibitemShut {NoStop}%
\bibitem [{\citenamefont {Kim}\ \emph {et~al.}(1995)\citenamefont {Kim},
  \citenamefont {Lee},\ and\ \citenamefont {Wen}}]{kimPRB95}%
  \BibitemOpen
  \bibfield  {author} {\bibinfo {author} {\bibfnamefont {Y.~B.}\ \bibnamefont
  {Kim}}, \bibinfo {author} {\bibfnamefont {P.~A.}\ \bibnamefont {Lee}}, \ and\
  \bibinfo {author} {\bibfnamefont {X.-G.}\ \bibnamefont {Wen}},\ }\href
  {\doibase 10.1103/PhysRevB.52.17275} {\bibfield  {journal} {\bibinfo
  {journal} {Phys. Rev. B}\ }\textbf {\bibinfo {volume} {52}},\ \bibinfo
  {pages} {17275} (\bibinfo {year} {1995})}\BibitemShut {NoStop}%
\bibitem [{\citenamefont {Bertini}\ \emph {et~al.}(2021)\citenamefont
  {Bertini}, \citenamefont {Heidrich-Meisner}, \citenamefont {Karrasch},
  \citenamefont {Prosen}, \citenamefont {Steinigeweg},\ and\ \citenamefont
  {\ifmmode \check{Z}\else \v{Z}\fi{}nidari\ifmmode~\check{c}\else
  \v{c}\fi{}}}]{bertiniRMP21}%
  \BibitemOpen
  \bibfield  {author} {\bibinfo {author} {\bibfnamefont {B.}~\bibnamefont
  {Bertini}}, \bibinfo {author} {\bibfnamefont {F.}~\bibnamefont
  {Heidrich-Meisner}}, \bibinfo {author} {\bibfnamefont {C.}~\bibnamefont
  {Karrasch}}, \bibinfo {author} {\bibfnamefont {T.}~\bibnamefont {Prosen}},
  \bibinfo {author} {\bibfnamefont {R.}~\bibnamefont {Steinigeweg}}, \ and\
  \bibinfo {author} {\bibfnamefont {M.}~\bibnamefont {\ifmmode \check{Z}\else
  \v{Z}\fi{}nidari\ifmmode~\check{c}\else \v{c}\fi{}}},\ }\href {\doibase
  10.1103/RevModPhys.93.025003} {\bibfield  {journal} {\bibinfo  {journal}
  {Rev. Mod. Phys.}\ }\textbf {\bibinfo {volume} {93}},\ \bibinfo {pages}
  {025003} (\bibinfo {year} {2021})}\BibitemShut {NoStop}%
\bibitem [{\citenamefont {Hirobe}\ \emph {et~al.}(2017)\citenamefont {Hirobe},
  \citenamefont {Sato}, \citenamefont {Kawamata}, \citenamefont {Shiomi},
  \citenamefont {Uchida}, \citenamefont {Iguchi}, \citenamefont {Koike},
  \citenamefont {Maekawa},\ and\ \citenamefont {Saitoh}}]{hirobeNATP16}%
  \BibitemOpen
  \bibfield  {author} {\bibinfo {author} {\bibfnamefont {D.}~\bibnamefont
  {Hirobe}}, \bibinfo {author} {\bibfnamefont {M.}~\bibnamefont {Sato}},
  \bibinfo {author} {\bibfnamefont {T.}~\bibnamefont {Kawamata}}, \bibinfo
  {author} {\bibfnamefont {Y.}~\bibnamefont {Shiomi}}, \bibinfo {author}
  {\bibfnamefont {K.-i.}\ \bibnamefont {Uchida}}, \bibinfo {author}
  {\bibfnamefont {R.}~\bibnamefont {Iguchi}}, \bibinfo {author} {\bibfnamefont
  {Y.}~\bibnamefont {Koike}}, \bibinfo {author} {\bibfnamefont
  {S.}~\bibnamefont {Maekawa}}, \ and\ \bibinfo {author} {\bibfnamefont
  {E.}~\bibnamefont {Saitoh}},\ }\href@noop {} {\bibfield  {journal} {\bibinfo
  {journal} {Nature Phys.}\ }\textbf {\bibinfo {volume} {13}},\ \bibinfo
  {pages} {30} (\bibinfo {year} {2017})}\BibitemShut {NoStop}%
\bibitem [{\citenamefont {Jordan}\ and\ \citenamefont
  {Wigner}(1928)}]{jordanZP28}%
  \BibitemOpen
  \bibfield  {author} {\bibinfo {author} {\bibfnamefont {P.}~\bibnamefont
  {Jordan}}\ and\ \bibinfo {author} {\bibfnamefont {E.}~\bibnamefont
  {Wigner}},\ }\href {\doibase 10.1007/BF01331938} {\bibfield  {journal}
  {\bibinfo  {journal} {Zeit. Phys.}\ }\textbf {\bibinfo
  {volume} {47}},\ \bibinfo {pages} {631} (\bibinfo {year} {1928})}\BibitemShut
  {NoStop}%
\bibitem [{Note2()}]{Note2}%
  \BibitemOpen
  \bibinfo {note} {A representative quantum spin chain, a copper-oxide material
  Sr$_2$CuO$_3$, has a lattice constant of $a\approx 4\r A$.}\BibitemShut
  {Stop}%
\bibitem [{\citenamefont {Johnson}\ \emph {et~al.}(1973)\citenamefont
  {Johnson}, \citenamefont {Krinsky},\ and\ \citenamefont
  {McCoy}}]{johnsonPRA73}%
  \BibitemOpen
  \bibfield  {author} {\bibinfo {author} {\bibfnamefont {J.~D.}\ \bibnamefont
  {Johnson}}, \bibinfo {author} {\bibfnamefont {S.}~\bibnamefont {Krinsky}}, \
  and\ \bibinfo {author} {\bibfnamefont {B.~M.}\ \bibnamefont {McCoy}},\ }\href
  {\doibase 10.1103/PhysRevA.8.2526} {\bibfield  {journal} {\bibinfo  {journal}
  {Phys. Rev. A}\ }\textbf {\bibinfo {volume} {8}},\ \bibinfo {pages} {2526}
  (\bibinfo {year} {1973})}\BibitemShut {NoStop}%
\bibitem [{\citenamefont {Hikihara}\ and\ \citenamefont
  {Furusaki}(2004)}]{hikiharaPRB04}%
  \BibitemOpen
  \bibfield  {author} {\bibinfo {author} {\bibfnamefont {T.}~\bibnamefont
  {Hikihara}}\ and\ \bibinfo {author} {\bibfnamefont {A.}~\bibnamefont
  {Furusaki}},\ }\href@noop {} {\bibfield  {journal} {\bibinfo  {journal}
  {Phys. Rev. B}\ }\textbf {\bibinfo {volume} {69}},\ \bibinfo {pages} {064427}
  (\bibinfo {year} {2004})}\BibitemShut {NoStop}%
\bibitem [{\citenamefont {Keselman}\ \emph {et~al.}(2020)\citenamefont
  {Keselman}, \citenamefont {Balents},\ and\ \citenamefont
  {Starykh}}]{keselmanPRL20}%
  \BibitemOpen
  \bibfield  {author} {\bibinfo {author} {\bibfnamefont {A.}~\bibnamefont
  {Keselman}}, \bibinfo {author} {\bibfnamefont {L.}~\bibnamefont {Balents}}, \
  and\ \bibinfo {author} {\bibfnamefont {O.~A.}\ \bibnamefont {Starykh}},\
  }\href {\doibase 10.1103/PhysRevLett.125.187201} {\bibfield  {journal}
  {\bibinfo  {journal} {Phys. Rev. Lett.}\ }\textbf {\bibinfo {volume} {125}},\
  \bibinfo {pages} {187201} (\bibinfo {year} {2020})}\BibitemShut {NoStop}%
\bibitem [{\citenamefont {Affleck}\ and\ \citenamefont
  {Haldane}(1987)}]{affleckPRB87}%
  \BibitemOpen
  \bibfield  {author} {\bibinfo {author} {\bibfnamefont {I.}~\bibnamefont
  {Affleck}}\ and\ \bibinfo {author} {\bibfnamefont {F.~D.~M.}\ \bibnamefont
  {Haldane}},\ }\href {\doibase 10.1103/PhysRevB.36.5291} {\bibfield  {journal}
  {\bibinfo  {journal} {Phys. Rev. B}\ }\textbf {\bibinfo {volume} {36}},\
  \bibinfo {pages} {5291} (\bibinfo {year} {1987})}\BibitemShut {NoStop}%
\bibitem [{\citenamefont {Smith}\ \emph {et~al.}(2003)\citenamefont {Smith},
  \citenamefont {De~Soto}, \citenamefont {Slichter}, \citenamefont {Schlueter},
  \citenamefont {Kini},\ and\ \citenamefont {Daugherty}}]{smithPRB03}%
  \BibitemOpen
  \bibfield  {author} {\bibinfo {author} {\bibfnamefont {D.~F.}\ \bibnamefont
  {Smith}}, \bibinfo {author} {\bibfnamefont {S.~M.}\ \bibnamefont {De~Soto}},
  \bibinfo {author} {\bibfnamefont {C.~P.}\ \bibnamefont {Slichter}}, \bibinfo
  {author} {\bibfnamefont {J.~A.}\ \bibnamefont {Schlueter}}, \bibinfo {author}
  {\bibfnamefont {A.~M.}\ \bibnamefont {Kini}}, \ and\ \bibinfo {author}
  {\bibfnamefont {R.~G.}\ \bibnamefont {Daugherty}},\ }\href {\doibase
  10.1103/PhysRevB.68.024512} {\bibfield  {journal} {\bibinfo  {journal} {Phys.
  Rev. B}\ }\textbf {\bibinfo {volume} {68}},\ \bibinfo {pages} {024512}
  (\bibinfo {year} {2003})}\BibitemShut {NoStop}%
\bibitem [{\citenamefont {Winter}\ \emph {et~al.}(2017)\citenamefont {Winter},
  \citenamefont {Riedl},\ and\ \citenamefont {Valent\'{\i}}}]{winterPRB17}%
  \BibitemOpen
  \bibfield  {author} {\bibinfo {author} {\bibfnamefont {S.~M.}\ \bibnamefont
  {Winter}}, \bibinfo {author} {\bibfnamefont {K.}~\bibnamefont {Riedl}}, \
  and\ \bibinfo {author} {\bibfnamefont {R.}~\bibnamefont {Valent\'{\i}}},\
  }\href {\doibase 10.1103/PhysRevB.95.060404} {\bibfield  {journal} {\bibinfo
  {journal} {Phys. Rev. B}\ }\textbf {\bibinfo {volume} {95}},\ \bibinfo
  {pages} {060404} (\bibinfo {year} {2017})}\BibitemShut {NoStop}%
\bibitem [{\citenamefont {Ma}\ \emph {et~al.}(2021)\citenamefont {Ma},
  \citenamefont {Dong}, \citenamefont {Wang}, \citenamefont {Zheng},
  \citenamefont {Ran}, \citenamefont {Bao}, \citenamefont {Cai}, \citenamefont
  {Shangguan}, \citenamefont {Wang}, \citenamefont {Boehm}, \citenamefont
  {Steffens}, \citenamefont {Regnault}, \citenamefont {Wang}, \citenamefont
  {Su}, \citenamefont {Yu}, \citenamefont {Liu}, \citenamefont {Li},\ and\
  \citenamefont {Wen}}]{maPRB21}%
  \BibitemOpen
  \bibfield  {author} {\bibinfo {author} {\bibfnamefont {Z.}~\bibnamefont
  {Ma}}, \bibinfo {author} {\bibfnamefont {Z.-Y.}\ \bibnamefont {Dong}},
  \bibinfo {author} {\bibfnamefont {J.}~\bibnamefont {Wang}}, \bibinfo {author}
  {\bibfnamefont {S.}~\bibnamefont {Zheng}}, \bibinfo {author} {\bibfnamefont
  {K.}~\bibnamefont {Ran}}, \bibinfo {author} {\bibfnamefont {S.}~\bibnamefont
  {Bao}}, \bibinfo {author} {\bibfnamefont {Z.}~\bibnamefont {Cai}}, \bibinfo
  {author} {\bibfnamefont {Y.}~\bibnamefont {Shangguan}}, \bibinfo {author}
  {\bibfnamefont {W.}~\bibnamefont {Wang}}, \bibinfo {author} {\bibfnamefont
  {M.}~\bibnamefont {Boehm}}, \bibinfo {author} {\bibfnamefont
  {P.}~\bibnamefont {Steffens}}, \bibinfo {author} {\bibfnamefont {L.-P.}\
  \bibnamefont {Regnault}}, \bibinfo {author} {\bibfnamefont {X.}~\bibnamefont
  {Wang}}, \bibinfo {author} {\bibfnamefont {Y.}~\bibnamefont {Su}}, \bibinfo
  {author} {\bibfnamefont {S.-L.}\ \bibnamefont {Yu}}, \bibinfo {author}
  {\bibfnamefont {J.-M.}\ \bibnamefont {Liu}}, \bibinfo {author} {\bibfnamefont
  {J.-X.}\ \bibnamefont {Li}}, \ and\ \bibinfo {author} {\bibfnamefont
  {J.}~\bibnamefont {Wen}},\ }\href {\doibase 10.1103/PhysRevB.104.224433}
  {\bibfield  {journal} {\bibinfo  {journal} {Phys. Rev. B}\ }\textbf {\bibinfo
  {volume} {104}},\ \bibinfo {pages} {224433} (\bibinfo {year}
  {2021})}\BibitemShut {NoStop}%
\bibitem [{\citenamefont {Yuan}\ \emph {et~al.}(2021)\citenamefont {Yuan},
  \citenamefont {Cao}, \citenamefont {Kamra}, \citenamefont {Duine},\ and\
  \citenamefont {Yan}}]{yuanCM21}%
  \BibitemOpen
  \bibfield  {author} {\bibinfo {author} {\bibfnamefont {H.~Y.}\ \bibnamefont
  {Yuan}}, \bibinfo {author} {\bibfnamefont {Y.}~\bibnamefont {Cao}}, \bibinfo
  {author} {\bibfnamefont {A.}~\bibnamefont {Kamra}}, \bibinfo {author}
  {\bibfnamefont {R.~A.}\ \bibnamefont {Duine}}, \ and\ \bibinfo {author}
  {\bibfnamefont {P.}~\bibnamefont {Yan}},\ }\href@noop {} {\  (\bibinfo {year}
  {2021})},\ \Eprint {http://arxiv.org/abs/2111.14241} {arXiv:2111.14241
  [quant-ph]} \BibitemShut {NoStop}%
\bibitem [{\citenamefont {Jakobi}\ \emph {et~al.}(2017)\citenamefont {Jakobi},
  \citenamefont {Neumann}, \citenamefont {Wang}, \citenamefont {Dasari},
  \citenamefont {El~Hallak}, \citenamefont {Bashir}, \citenamefont {Markham},
  \citenamefont {Edmonds}, \citenamefont {Twitchen},\ and\ \citenamefont
  {Wrachtrup}}]{jakobiNATN17}%
  \BibitemOpen
  \bibfield  {author} {\bibinfo {author} {\bibfnamefont {I.}~\bibnamefont
  {Jakobi}}, \bibinfo {author} {\bibfnamefont {P.}~\bibnamefont {Neumann}},
  \bibinfo {author} {\bibfnamefont {Y.}~\bibnamefont {Wang}}, \bibinfo {author}
  {\bibfnamefont {D.~B.~R.}\ \bibnamefont {Dasari}}, \bibinfo {author}
  {\bibfnamefont {F.}~\bibnamefont {El~Hallak}}, \bibinfo {author}
  {\bibfnamefont {M.~A.}\ \bibnamefont {Bashir}}, \bibinfo {author}
  {\bibfnamefont {M.}~\bibnamefont {Markham}}, \bibinfo {author} {\bibfnamefont
  {A.}~\bibnamefont {Edmonds}}, \bibinfo {author} {\bibfnamefont
  {D.}~\bibnamefont {Twitchen}}, \ and\ \bibinfo {author} {\bibfnamefont
  {J.}~\bibnamefont {Wrachtrup}},\ }\href {\doibase 10.1038/nnano.2016.163}
  {\bibfield  {journal} {\bibinfo  {journal} {Nature Nanotech.}\ }\textbf
  {\bibinfo {volume} {12}},\ \bibinfo {pages} {67} (\bibinfo {year}
  {2017})}\BibitemShut {NoStop}%
\bibitem [{\citenamefont {Yamashita}\ \emph
  {et~al.}(2008{\natexlab{a}})\citenamefont {Yamashita}, \citenamefont
  {Nakazawa}, \citenamefont {Oguni}, \citenamefont {Oshima}, \citenamefont
  {Nojiri}, \citenamefont {Shimizu}, \citenamefont {Miyagawa},\ and\
  \citenamefont {Kanoda}}]{yamashitaNATP08a}%
  \BibitemOpen
  \bibfield  {author} {\bibinfo {author} {\bibfnamefont {S.}~\bibnamefont
  {Yamashita}}, \bibinfo {author} {\bibfnamefont {Y.}~\bibnamefont {Nakazawa}},
  \bibinfo {author} {\bibfnamefont {M.}~\bibnamefont {Oguni}}, \bibinfo
  {author} {\bibfnamefont {Y.}~\bibnamefont {Oshima}}, \bibinfo {author}
  {\bibfnamefont {H.}~\bibnamefont {Nojiri}}, \bibinfo {author} {\bibfnamefont
  {Y.}~\bibnamefont {Shimizu}}, \bibinfo {author} {\bibfnamefont
  {K.}~\bibnamefont {Miyagawa}}, \ and\ \bibinfo {author} {\bibfnamefont
  {K.}~\bibnamefont {Kanoda}},\ }\href {http://dx.doi.org/10.1038/nphys942}
  {\bibfield  {journal} {\bibinfo  {journal} {Nature Phys.}\ }\textbf {\bibinfo
  {volume} {4}},\ \bibinfo {pages} {459} (\bibinfo {year}
  {2008}{\natexlab{a}})}\BibitemShut {NoStop}%
\bibitem [{\citenamefont {Yamashita}\ \emph {et~al.}(2011)\citenamefont
  {Yamashita}, \citenamefont {Yamamoto}, \citenamefont {Nakazawa},
  \citenamefont {Tamura},\ and\ \citenamefont {Kato}}]{yamashitaNATC11}%
  \BibitemOpen
  \bibfield  {author} {\bibinfo {author} {\bibfnamefont {S.}~\bibnamefont
  {Yamashita}}, \bibinfo {author} {\bibfnamefont {T.}~\bibnamefont {Yamamoto}},
  \bibinfo {author} {\bibfnamefont {Y.}~\bibnamefont {Nakazawa}}, \bibinfo
  {author} {\bibfnamefont {M.}~\bibnamefont {Tamura}}, \ and\ \bibinfo {author}
  {\bibfnamefont {R.}~\bibnamefont {Kato}},\ }\href
  {http://dx.doi.org/10.1038/ncomms1274} {\bibfield  {journal} {\bibinfo
  {journal} {Nature Commun.}\ }\textbf {\bibinfo {volume} {2}},\ \bibinfo {pages}
  {275} (\bibinfo {year} {2011})}\BibitemShut {NoStop}%
\bibitem [{\citenamefont {Yamashita}\ \emph
  {et~al.}(2008{\natexlab{b}})\citenamefont {Yamashita}, \citenamefont
  {Nakata}, \citenamefont {Kasahara}, \citenamefont {Sasaki}, \citenamefont
  {Yoneyama}, \citenamefont {Kobayashi}, \citenamefont {Fujimoto},
  \citenamefont {Shibauchi},\ and\ \citenamefont {Matsuda}}]{yamashitaNATP08b}%
  \BibitemOpen
  \bibfield  {author} {\bibinfo {author} {\bibfnamefont {M.}~\bibnamefont
  {Yamashita}}, \bibinfo {author} {\bibfnamefont {N.}~\bibnamefont {Nakata}},
  \bibinfo {author} {\bibfnamefont {Y.}~\bibnamefont {Kasahara}}, \bibinfo
  {author} {\bibfnamefont {T.}~\bibnamefont {Sasaki}}, \bibinfo {author}
  {\bibfnamefont {N.}~\bibnamefont {Yoneyama}}, \bibinfo {author}
  {\bibfnamefont {N.}~\bibnamefont {Kobayashi}}, \bibinfo {author}
  {\bibfnamefont {S.}~\bibnamefont {Fujimoto}}, \bibinfo {author}
  {\bibfnamefont {T.}~\bibnamefont {Shibauchi}}, \ and\ \bibinfo {author}
  {\bibfnamefont {Y.}~\bibnamefont {Matsuda}},\ }\href
  {https://doi.org/10.1038/nphys1134} {\bibfield  {journal} {\bibinfo
  {journal} {Nature Phys.}\ }\textbf {\bibinfo {volume} {5}},\ \bibinfo
  {pages} {44 EP } (\bibinfo {year} {2008}{\natexlab{b}})}\BibitemShut
  {NoStop}%
\bibitem [{\citenamefont {Florens}\ and\ \citenamefont
  {Georges}(2004)}]{florensPRB04}%
  \BibitemOpen
  \bibfield  {author} {\bibinfo {author} {\bibfnamefont {S.}~\bibnamefont
  {Florens}}\ and\ \bibinfo {author} {\bibfnamefont {A.}~\bibnamefont
  {Georges}},\ }\href {\doibase 10.1103/PhysRevB.70.035114} {\bibfield
  {journal} {\bibinfo  {journal} {Phys. Rev. B}\ }\textbf {\bibinfo {volume}
  {70}},\ \bibinfo {pages} {035114} (\bibinfo {year} {2004})}\BibitemShut
  {NoStop}%
\bibitem [{\citenamefont {Gordon~Baym}(1991)}]{baymBOOK91}%
  \BibitemOpen
  \bibfield  {author} {\bibinfo {author} {\bibfnamefont {C.~P.}\ \bibnamefont
  {Gordon~Baym}},\ }\href@noop {} {\emph {\bibinfo {title} {Landau
  Fermi‐Liquid Theory: Concepts and Applications}}}\ (\bibinfo  {publisher}
  {Wiley-VCH Verlag GmbH},\ \bibinfo {year} {1991})\BibitemShut {NoStop}%
\bibitem [{Note3()}]{Note3}%
  \BibitemOpen
  \bibinfo {note} {The prefix `U(1)' in U(1) quantum spin liquid does not
  correspond to any microscopic symmetries of the underlying spin system, which
  we assume throughout to have full SU(2) symmetry. In the spinon Fermi surface
  state, the spin-1/2 operator is expressed using the Abrikosov representation,
  $s_\alpha =\psi ^\protect \dag _{\sigma }\sigma ^\alpha _{\sigma \sigma
  '}\psi _{\sigma '}/2$, which introduces a symmetry under arbitrary local
  phase rotations of the fermions, i.e., $\psi _\sigma \rightarrow e^{i\lambda
  }\psi _\sigma $. This symmetry is identical to the regular gauge symmetry one
  encounters in, e.g., quantum electrodynamics. Requiring physical states to be
  invariant under the associated gauge transformation, Eq.~(\ref {fullS3}) is
  written in an explicitly gauge-invariant form. Since $\lambda $ is an
  arbitrary phase, the state is referred to as a U(1) state.}\BibitemShut
  {Stop}%
\bibitem [{\citenamefont {Polchinski}(1994)}]{polchinskiNPB94}%
  \BibitemOpen
  \bibfield  {author} {\bibinfo {author} {\bibfnamefont {J.}~\bibnamefont
  {Polchinski}},\ }\href {\doibase
  https://doi.org/10.1016/0550-3213(94)90449-9} {\bibfield  {journal} {\bibinfo
   {journal} {Nucl. Phys. B}\ }\textbf {\bibinfo {volume} {422}},\ \bibinfo
  {pages} {617 } (\bibinfo {year} {1994})}\BibitemShut {NoStop}%
\bibitem [{\citenamefont {Mukherjee}\ \emph {et~al.}(2018)\citenamefont
  {Mukherjee}, \citenamefont {Kundu},\ and\ \citenamefont
  {Fertig}}]{mukherjeePRB18}%
  \BibitemOpen
  \bibfield  {author} {\bibinfo {author} {\bibfnamefont {D.~K.}\ \bibnamefont
  {Mukherjee}}, \bibinfo {author} {\bibfnamefont {A.}~\bibnamefont {Kundu}}, \
  and\ \bibinfo {author} {\bibfnamefont {H.~A.}\ \bibnamefont {Fertig}},\
  }\href {\doibase 10.1103/PhysRevB.98.184413} {\bibfield  {journal} {\bibinfo
  {journal} {Phys. Rev. B}\ }\textbf {\bibinfo {volume} {98}},\ \bibinfo
  {pages} {184413} (\bibinfo {year} {2018})}\BibitemShut {NoStop}%
\bibitem [{\citenamefont {Kamenev}(2011)}]{kamenevBOOK11}%
  \BibitemOpen
  \bibfield  {author} {\bibinfo {author} {\bibfnamefont {A.}~\bibnamefont
  {Kamenev}},\ }\href@noop {} {\emph {\bibinfo {title} {Field Theory of
  Non-Equilibrium Systems}}}\ (\bibinfo  {publisher} {Cambridge University
  Press},\ \bibinfo {address} {Cambridge},\ \bibinfo {year} {2011})\BibitemShut
  {NoStop}%
\bibitem [{\citenamefont {Witten}(1984)}]{wittenCMP84}%
  \BibitemOpen
  \bibfield  {author} {\bibinfo {author} {\bibfnamefont {E.}~\bibnamefont
  {Witten}},\ }\href {\doibase cmp/1103940923} {\bibfield  {journal} {\bibinfo
  {journal} {Commun. Math. Phys.}\ }\textbf {\bibinfo {volume} {92}},\ \bibinfo
  {pages} {455} (\bibinfo {year} {1984})}\BibitemShut {NoStop}%
\end{thebibliography}
\end{document}